\documentclass[11pt,english,twoside]{article}

\usepackage[T1]{fontenc}
\usepackage[latin1]{inputenc}
\usepackage[english]{babel}
\usepackage{lmodern}
\usepackage{amsmath,amssymb}
\usepackage{float}
\usepackage{graphicx}
\usepackage[dvips]{epsfig}
\usepackage{psfrag}

\usepackage{lscape}
\usepackage[all]{xy}

\usepackage{hyperref}

\newcommand{\beq}{\begin{equation}}
\newcommand{\eeq}{\end{equation}}
\newcommand{\bea}{\begin{eqnarray}}
\newcommand{\eea}{\end{eqnarray}}
\newcommand{\nn}{\nonumber}
\newcommand{\w}{\wedge}
\newcommand{\del}{\partial}
\newcommand{\aaa}{{\cal A}}
\newcommand{\bbb}{{\cal B}}

\newcommand{\fff}{{\cal F}}
\newcommand{\mmm}{{\cal M}}

\newcommand{\ov}{\overline}
\newcommand{\wt}{\widetilde}

\DeclareMathOperator{\re}{Re}
\DeclareMathOperator{\im}{Im}

\newcommand{\Gg}{\mathfrak{g}}
\newcommand{\Gh}{\mathfrak{h}}
\newcommand{\Gn}{\mathfrak{n}}

\def\d {{\rm d}}

\makeatletter
\@addtoreset{equation}{section}
\makeatother

\begin{document}

\begin{titlepage}
\setcounter{page}{0}
\begin{center}

\rightline{\small IPhT-T10/022}

\vskip 2cm

\begin{LARGE}
   \textbf{Supersymmetry breaking branes on solvmanifolds}
   \vskip 0.2cm
   \textbf{  and }
   \vskip 0.3cm
  \textbf{de Sitter vacua in string theory}
\end{LARGE}

\vskip 1.2cm

\textbf{David Andriot$^{a}$, Enrico Goi$^{b}$, Ruben Minasian$^{b}$, Michela Petrini$^{a}$}

\vskip 0.8cm
{}$^a$ \textit{LPTHE, CNRS, UPMC Univ Paris 06\\
Bo\^ite 126, 4 Place Jussieu\\
F-75252 Paris cedex 05, France}\\
\vskip 0.4cm
{}$^{b}$\textit{Institut de Physique Th\'eorique, CEA/Saclay \\
91191 Gif-sur-Yvette Cedex, France}\\
\vskip 0.4cm
andriot@lpthe.jussieu.fr, enrico.goi@cea.fr,\\
ruben.minasian@cea.fr, petrini@lpthe.jussieu.fr

\end{center}

\vskip 1cm

\begin{center}
{\bf Abstract}
\end{center}

We consider IIA compactifications on solvmanifolds with O6/D6 branes and study the conditions for obtaining de Sitter vacua in ten dimensions. While this
 is a popular set-up for searching de Sitter vacua, we propose a new method to include supersymmetry breaking sources. For space-time filling branes
 preserving bulk supersymmetry, the energy density can easily be extremized with respect to all fields, thanks to the replacement of the DBI action
 by a pullback of a special form given by a pure spinor. For sources breaking bulk supersymmetry, we propose to replace the DBI action by the pullback of a more
 general polyform, which is no longer pure. This generalization provides corrections to the energy-momentum tensor which give a positive   
contribution to the cosmological constant. We find a de Sitter solution to all (bulk and world-volume) equations derived from this action. We argue it
 solves the equations derived from the standard source action. The paper also contains a review of solvmanifolds.

\vfill

\end{titlepage}

\tableofcontents

\newpage

\section{Introduction}

In recent years much progress has been achieved in the classification and construction of supersymmetric
flux compactifications \cite{Flux}. This is largely due to the fact that, at least for type II supergravities, supersymmetry allows
to look at  first order differential equations, which together with the Bianchi identities for the fluxes, imply
the solutions of the ten-dimensional equations of motion \cite{LT, GMSW, KT}.  A natural frame to analyse
backgrounds with ${\mathcal N}=1$ supersymmetry  is provided by
generalized complex geometry (GCG) \cite{Hit, G},  which was developed concurrently with the progress on
the physics side of  the problem. In this formalism,
the natural variables are a certain combination of globally defined even or odd differential forms, $\Phi_\pm$,
called pure spinors, and the
supersymmetry conditions amount to a set of differential equations for such spinors.
For ${\mathcal N}=1$ compactifications
to four--dimensional Minkowski space these are
\bea
\label{eq:susyeqIIintro}
&& {\rm d}_H (e^{2A - \phi} \Phi_1)= 0 \, , \nn\\
&& {\rm d}_H (e^{A -\phi} \re \Phi_2) =  0 \, ,\nn\\
&& {\rm d}_H (e^{3A -\phi} \im \Phi_2) = \frac{|a|^2}{8} e^{3A} \ast \lambda(F)  \, .
\eea
$\Phi_1=\Phi_{\pm}$ and $  \Phi_2=\Phi_{\mp}$  for IIA/IIB , where $+$ and $-$ denote even and odd forms,
respectively.  $\phi$ is the dilaton and  $|a|^2=||\Phi_{\pm}||$ is the norm of the pure spinors\footnote{\label{Hodge}
To define the norm of the pure spinors
we introduce the Mukai pairing of two polyforms as the top form:
\beq
\langle X_1,X_2 \rangle=(X_1\w \lambda(X_2))|_{{\rm top}} \ ,
\eeq
where $\lambda$ acts on any $p$-form $A_p$  as the complete reversal of its indices:
$\lambda(A_p)=(-1)^{\frac{p(p-1)}{2}}A_p$. Then we can define the norm of $\Phi_\pm$ as
$8 \langle \Phi_{\pm},\ov{\Phi}_{\pm} \rangle=-i||\Phi_{\pm}||^2 {\rm vol}$.
We take the following convention for the Hodge star:
\beq *(\d x^{\mu_1} \w ... \w \d x^{\mu_p})= \frac{\sqrt{|g|}}{(d-p) !} (-1)^{(d-p)p}\ \epsilon^{\mu_1 .. \mu_p\ \mu_{p+1} .. \mu_d}\ g_{\mu_{p+1} \nu_{p+1}} .. g_{\mu_d \nu_d}\ \d x^{\nu_{p+1}} \w ... \w \d x^{\nu_d} \ , \eeq
with $d$ the dimension of the space, $|g|$ the determinant of the metric. For $\epsilon$ we choose the convention
$\epsilon_{1 \dots d}=1$.}, which is fixed to  $|a|^2=e^A$.
$F$ denotes the sum of the RR fluxes on the internal manifold
\bea
\textrm{IIA}&:&\ F=F_0+F_2+F_4+F_6 \ , \\
\textrm{IIB}&:&\ F=F_1+F_3+F_5 \ ,
\eea
and is related to the total ten-dimensional
RR field-strength $F^{(10)}$ by
\beq
\label{RRans}
F^{(10)}=F+\textrm{vol}_{(4)}\w \lambda(*F) \ ,
\eeq
where $\textrm{vol}_{(4)}$ is the warped four-dimensional volume form with warp factor
$e^{2A}$. The NS flux $H$ enters the equations through the differential ${\rm d}_H=  {\rm d}-  H\wedge$. \\

The first equation in \eqref{eq:susyeqIIintro} requires the existence on the manifold of a
closed pure spinor, and
the integrability of the associated generalized complex structure \cite{GMPT4,GMPT5}.
Spaces admitting a closed pure spinor are generalized Calabi-Yau (GCY).
This is, therefore, a necessary condition for preserving supersymmetry.
In addition, we should  require the existence of a second compatible pure spinor\footnote{The existence of
a pure spinor reduces the structure group on
$TM \oplus T^*M$ to SU(3,3). If the manifold admits a second compatible pure spinor
the structure group is further reduced to SU(3)$\times$SU(3).} whose real part is closed, and whose imaginary part is the RR field.
The metric in the internal space is determined by the two pure spinors.

One can see easily that the RR equations of motion
automatically follow from the supersymmetry conditions, provided that no NS source is present ($\d H=0$). Differentiating the last equation in \eqref{eq:susyeqIIintro},
one indeed recovers the RR flux equations of motion
\beq
\label{rreom}
(\d +H \wedge ) (e^{4A} *F)=0 \ .
\eeq

In a sense, up to this point, finding a supersymmetric string
background is a perfectly algorithmic procedure. Indeed, starting from a
generalized CY structure, i.e. a twisted closed pure spinor, one has to
find a second compatible pure spinor, and calculate the RR flux by acting on
the latter with $(\d-H\wedge)$. In order to promote a configuration satisfying the supersymmetry conditions to a
full solution, one has to check the Bianchi Identities (BI) for all  fluxes
\bea
\label{eq:BIeqII}
&& {\rm d}(F)=   \delta(\mathrm{source}) \, , \nn\\
&& \d H=0 \, .
\eea
Because of tadpole cancellation, the sources charged with respect to RR fields need to have an overall negative tension,
and hence the dominant charge is that of an O-plane. This is the final step in the search for $\mathcal{N}=1$
vacua on Minkowski. For $AdS_4$ supersymmetric solution, a similar procedure can be defined.

\vskip .5cm
\noindent
For non-supersymmetric backgrounds, the situation is much more complicated, since, a priori, first order equations
such as (\ref{eq:susyeqIIintro}) are not anymore valid. Recently a procedure has been proposed in \cite{LMMT}
that generalizes to non-supersymmetric backgrounds the first order pure spinor equations \eqref{eq:susyeqIIintro}.
The idea of \cite{LMMT} is to decompose the supersymmetry breaking terms in \eqref{eq:susyeqIIintro}
in the Spin(6,6) basis constructed from the pure spinors. For instance, for Minkowski compactifications, the modified
first order equations are
\bea
\label{eq:nonsusyeqIIintro}
&& {\rm d}_H (e^{2A - \phi} \Phi_1)=  \Upsilon \, , \nn\\
&& {\rm d}_H (e^{A -\phi} \re \Phi_2) = \re \Xi   \, ,\nn\\
&& {\rm d}_H (e^{3A -\phi} \im \Phi_2) - \frac{|a|^2}{8} e^{3A} \ast \lambda(F) = \im \Xi \, ,
\eea
where schematically
\bea
\Upsilon &=&  a_0 \Phi_2 +\wt{a}_0 \ov{\Phi}_2 + a^1_m \gamma^m \Phi_1 + a^2_m  \Phi_1 \gamma^m +
 \wt{a}^1_m \gamma^m \ov{\Phi}_1 + \wt{a}_m^2  \ov{\Phi}_1 \gamma^m  \nn \\
&& + a_{mn} \gamma^m \Phi_2 \gamma^n  + \wt{a}_{mn}  \gamma^n \ov{\Phi}_2 \gamma^m \, , \\
\Xi &=&  b_0 \, \Phi_1 +\wt{b}_0  \, \ov{\Phi}_1 + b^1_m \gamma^m \Phi_2 + b^2_m  \Phi_2 \gamma^m
+ b_{mn} \gamma^m \Phi_1 \gamma^n  + \wt{b}_{mn}  \gamma^n \ov{\Phi}_1 \gamma^m \, .
\eea

In the particular case of an SU(3) structure, this decomposition is equivalent to the expansion of
(\ref{eq:susyeqIIintro}) in the SU(3) torsion classes.

Equations \eqref{eq:nonsusyeqIIintro} rely on the assumption that the four-dimensional space-time admits
Killing spinors and that the supersymmetry breaking is due to the internal spinors only.
This applies of course to Minkowski and Anti de Sitter backgrounds, but not for  de Sitter solutions or cases when
supersymmetry is broken in four-dimensions. \\

The purpose of the paper is twofold. On one side, we would like to make some first steps towards determining
a set of first order equations also for configurations where four-dimensional supersymmetry is broken.
In particular, we shall propose a first order equation similar to the last equation in \eqref{eq:susyeqIIintro}, 
so that the flux equations of motion \eqref{rreom} follow automatically. On the other side,
we would like to reexamine the problem of finding de Sitter vacua directly in ten-dimensions and focus
only on simple conservative compactifications (i.e.  ``geometric'' set-up).


We will consider de Sitter vacua in IIA supergravity. In this context, several no-go theorems
and ways of circumventing them have been
proposed \cite{MN, IRW, HKTT, Si, HSUV, DHSV, CKKLWZ, FPRW, DKV}. In particular, in presence of O6/D6 sources, a minimal requirement to evade the no-go theorem \cite{HKTT} is to have a negatively curved internal manifold and a non-zero $F_0$ (Romans mass parameter) \cite{MN, HSUV, CKKLWZ}. Therefore, we will focus on type IIA configurations with non-zero NS three-form and  RR zero and two-forms. Moreover,
we assume that all the sources (there may be intersecting ones)
are space-time filling and are of the same dimension $p=6$.

Tracing the four-dimensional part of Einstein equation and using the dilaton equation of motion, one can show that the four-dimensional curvature and the ``source term'' can be written as
\bea
\label{eq:curv1}
R_4&=&\frac{2}{3}\left[-R_6-\frac{g_s^2}{2}|F_2|^2+\frac{1}{2}(|H|^2-g_s^2 |F_0|^2)  \right] \, ,\\
\label{eq:curv2}
g_s \frac{T_{0}}{p+1} &=&\frac{1}{3} \left[-2R_6+|H|^2+2 g_s^2 (|F_0|^2 +|F_2|^2)\right] \, ,
\eea
where, for simplicity, we have taken constant dilaton, $e^\phi = g_s$,
and no-warping\footnote{In general, with non-trivial dilaton and a ten-dimensional metric of the form
\beq
ds^{2}=e^{2A(y)}g_{\mu\nu}(x) \d x^{\mu}\d x^{\nu}+g_{mn}(y) \d y^{m} \d y^{n} \, ,
\eeq
equation  \eqref{eq:curv1} becomes
\bea
e^{-2A}R_{4}&=&\frac{2}{3} [-R_{6}-\frac{e^{2\phi}}{2}|F_{2}|^{2}+\frac{1}{2}(|H|^{2}-e^{2\phi}|F_{0}|^{2}) ] \nn\\
 &&-8\nabla^{2} A+20|\partial_{m}A|^{2}-\frac{8}{3}\nabla^{2} \phi+\frac{8}{3}|\partial_{m}\phi|^{2}-\frac{32}{3}g^{mn}\partial_{m}A\partial_{n}\phi \,.
 \eea
All derivatives are taken with respect to the coordinates on $M$.
We shall return to the discussion of the warp factor and the dilaton.}.

Further simplifications are possible when one assumes that the sources preserve the supersymmetry of the bulk; this condition is usually expressed in terms of an equation involving the bulk supersymmetry parameters and  the world-volume chiral operator entering the $\kappa$-symmetry transformations. Up to terms quadratic in the $\kappa$-symmetry condition, one can always rewrite the brane world-volume action
 as the pullback of the non-integrable pure spinor
\bea
\label{calibr}
\left(i^*[\im\Phi_2] \w e^{\fff}\right) = \frac{|a|^2}{8}\sqrt{|i^*[g] + \fff|} \d^\Sigma x \, ,
\eea
where $i$ denotes the embedding of the world-volume into the internal manifold $M$, $g$ is the internal metric and $\fff$ is the gauge invariant combination of the field strength of the world-volume gauge field and the pullback of $B$. For sources preserving the supersymmetry of the bulk, one can then replace the DBI action by the left-hand side of
 (\ref{calibr}). The equations of motion derived  from both actions are the same, since the corrections would be linear in the $\kappa$-symmetry condition, and then vanishing in the supersymmetric case. In particular,
one can show that the world-volume equations of motion are then automatically implied by the last equation of (\ref{eq:susyeqIIintro}). So the
condition (\ref{calibr}) together with the last equation of (\ref{eq:susyeqIIintro}) give (generalized) calibrated sources, i.e. their energy density
 is minimized \cite{K, MS, KM, KT}.

For such supersymmetric configurations, the four- and six-dimensional traces of the source energy-momentum tensor and the source term in the dilaton equation are all proportional to each other, and one arrives at
\bea
&& R_4=\frac{2}{3} (g_s^2|F_0|^2-|H|^2
) \ , \label{curv46}\\
&& R_6+\frac{1}{2}g_s^2 |F_2|^2 + \frac{3}{2}(g_s^2|F_0|^2-|H|^2)
 =0 \, ,
\eea
together with eq. \eqref{eq:curv2}. The last equation is just a constraint on internal quantities, while the two others fix $R_4$ and the source term $T_0$.  From these two equations we recover  the minimal requirement of having $F_0 \neq 0$ and $R_6 <0$. In practice, however, this is not enough to find a de Sitter vacuum. In particular, we can see that $F_0$ alone can give a positive value to the cosmological constant, and adding more fluxes, $F_4$ and $F_6$,  does not help since they give negative contributions.  Indeed, up to date, all known examples of stable
de Sitter vacua require some additional ingredients such as KK monopoles and Wilson lines \cite{Si}, non-geometric fluxes  \cite{CGM}, or $\alpha^\prime$ corrections and probe D6 branes \cite{PTW}.
In this paper, we will work in ten dimensions and mainly focus on classical geometric compactifications. \\

Since we are interested in non-supersymmetric backgrounds, there is a priori no reason to impose that the sources preserve the bulk
 supersymmetry. The condition (\ref{calibr}) could therefore be violated. To do so, we make in this paper the following proposal: we replace
 (\ref{calibr}), in analogy with (\ref{eq:nonsusyeqIIintro}) where the violation of the bulk supersymmetry conditions is encoded in the general
 polyforms $\Upsilon$ and $\Xi$, by
\bea
\label{calibr-n}
\left(i^*[\im X_-] \w e^{\fff}\right)= \sqrt{|i^*[g] + \fff|} \d^\Sigma x  \, ,
\eea
where $X_-$ is an odd polyform given by a general expansion similar to $\Upsilon$.
For supersymmetric configurations, $X_-$ reduces to $8 \Phi_-$, but in general, it is no longer a pure spinor. As discussed, such a replacement for sources
preserving bulk supersymmetry is correct up to quadratic terms in the $\kappa$-symmetry condition, and corrections to the equations of motion derived
 from it will vanish linearly if the condition holds. In our case the structure of the corrections is not explicit, and we cannot conclude that the equations
 of motion derived from left-hand side of (\ref{calibr-n}) are the same as those derived from DBI. We will thus proceed as follows: we first find solutions
 using the equations derived from the left-hand side of (\ref{calibr-n}) and then we will check whether these are solutions to the equations derived from the
 standard DBI action.

An advantage of replacing DBI by the pullback of a form from the bulk is that it is actually easier to take the variation with respect to the various
 fields, in particular the bulk ones. Moreover the variation of the left-hand side of (\ref{calibr-n}) with respect to the metric will lead to interesting consequences for de Sitter solutions:  new terms are generated in the
 energy momentum tensor which help to lift the cosmological constant to positive values (see further in (\ref{nonsusyR4})). This is the main 
motivation for using this proposal, but a full understanding of it should be provided in future work. One possible interpretation is that such sources could be thought as standard D-branes or O-planes but their embedding into space-time
 (here into $M$) is modified. While for supersymmetric configurations the geometry of the subspace wrapped by the source is encoded in $\im\Phi_2$,
 here it would be encoded in the more general expansion $\im X_-$, of which $\im\Phi_2$ is only one possible term. Therefore, the breaking of
bulk supersymmetry seems to come from allowing more general geometries for the wrapped subspaces, and the new terms in the energy momentum tensor could
come from the non standard embedding, in particular a dependence of the embedding functions on the metric moduli.

Since the bulk supersymmetry is broken, we could as well modify (\ref{eq:susyeqIIintro}) and, in view of (\ref{calibr-n}), we propose here the following generalization of the
first order conditions:
\bea
&& {\rm d}_H(e^{2A -\phi}  \re X_- ) =  0 \, ,\\
&& {\rm d}_H(e^{4A -\phi} \im X_-  ) =  c_0 e^{4A} \ast \lambda(F)  \, , \label{xxeom}
\eea
where $c_0 $ is a positive constant fixed by the parameters of the solution. Hence the introduction of $X_-$ allows,
as for the supersymmetric case, to trade the RR equations of motion 
for first order equations (clearly \eqref{rreom} follows by differentiating \eqref{xxeom}), while, in addition, it
helps via (\ref{calibr-n}) to solve the internal Einstein equation. This is a first step towards developing a more
systematic procedure to find non-supersymmetric backgrounds.


 For the NSNS fields, we will check explicitly that our solution is a solution to the equations of motion derived from DBI, making use of a
 dependence of the embedding functions on the metric moduli. What remains are the world-volume
fields (note the $\fff$ will be trivial for us). Let us comment on their equations. As mentioned previously, for sources preserving bulk supersymmetry,
 a world-volume equation of motion, obtained by varying \eqref{calibr} augmented by the WZ terms,
 turns out to follow simply from a partial pullback of the bulk pure spinor equation \eqref{eq:susyeqIIintro}. Then the minimization
 of the world-volume energy is automatic \cite{K,MS}. The equations of motion derived from the left-hand side of \eqref{calibr-n} should also be compared
 with the partial pullback of \eqref{xxeom}. We shall denote the transverse differentiation  by $\partial_{\alpha}$ and a flux with all but one index pulled
 back to the world-volume  by $i^*[F]_{\alpha}$. Neglecting the world-volume gauge fields, we can write the resulting equation as
\beq
\partial_{\alpha} \left(i^*[e^{4A -\phi} \im (e^{-B} X_-  )]\right) -  i^*[e^{4A}e^{-B} \ast \lambda(F)]_{\alpha}=0 \ .
\eeq
Comparison with the components of \eqref{xxeom} gives
\beq
 (c_0-1)\ i^*[e^{-B} \ast \lambda(F)]_{\alpha}=0 \ . \label{condstab}
 \eeq
In the supersymmetric case, where we replace $X_-$ by $\Phi_-$, $c_0=1$ and the equation is automatically satisfied. Here we will consider solutions with
 a vanishing partial pullback $i^*[e^{-B} \ast \lambda(F)]_{\alpha}$, so the world-volume equations derived from the left-hand side of \eqref{calibr-n}
will be satisfied, making the energy of our sources extremized. We will also check that our solution satisfies the equations of motion obtained by
 the variation of the standard DBI+WZ action.

\vskip .5cm
\noindent
The strategy  to find a non-supersymmetric solution to our proposed action is the following. We start with
one particular solution to \eqref{eq:susyeqIIintro}, which is a supersymmetric compactification of IIA on a solvmanifold labeled $s \, 2.5$.
The solvable  algebra is given by ($q_1 35, q_2 45, -q_2 15, - q_1 25, 0, 0$). The solution involves intersecting O6 planes
(and possibly
 D6 branes - depending on the choice of parameters). Due to the general problems in constructing localized intersecting
branes, the sources are smeared, and hence the model would suffer from general criticism \cite{DK}. It does have some
convenient features though, and it serves as a good illustration to the method we would like to propose. We shall
return to the question of localization and to the possibility of incorporating warp factor and non-constant dilaton
in the paper. For now, without further apologies, we shall use  the $s \, 2.5$ model as a point of
departure for our non-supersymmetric construction.

$s \, 2.5$ is a special case of a more general solvable algebra ($ q_1 (p 25 + 35), q_2 (p 15 + 45), q_2 (p 45 -15),
q_1 (p 35 - 25), 0,0$). It is then natural to see whether the corresponding solvmanifold also admits
solutions. A natural ansatz would be generalizations of  the supersymmetric solutions on $s \, 2.5$.
To do so, we shall extend to solvmanifolds the twist transformation worked out in \cite{AMP} for nilmanifolds. It turns out that the first two equations of (\ref{eq:susyeqIIintro}) can be satisfied for $p\neq 0$ provided a certain combination of moduli, which we call $\lambda$, takes value  $1$.
In other words, for generic $p$ and $\lambda = 1$ we find supersymmetric solutions (corresponding
to a vanishing four-dimensional curvature). For generic $\lambda$, the pure spinor equations are not satisfied and
supersymmetry is broken. It is certainly of great practical importance to have a convenient limit in which our construction can be tested.

The proposed source action (\ref{calibr-n}) allows to rewrite (\ref{curv46}) for the four dimensional Ricci tensor as
\beq
\label{nonsusyR4}
R_{4}=\frac{2}{3}\left(\frac{g_s}{2} (T_0-T) +g_s^2 |F_{0}|^2 - |H|^2 \right) \ ,
\eeq
where the source term $T_0$ is different from the trace of the energy momentum tensor $T$.  As can be seen from (\ref{eq:curv2}), $T_0$ gives a positive contribution to $R_4$ and in our case, it turns out that $T_0-T$ is also positive.
Thus, with our proposal \eqref{calibr-n} we are indeed able to find a ten-dimensional de Sitter solution. Checking that it also satisfies the equations
of motion derived from the standard source action (with a dependence of the embedding functions on the metric moduli) will make it a solution of type IIA supergravity.

\vskip .5cm
\noindent
The details of the solution, as well as the treatment of our proposal for supersymmetry breaking branes, are presented in
Section 3. This discussion is complemented by the analysis of the four-dimensional effective potential. In particular,
we will discuss how the supersymmetry breaking proposal for the sources provides new terms in the potential. Also we will
perform an analysis of stability of the solution in the volume and dilaton moduli.

While our discussion for de Sitter solution is based on a specific example, the construction is more general,
and we present much of the
technical machinery in Section 2. This contains a discussion of supersymmetric solutions, and the twist construction
of solvmanifolds which serve as internal spaces. The construction has been used previously for nilmanifolds (which are
an iteration of torus bundles over a base manifold being a torus itself) \cite{AMP}. It is extended here to the case of
 solvable algebras. Our basic example is based on a solvable group that admits a lattice and hence yields a compact
six-dimensional solvmanifold. As we shall see the construction can be applied also to algebras that admit no such lattice,
 and it may lead to non-geometric backgrounds. A more formal presentation of the solvable algebras and the geometry of
 (compact) solvmanifolds is given in Appendix \ref{apmath}.
In Appendix \ref{T-dualsolv} we discuss some global aspects of T--duality on solvmanifolds.

\section{Supersymmetric backgrounds, solvmanifolds and twist transformations}\label{susyback}

In this paper we are interested in string backgrounds where the internal compactification manifold is a
solvmanifold.
Nil- and solvmanifolds have been extensively used in type II compactifications, both to four-dimensional
Minkowski or Anti de Sitter,
and appear to be good candidates for possible de Sitter vacua as well.
Indeed their geometry is pretty well understood (for instance all nilmanifolds are generalized Calabi-
Yaus \cite{CG}) and, in particular, they can have negative curvature and therefore support internal fluxes
(as well as  D-branes and O-plane sources). \\

Nil- and solvmanifolds are homogeneous spaces constructed from
nilpotent or solvable groups $G$, nilpotent being actually a particular case of solvable. When the group $G$ is
not compact, the manifold can be made compact by  quotienting  $G$ by a lattice $\Gamma$, i.e. a discrete
co-compact subgroup of $G$. The dimension of the resulting manifold\footnote{This definition of solvmanifold it is not the most general:
 one could consider cases where the
$d$-dimensional solvmanifold is the quotient of a higher dimensional group with a continuous
subgroup $\Gamma$. This is the case for the Klein bottle, for instance.} is the same as that of the group
$G$.
In this paper we will focus on manifolds of dimension six. It can be proven \cite{M} that a lattice
$\Gamma$ can always be found for nilmanifolds, while for solvmanifolds its existence is harder to establish.
We refer to Appendix  \ref{apmath} for a detailed discussion of the algebraic
aspects and the compactness properties of nil and solvmanifolds. Here, we focus on their geometry.

Given a $d$-dimensional Lie algebra $\Gg$ expressed
in some vector  basis $\{E_1,\ \dots,\ E_d \}$ as
\beq
[E_b, E_c]= f^a_{\ \ bc} E_a \ ,
\eeq
where $f^a_{\ \ bc}$ are the structure constants, we can define the dual space of one-forms $\Gg^*$
 with basis $\{e^1,\ \dots,\ e^d\}$. They satisfy the Maurer-Cartan equation
\beq
\label{MC}
{\rm d}e^a =-\frac{1}{2} f^a_{\ \ bc} e^b\w e^c = - \sum_{b<c} f^a_{\ \ bc} \ e^b \wedge e^c\ ,
\eeq
with the exterior derivative ${\rm d}$.  Since $\Gg^{*}\approx T_{e}G^{*}$, $\{e^1,\ \dots,\ e^d\}$
provide, by left invariance, a basis for the cotangent space $T_{x}G^{*}$ at every point $x\in G$ and,
thus, are globally defined one-forms on the manifold.
When the manifold is obtained as a quotient with a lattice $\Gamma$, the one-forms will have non
trivial identification through the lattice action\footnote{\label{footchev} In general there is a natural inclusion
$(\Lambda\Gg^{*},\delta)\rightarrow(\Lambda(G/\Gamma),d)$ between the
Chevalley-Eilenberg complex on $G$ and the de Rham complex of differential forms on $G/\Gamma$.
This inclusion induces an injection map between cohomology groups $H^{*}(\Gg)\rightarrow H_{dR}^{*}(G/\Gamma)$
which turns out to be an isomorphism for completely solvable groups. We recall that a Lie group $G$ with Lie
algebra $\Gg$ is said to be completely solvable if the linear map $ad_X:\Gg\rightarrow\Gg$ only has real roots
$\forall X\in\Gg$. Note that all nilmanifolds are completely solvable and thus the injection is an isomorphism
(Nomizu's theorem \cite{N}), the extension to non-nilpotent completely solvable groups being the so-called Hattori
theorem \cite{OT}. For more details and for a list of Betti numbers of solvmanifolds up to dimension six
see \cite{B}.}.
Nil and solvmanifolds, as we define them in this paper, are always parallelizable \cite{OT}, even if
they are not necessarily Lie groups.

The Maurer-Cartan equations reflect the topological structure of the corresponding manifolds. For example,
nilmanifolds all consist of
iterated  fibrations of circles over tori, where the iterated structure is related to the descending or
ascending series of the algebra (see \cite{CG,B,R}). This can be easily seen on a very simple example, the
nilmanifold obtained from the three-dimensional Heisenberg algebra
\beq
[E_2, E_3]= E_1    \qquad \Leftrightarrow  \qquad \d e^1=-e^2\w e^3 \, .
\eeq
The  Maurer-Cartan equation is solved by the one-forms
\beq
e^1=\d x^1 -x^2 \d x^3 \ ,\ e^2=\d x^2,\ e^3=\d x^3 \ .
\eeq
From the connection form, $-x^2 \d x^3$ , one can read the topology of the nilmanifold in question, which is
a non-trivial fibration of the circle in direction $1$ on the two-torus in directions $2,3$:
\beq
\begin{array}{ccc}
 S_{\{1\}}^1  & \hookrightarrow & H/\Gamma_1 \\
 & & \downarrow \\
 & & T_{\{23\}}^2 \end{array}
\eeq

Solvmanifolds are classified according to the dimension of the  nilradical $\Gn$ (the largest nilpotent ideal)
of the corresponding algebra.
In six dimensions, $\Gn$ can have dimension from 3 to  6.
If $\textrm{dim}\ \Gn= 6$, then $\Gn=\Gg$ and the algebra is nilpotent.  At the level of the group\footnote{We denote by $\Gn$ the ideal in the algebra and with $N$ the corresponding subgroup.} we have that,
if $\textrm{dim}  N < 6$, then $G$ contains an abelian subgroup of dimension $k$ \cite{Au,CS}.
This means we have
 $G/N=\mathbb{R}^k$. When the group admits a lattice $\Gamma$, one can show that $\Gamma_N= \Gamma \cap N$
is a lattice in $N$, $\Gamma N=N \Gamma$ is a closed subgroup of $G$,
and so $G/(N\Gamma)=T^k$ is a torus. The solvmanifold is a
non-trivial fibration of a nilmanifold over the torus $T^k$
\beq
\begin{array}{ccc} \label{Mostowbundle}
N/\Gamma_N =(N\Gamma)/ \Gamma & \hookrightarrow & G/\Gamma \\
 & & \downarrow \\
 & & T^k=G/(N\Gamma) \end{array}
\eeq
This bundle is called the Mostow bundle \cite{Mostow}. As we shall see, the corresponding fibration can be more complicated than
in the nilmanifold case. In general, Mostow bundles are not  principal. \\

In the following we will restrict to almost abelian solvable groups,  for which the construction of the
Mostow bundle is particularly simple.
Consider first almost nilpotent solvable groups. These are solvable groups that have nilradical of
dimension $\textrm{dim}\ N=\textrm{dim}\ G -1$. As discussed in Appendix \ref{apalg}, the group is then given by the semi-direct product
\beq
G=\mathbb{R} \ltimes_{\mu} N
\eeq
of its nilradical with $\mathbb{R}$, where $\mu$ is some action on $N$ depending on the direction
$\mathbb{R}$
\beq
 (t_1,n_1) \cdot (t_2,n_2) = (t_1 \cdot t_2, n_1\cdot \mu_{t_1}(n_2))
\qquad \forall  (t,n )\in \mathbb{R} \times N \ .
\eeq
In general, we label by $t$ the coordinate on  $\mathbb{R}$ and by $\partial_t$ the corresponding
vector of the algebra.
From a geometrical point of view, $\mu(t)$ encodes the fibration of the Mostow bundle.

An almost abelian solvable group is an almost nilpotent group whose nilradical is abelian
\beq
N=\mathbb{R}^{\textrm{dim}G -1} \, .
\eeq
In this case, the action of $\mathbb{R}$ on $N$ is given by
\beq
\mu(t)=Ad_{\partial_t }(\Gn)=e^{t\ ad_{\partial_t}(\Gn)}.
\eeq
Another nice feature of almost abelian solvable groups is that
a simple criterion exists to determine whether the associated solvmanifold is compact:
the group admits a lattice if and only if there exists a $ t_0 \neq 0$ for which $\mu(t_0)$ can be conjugated
to an integer matrix. \\

As an example, we can consider two three-dimensional almost abelian solvable algebras
\bea
\varepsilon_2  &: &     [ E_2, E_3 ] = E_1\quad \qquad \Leftrightarrow \qquad
\d e^1 = - e^2 \w e^3  \nn \\
& & [E_1, E_3 ] = - E_2  \, \, \qquad \Leftrightarrow \qquad   \d e^2 = e^1 \w e^3  \, \\
& &   \nn \\
\varepsilon_{1,1}  &: &
[E_1, E_3] = E_1   \quad  \qquad \Leftrightarrow \qquad
\d e^1 = - e^1 \w e^3 \nn \\
& & [E_2,E_3] = -E_2    \,\,  \qquad \Leftrightarrow \qquad  \d e^2 =  e^2 \w e^3 \, . \label{E1basis2}
\eea
In the following, we will label the algebras according to their Maurer-Cartan equations. For instance,
$\varepsilon_2$  is denoted by $(-23,13,0)$, where each entry $i$ gives the result of $\d e^i$.

For the algebra $\varepsilon_2 : (-23,13,0)$, the nilradical is given by $\Gn = \{E_1,E_2\}$ and
$\partial_t=E_3$. In this basis, the restriction of the adjoint representation to the nilradical is
\beq
ad_{\partial_t}(\Gn)=\left( \begin{array}{cc} 0 & -1 \\ 1 & 0 \end{array} \right) \ ,
\eeq
which gives a $\mu$ matrix of the form
\beq
\mu(t)=e^{t\ ad_{\partial_t}(\Gn)}=\left( \begin{array}{cc} \cos(t) & -\sin(t) \\ \sin(t) & \cos(t) \end{array} \right) \ .\label{example1}
\eeq
It is easy to see that, for $t_0=n\frac{\pi}{2}$, with $n \in \mathbb{Z}^*$, $\mu(t_0)$ is an integer
matrix and hence the corresponding manifold is compact.

For the algebra $\varepsilon_{1,1} : (-13, 23, 0)$ the analysis is less straightforward. The nilradical is
$\Gn= \{E_1, E_2 \}$ and again $\partial_t=E_3$. Then, in the $(E_1,E_2)$ basis,
\beq
ad_{\partial_t}(\Gn)=\left( \begin{array}{cc} -1 & 0 \\ 0 & 1 \end{array} \right) \ , \
\mu(t)=e^{t\ ad_{\partial_t}(\Gn)}=\left( \begin{array}{cc} e^{-t} & 0 \\ 0 & e^t \end{array} \right) \ , \label{example2}
\eeq
and it is clearly not possible to find a $t_0 \neq 0$ such that $\mu(t_0)$ is an integer. To see whether
the group admits a lattice, we then have to go to another basis. In other words, $\mu(t_0)$ will be conjugated to an
integer matrix.  As in  \cite{HSUV}, we can define a new basis
\beq
E_1 \rightarrow \sqrt{\frac{q_2}{q_1}} \frac{E_1-E_2}{\sqrt{2}}  \ , \ E_2 \rightarrow \frac{E_1+E_2}{\sqrt{2}}  \ , \ E_3 \rightarrow \sqrt{q_1 q_2} \,  E_3 \ , \label{isomo}
\eeq
with $q_1,q_2$ strictly positive constants, such that the algebra reads
\beq
[E_1,E_3]=q_2 E_2 \qquad \quad [E_2,E_3]=q_1 E_1 \, .  \label{apE1basis2}
\eeq
In this new basis
\beq
\label{e1adj}
ad_{\partial_t}(\Gn)=\left( \begin{array}{cc} 0 & -q_1 \\ -q_2 & 0 \end{array} \right) \ , \
\mu(t)=\left( \begin{array}{cc} \cosh (\sqrt{q_1 q_2} t) & -\sqrt{\frac{q_1}{q_2}}\sinh (\sqrt{q_1 q_2} t) \\
-\sqrt{\frac{q_2}{q_1}}\sinh (\sqrt{q_1 q_2} t) & \cosh (\sqrt{q_1 q_2} t) \end{array} \right) \, ,
\eeq
so that $\mu(t)$ can be made integer  with the choice of parameters
\beq
\label{intcon}
 t_0 \neq 0 \ , \  \cosh (\sqrt{q_1 q_2} t_0)=n_1 \ , \ \frac{q_1}{q_2}=\frac{n_2}{n_3} \ , \ n_2 n_3=n_1^2-1 \ ,
\ n_{1,2,3} \in \mathbb{Z}^* \ .
\eeq
Thus also the algebra $\varepsilon_{1,1}$ can be used to construct compact solvmanifolds.  Notice that
the values $q_1=q_2=1$ are  not allowed by the integer condition \eqref{intcon}. \\

\begin{figure}[H]
\begin{center}
\includegraphics[height=6cm]{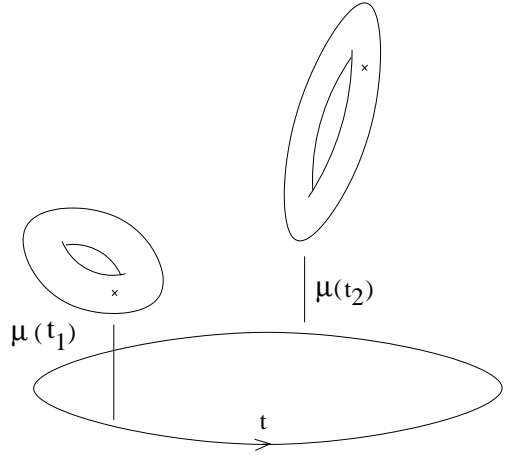}
\caption{Mostow bundle for the solvmanifolds $\epsilon_2$ and $\epsilon_{1,1}$. The base is the
circle in the $t$ direction, and due to the nilradical being abelian the fiber is $T^2$. The fibration is encoded
in $\mu(t)$ which is either a rotation or a ``hyperbolic rotation''  twisting the $T^2$ moving along the base.}
\end{center}
\end{figure}

\subsection{Twist construction of (almost abelian) solvmanifolds} \label{twistc}

In the previous section we showed how to obtain explicitly,
at least for almost abelian solvmanifolds, the operator $\mu(t)$ giving the structure of the Mostow bundle
and what condition it has to satisfy in order for the manifold to be compact.

In this section we focus on six-dimensional almost abelian algebras and the corresponding compact
solvmanifolds, and we discuss how to use the adjoint action $\mu(t)$ to construct the
globally defined one-forms of the solvmanifolds from those on $T^6$.
In Appendix \ref{apA} we show how the construction works in general for  almost nilpotent and nilpotent algebras.
For nilmanifolds the construction proposed in \cite{AMP} is recovered. \\

Let us first discard global issues related to the compactness of the manifolds. Then, given an almost
abelian solvable group $G$, we want to relate one-forms on $T^*\mathbb{R}^6$ to those of $T^*G=\Gg^*$
\beq
\label{twistgen}
A \ \left( \begin{array}{c} {\rm d}x^1  \\ \vdots  \\ {\rm d}x^6  \end{array} \right) =
\left( \begin{array}{c} e^1  \\ \vdots  \\ e^6  \end{array} \right) \ .
\eeq
Here $A$ is a local matrix that should contain the bundle structure of $G$. From the Mostow bundle,
\eqref{Mostowbundle}, it is natural to identify $x^6$ with the coordinate $t$ parametrising the $\mathbb{R}$
subalgebra and to take the corresponding one-form as ${\rm d}x^6= {\rm d}t$.
Then the matrix $A$ takes the form
\beq
A=  \begin{pmatrix} A_M & 0 \\  0 & 1\end{pmatrix}  \ ,
\eeq
where $A_M$ is a five-dimensional matrix given by
\beq
A_M=\mu(-t)=\mu(t)^{-1}= e^{-t \ ad_{\partial_t}(\Gn)}  \ . \label{Twistalmab}
\eeq
It is straightforward to show that  the forms constructed this way verify the Maurer-Cartan equation
(see (\ref{apMC})):
\beq
\d e^i= \d (e^{-t\ ad_{\partial_t}})^i_{\ \ k} \w \d x^k= \dots = -f^i_{\ \ tj}\ \d t \w e^j \ .
\eeq
Note that taking, for instance, $\mu(t)$ as in  (\ref{example1}), the corresponding $A$
is not a diffeomorphism and therefore can change topology.\\

We now come back to the consistency of this construction and the question of compactness. To this end we need to investigate the monodromy properties of the matrix $A_{M}$ and the related one-forms under a complete turn around the base circle.

Let us consider the following identification: $t\sim t+t_{0}$ where $t_{0}$ is the periodicity of the base circle. To obtain a consistent construction (having globally defined one-forms) we must preserve the structure of the torus we are fibering over the $t$ direction. This amounts to asking that an arbitrary point of the torus is sent to an equivalent one after we come back to the point $t$ from which we started. The monodromies of the fiber are fixed, thus the only allowed shifts are given by their integer multiples. The way points in the torus are transformed when we go around the base circle is encoded in a matrix $M_{\fff}$ which has to be integer valued. The identification along the $t$ direction is given by
\beq
\label{id6}
T_{6}:\left\{
    \begin{array}{l}
        t  \rightarrow t+t_{0}\\
        x^{i}  \rightarrow (M_{\fff})^{i}_{\ \ j}x^{j}
    \end{array} \right.
\qquad i,j=1,\ldots,5 \, ,
\eeq
while those along the remaining directions are trivial
\beq
\label{otherid}
T_{i}:\left\{
    \begin{array}{l}
        x^{i}  \rightarrow x^{i}+1\\
        x^{j}  \rightarrow x^{j}\\
        t   \rightarrow t
    \end{array} \right.
\qquad i,j=1,\ldots,5\,\mathrm{;}\,\,\,i\neq j \, .
\eeq

Let us now consider the one-forms \eqref{twistgen} we have constructed via the twist $A_{M}$. It is straightforward to see that \eqref{twistgen} are invariant under the trivial identifications, while under the non-trivial $T_6$, we have for $i,j=1,\ldots,5$
\beq
\tilde{e}^{i} = A_M(t+t_{0})^{i}_{\phantom{1}j} \d \tilde{x}^{j} = [A_M(t) A_M(t_{0}) M_{\fff}]^{i}_{\ \ j} \d x^{j} \ .
\eeq
The one-forms are globally defined if they are invariant under this identification:
\beq
\tilde{e}^{i} =e^{i}= A_M(t)^{i}_{\ \ j} \d x^{j} \, .
\eeq
Therefore, in the construction, we have to satisfy the following condition:
\beq \label{idoneform}
A_M(t_{0}) M_{\fff} = \mathbb{I}_5 \Leftrightarrow M_{\fff}=A_{M}^{-1}(t_{0})=A_{M}(-t_{0}) \ .
\eeq
Consistency requires the matrix $A_{M}$ to be such that $A_{M}(-t_{0})$ is integer valued for at least one $t_{0}\neq 0$. This will impose a quantization condition on the period of the base circle, which can take only a discrete set of values (in general it will be a numerable set, as we will see in the examples). Once we fix $t_{0}$, the integer entries of $A_{M}(-t_{0})$ will provide the set of identifications.

It is worth stressing that being able to give the correct identifications of the one-forms of the manifold is the same as having a lattice: the identifications \eqref{idoneform} express the lattice action, and give globally defined one-forms only if $A_{M}(-t_0)=\mu(t_0)$ is integer valued for some $t_{0}$. As already discussed, this is the condition to have a lattice (as stated in \cite{B}, see also Appendix \ref{apcomp}). Let us emphasize that the one-forms \eqref{twistgen}, constructed via the twist, are globally defined only if we start from a basis of the Lie algebra where $A_M(t)$ is integer valued for some value of $t$. We give a list of algebras in such a basis in Appendix \ref{apglobdef}.

Note that obtaining a set of globally defined one-forms is an expected result, since we are transforming a six-torus into a
solvmanifold, which we know to be parallelizable. Moreover, we also know that, with a consistent twist, we are not leaving the geometrical framework.\\

As an example, we write the explicit form of the twist matrix for the
two almost abelian six-dimensional algebras we need in this paper\footnote{We use the same
notation as in the standard classification of solvable algebras \cite{M1,T,B}: the number 5 indicates
the dimension of the (indecomposable) algebra, while the second simply gives its position in the list of indecomposable
algebras of dimension $5$.}. In the basis where the one-forms are globally defined the two algebras are
\bea
\label{g57}
\Gg_{5.7}^{1,-1,-1} \oplus \mathbb{R}   &:&  (q_1 25 , q_2 15, q_2 45, q_1 35,0,0)  \, ,  \\
\label{g517}
\Gg_{5.17}^{p,-p,\pm 1} \oplus \mathbb{R}  &:&
(q_1 (p 25 + 35), q_2 (p 15 + 45), q_2 (p 45 - 15), q_1 (p 35 - 25), 0,0) \, .
\eea
In both cases the parameters $q_1$ and  $q_2$ are strictly positive.
This is not the most general form of these algebras, which in general\footnote{The general form for $\Gg_{5.7}^{p,q,r}$ is
\beq
\frac{1}{2}\Big(-\beta (1+r) 15 + q_1 (1-r) 25 ,
- \beta (1+r) 25 + q_2 (1-r) 15, - \beta (q+p) 35 + q_2 (p-q) 45, - \beta (q+p) 45 + q_1 (p-q) 35, 0 \Big) \ , \nn
\eeq
where we set $\beta = \sqrt{q_1 q_2}$. Similarly, for $ \Gg_{5.17}^{p,-p,r}$ we have
\beq
\left(q_1 p 25 + \frac{1}{2}[q_1(r^2 + 1) 35 + \beta (r^2-1) 45],
q_2 p 15 + \frac{1}{2}[q_2(r^2 + 1) 45 + \beta (r^2-1) 35],
q_2 (- 15 + p 45), q_1 ( - 25 + p 35), 0 \right)  \, . \nn
\eeq} contain some free parameters $p$, $q$ and
$r$.
Here we wrote the values of the parameters for which we were able to find a lattice:
$p=-q=-r=1$ for the first algebra and $r= \pm 1$ for the second.

In the rest of the paper, by abuse of notation, we will denote  the algebra and
the corresponding solvmanifold with the same name.\\

For \eqref{g57},  a type IIA solution with O6 planes was found in \cite{CFI}.
The algebra being a direct product of a trivial direction and a five-dimensional indecomposable algebra,
the adjoint matrix $ad_{\partial_{x^{5}}}(\Gn)$ is block-diagonal, with the non-trivial blocks given by $-ad_{\partial_{t}}(\Gn)$ in \eqref{e1adj} and its transpose. Then the
twist matrix is
\beq
\label{twist1}
A= \begin{pmatrix}
 A_M &   \\
 & \mathbb{I}_2
\end{pmatrix}
\qquad
A_M = \left(\begin{array}{cc|cc} \alpha & - \beta
& & \\ - \gamma  & \alpha  & & \\ \hline & & \alpha
 & - \gamma  \\ & & -\beta &  \alpha \end{array} \right) \, ,
\eeq
where, not to clutter notation, we defined
\bea
&& \alpha =\cosh(\sqrt{q_1 q_2} x^5)  \, ,\nn \\
&& \beta = \sqrt{\frac{q_1}{q_2}} \sinh(\sqrt{q_1 q_2} x^5) \, ,\nn \\
&& \gamma= \sqrt{\frac{q_2}{q_1}} \sinh(\sqrt{q_1 q_2} x^5) \, .
\eea

The forms obtained by the twist \eqref{twist1} are globally defined \cite{HSUV}. Indeed they are invariant under
constant shifts of each  $x^i$ for  $i=1,2,3,4$ and $6$,  with the other variables fixed,
and the following non-trivial identification under shifts for $x^5$
\beq
(x^1, \dots, x^6)=( \alpha x^1 + \beta x^2, \gamma x^1 + \alpha x^2, \alpha x^3 + \gamma x^4, \beta x^3+ \alpha x^4, x^5 + l, x^6) \, ,
\eeq
where in $\alpha, \beta, \gamma$ we took $x^5=l$. For the above identifications to be discrete \cite{HSUV} $\alpha$, $\beta$, and $\gamma$ must be all integers. This is equivalent to having the matrix $\mu(x^5 = l)$ integer and, hence, it is the
same as the compactness criterion. The existence of a lattice for the solution in \cite{CFI} was also discussed in
\cite{GMPT6}. In that case the parameters $\alpha$, $\beta$ and $\gamma$ were set to $\alpha=2, \ \beta=3, \ \gamma=1$. \\

For the second algebra, $\Gg_{5.17}^{p,-p,r} \oplus \mathbb{R}$, we will consider separately the cases
$p=0$ and $p \neq 0$. For $p=0$ it reduces to $(q_1 35 , q_2 45 ,  - q_2 15 , - q_1 25 , 0 , 0)$ with $r^2 =1$. This algebra
and the associated manifold have been already considered in \cite{GMPT6}, where it was called $s$ 2.5.
For $p \neq 0$  the algebra can be seen as the direct sum
\beq
\Gg_{5.17}^{p,-p,r} \oplus \mathbb{R}\approx s\ 2.5 + p \ (\Gg_{5.7}^{1,-1,-1} \oplus \mathbb{R}) \, .
\eeq
The twist matrix is given by
\beq
\label{twist2}
A= \begin{pmatrix}
A_1 A_2 & \\
 & \mathbb{I}_2 \\
\end{pmatrix} \, .
\eeq
The two matrices $A_1$ and $A_2$ commute and give the two parts of the algebra
\beq
\label{Amatrix2}
A_1 = \left(\begin{array}{cc|cc} {\rm ch} & -\eta \,  {\rm sh}
& & \\ -\frac{1}{\eta}  {\rm sh}  & {\rm ch}  & & \\ \hline & & {\rm ch}
 & -\frac{1}{\eta}  {\rm sh} \\ & & -\eta \,   {\rm sh} &  {\rm ch} \end{array} \right)
\qquad
A_2 =
\left( \begin{array}{cc|cc} {\rm c} & & -\eta {\rm s} & \\  & {\rm c} & & -\frac{1}{\eta} {\rm s} \\ \hline \frac{1}{\eta} {\rm s} & & {\rm c} &  \\ & \eta {\rm s} & & {\rm c} \end{array} \right) \ ,
\eeq
where now we define $\eta=\sqrt{\frac{q_1}{q_2}}$ and
\bea
{\rm ch} =\cosh(p\sqrt{q_1 q_2} x^5)  & \quad & {\rm c}=\cos(\sqrt{q_1 q_2} x^5) \nn \\
{\rm sh}=\sinh(p\sqrt{q_1 q_2} x^5)   & \quad & {\rm s}=\sin(\sqrt{q_1 q_2} x^5) \, . \nn
\eea

In this case, imposing that the forms given by  the twist \eqref{twist2} are globally defined under discrete
identifications fixes the parameters in the twist to (with $x^5=l$)
\bea
&&{\rm ch \ c}=n_1 \ , \ \eta \  {\rm sh \ c}=n_2 \ , \ \frac{1}{\eta} {\rm sh \ c}=n_3 \nn\\
&&{\rm sh \ s}=n_4 \ , \ \eta \  {\rm ch \ s}=n_5 \ , \ \frac{1}{\eta} {\rm ch \ s}=n_6 \ , \quad n_i \in \mathbb{Z} \ .
\eea

The equations above have no solutions if the integers $n_i$ are all non-zero.
The only possibilities are either $n_1=n_2=n_3=0$ or $n_4=n_5=n_6=0$ (plus  the case where all
are zero, which is of no interest here). If one also imposes that the constraints must be solved
both for $p=0$ and $p\neq 0$, the first option, $n_1=n_2=n_3=0$, has to be discarded and
the only solution is
\bea
&&n_4=n_5=n_6=0 \ , \ s=0 \ , \ l=\frac{k\ \pi}{\sqrt{q_1q_2}} \ , \ c=(-1)^k \ , \ \tilde{n}_1=(-1)^k n_1 >0 \ , \ k\in \mathbb{Z} \nn\\
&&{\rm ch}=\tilde{n}_1 \ , \ {\rm sh}^2=n_2n_3 \ , \ n_3 \eta^2=n_2 \ , \ n_2n_3=\tilde{n}_1^2-1 \ , \
p=\frac{\cosh^{-1}(\tilde{n}_1)}{k\ \pi} \ .
\eea
$p$ is quantized by two integers, but one can show that it can be as close as we want to any real
value (the ensemble is dense in $\mathbb{R}$).\\

\subsection{Twist transformations in generalized geometry}

The twist defined in the previous section has a natural embedding in generalized geometry.
The basic idea of generalized geometry is to combine the tangent and cotangent bundle of
a given manifold $M$  (here the internal manifold of our compactification)
into a single object, the generalized tangent bundle $E$. This is an extension of $TM$ by $T^\ast M$.
Locally a section of $E$ is a  sum of a vector and a one-form
\beq
X = v + \xi  \, \in  \, TM \oplus T^\ast M \, ,
\eeq
which is glued on the overlap of two local patches,
$U_\alpha$ and $U_\beta$,  by
\beq
\label{eq:S-patch}
\begin{pmatrix}
v \\ \xi
\end{pmatrix}_{(\alpha)} = \begin{pmatrix} a & 0 \\ \omega a & a^{-T} \end{pmatrix}_{(\alpha\beta)}
\begin{pmatrix}
v \\ \xi
\end{pmatrix}_{(\beta)} \, .
\eeq
$a$ is an element of GL($d,\mathbb{R}$), and gives the
usual gluing of vectors and one-forms ($a^{-T}=(a^{-1})^T$), while $\omega$ is a two-form such that
$\omega_{(\alpha\beta)}=-\d \Lambda_{(\alpha\beta)}$. $\omega$ is related to the non-trivial fibration
of $T^*M$ over $TM$: this is encoded in a local two-form (the ``connective
structure'' of a gerbe) that is interpreted as the $B$-field, and $\omega$ corresponds to its gauge transformation.

For the backgrounds we will consider in this paper, $B=0$, meaning that
the generalized tangent bundle is trivial and can be identified with $TM \oplus T^\ast M$. \\

$E$ is endowed with two metrics
\beq
\label{metricI}
\mathcal{I} = \begin{pmatrix} 0 & \mathbb{I} \\
\mathbb{I} & 0 \end{pmatrix}  \qquad \qquad
\mathcal{H} =  \begin{pmatrix} g - B g^{-1} g  &  B g^{-1} \\
- g^{-1}  B  & g^{-1} \end{pmatrix}  \, ,
\eeq
where $\mathcal{I}$ is the natural metric on $E$ (which is used to derive the Clifford
algebra) while the generalized metric $\mathcal{H}$ encodes the information about the metric
and the $B$-field of the background.

The metric $\mathcal{I}$ is invariant under O($d,d$) transformations, which can be parametrised by
Gl($d$) transformations
\beq
X = v + \xi \mapsto X' =  A v + A^{-T} \xi \, ,
\eeq
plus shifts by a two-form $b$ and a two-vector $\beta$
\bea
&& X = v + \xi  \mapsto X' =  v + (\xi -i_v b) \, , \\
&& X = v + \xi \mapsto X' =   (v + \beta \cdot \xi)+  \xi \, .
\eea
These are the so-called  $B$- and $\beta$-transforms. \\

On $E$  one can define generalized vielbeine $\mathcal{E}$,  such that
\beq
\mathcal{I} =  \mathcal{E}^T \begin{pmatrix} 0 & \mathbb{I} \\
\mathbb{I} & 0 \end{pmatrix}  \mathcal{E} \qquad \qquad \mathcal{H} = \mathcal{E}^T \begin{pmatrix}
\mathbb{I} & 0  \\
0 & \mathbb{I} \end{pmatrix}  \mathcal{E} \, .
\eeq
Explicitly, the generalized vielbeine can be put in the form
\beq
\mathcal{E}^A_{\, \, \,  M} = \begin{pmatrix} e^a_{\, \, \, m} & 0 \\
                               - (\hat{e} B)_{a m} &  \hat{e}_a^{\, \, \, m}
\end{pmatrix} \,  , \label{genviel}
\eeq
where $e^a_{\, \, \, m}$ are the vielbeine on $M$, $\hat{e} = (e^T)^{-1}$, and $B$ is the $B$-field.
Comparing the O($d,d$) action on $\mathcal{E}$
\beq
\label{Oddactionv}
\mathcal{E} \mapsto \mathcal{E}' = \mathcal{E} O = \begin{pmatrix} e^a_{\, \, \, m} & 0 \\
                               - (\hat{e} B)_{a m} &  \hat{e}_a^{\, \, \, m}
\end{pmatrix} \begin{pmatrix} A^m_{\, \, \,  n} & B^{m n} \\
                                C_{m n}  &  D_m^{\, \, \, n} \end{pmatrix}\, ,
\eeq
with \eqref{twistgen}, it is natural to embed the twist transformation as
\beq
\label{twistactv}
O_{\rm tw} = \begin{pmatrix} A  & 0 \\
 0 &  (A^T)^{-1} \end{pmatrix} \, .
\eeq

\vspace{0.4cm}

The polyforms  $\Phi_\pm$ appearing in the supersymmetry conditions  \eqref{eq:susyeqIIintro},
correspond to ground states of the Clifford  algebra Cliff$(d,d)$, on $TM \oplus T^\ast M$.
More precisely they are  Majorana-Weyl Spin($d,d$) spinors, the positive (negative) chirality corresponding to
the even (odd) polyform.  We will follow the conventions of \cite{AMP}. We focus on manifolds of dimension six and
construct O($6,6$) bispinors in the Killing spinors on $M$, $\eta^{1,2}$,
\beq
\Phi_\pm = \eta^1_+ \otimes \eta^{2 \, \dagger}_\pm \, .
\eeq
Here we will consider the SU(3) structure manifolds, which admit
a single globally defined spinor $\eta_+$ of  unitary norm. Hence
\beq
\eta^1_+ = |a| \, e^{i \alpha} \eta_+  \, \ , \ \eta^2_+ = |b| \, e^{i \beta} \eta_+ \, , \nn
\eeq
where $|a|$ and $|b|$ are clearly the norms of $\eta^{1,2}$.
The corresponding pure spinors $\Psi_{\pm}$ on $E$ are
\bea
\Psi_+ &=&  e^{-\phi} e^{-B} \frac{8}{||\Phi_+||} \Phi_+  \, , \nn\\
\Psi_- &=& e^{-\phi} e^{-B} \frac{8}{||\Phi_-||} \Phi_-  \, , \nn
\eea
with $||\Phi_{\pm}||=|a|^2=|b|^2$.
The phases of the two pure spinor are $\theta_+ = \alpha - \beta$ and $\theta_-  = \alpha + \beta$.
$J$ is the K\"ahler form and $\Omega$ the holomorphic three-form on $M$. \\

Note that, while the bispinors are globally defined, the pure spinors $\Psi_\pm$ glue non-trivially on the double overlaps:
\beq
\Psi_\alpha = e^{\d \Lambda_{(\alpha \beta)}} \Psi_{\beta} \, .
\eeq
As already mentioned, in the backgrounds we will consider in this paper the generalized tangent bundle is trivial and
the dilaton is constant. In these cases, we can identify $\Psi_\pm$ and $\Phi_\pm$.\\

The O($d,d$) action on pure spinors is given by its spinorial representation
\beq
\label{Oddactionps}
O \cdot \Psi = e^{-\frac{1}{4} \Theta_{M N} [\Gamma^M,  \Gamma^N]} \cdot \Psi \, ,
\eeq
where $\Gamma^M$ are the Cliff$(d,d)$ gamma matrices ($\Gamma^m = \d x^m$ and $\Gamma_m = \iota_m$)
and $\Theta_{M N}$ are the O($d,d$) parameters
\beq
 \Theta_{M N} = \begin{pmatrix} a^m_{\, \, \,  n}  & \beta^{m n} \\
 b_{m n} &  - a_m^{\, \, \, n} \end{pmatrix} \, .
\eeq
Here $a^m_{\, \, \,  n}$, $b_{m n}$ and $\beta^{m n}$ parametrise the GL($d$)
transformations, $B$-transforms and $\beta$-transform, respectively.
Then the twist action (\ref{Twistalmab}) on the spinor reads \cite{AMP}
\beq
\label{twistactionps}
O_{\rm tw} \cdot \Psi =  \frac{1}{\sqrt{\det A}} \, \,
e^{- t\ [ad_{\partial_t}(\Gn)]^m_{\ \ n} e^n \wedge \, \iota_m}  \cdot \Psi \, ,
\eeq
where $e^m$ is a given basis of one-forms on $M$, and $\iota_m$ the associated contraction.

\subsection{Type IIA supersymmetric solutions from twist transformations}
\label{susysol}

Type IIA supersymmetric compactifications to four-dimensional Minkowski where the internal manifold is
the solvmanifold $\Gg_{5.17}^{0,0,\pm 1} \times S^1$ were found in \cite{CFI,GMPT6,A}.
As shown in Section \ref{twistc},
this manifold is related by twist to the more general manifold $\Gg_{5.17}^{p,-p,\pm 1} \times  S^1$.
It is then natural to ask what is the effect of twisting the solutions in \cite{GMPT6,A}. \\

We will take as starting point Model 3 of \cite{GMPT6}. This is an SU(3) structure solution
with smeared D6-branes and O6 planes in the directions (146) and (236).
For SU(3) structure, the two pure spinors are
\beq
\label{su3ps25}
\Phi_+ = \frac{e^{i\theta_+}}{8}  \, e^{-iJ} \qquad \qquad \Phi_- = -\frac{i}{8} \ \Omega \, .
\eeq
The phase in $\Phi_+$ is, in general, determined by the orientifold projection. For
O6 planes $\theta_+$ is actually free and we set it to zero. We take
\beq
\label{su3sol25}
\Omega = \sqrt{t_1 t_2 t_3} \, \chi^1 \wedge \chi^2 \wedge  \chi^3 \qquad
\quad J=\frac{i}{2} \sum_{k} t_k \chi^k \wedge \overline{\chi}^k \ ,
\eeq
with complex structure\footnote{$\Omega$ and $J$ are normalised as
\beq
\frac{4}{3} J^3 = i \Omega\w \ov{\Omega} =-8\ \mathrm{vol}_{(6)}
= -8\ \sqrt{|g|} \ e^1 \w e^2 \w e^3 \w e^4 \w e^5 \w e^6  \label{comp}
\eeq
where $\mathrm{vol}_{(6)}$ is the internal volume form.}
\bea
\label{css25}
&& \chi^1= e^1 + i \,\lambda \frac{\tau_3}{\tau_4}   \ e^2 \, , \nn\\
&& \chi^2= \tau_3 \, e^3 + i \tau_4 \ e^4 \, , \nn\\
&& \chi^3= e^5 - i \tau_6 \ e^6 \ .
\eea
For simplicity, we introduce $\lambda = \frac{t_2 \tau_4^2}{t_1}$.
$e^i$ are globally defined one-forms, obtained as in \eqref{twistgen}
\beq
e^m = (A_2)^m_{\, \, \, n} \d x^n \, ,
\eeq
with $A_2$ given by \eqref{Amatrix2}.
With this choice the metric is diagonal
\beq
\label{s2.5met}
g= \textrm{diag} \left( t_1, \lambda \, t_2 \, \tau_3^2, t_2 \, \tau_3^2, \lambda t_1, t_3, t_3 \tau_6^2  \right)  \, .
\eeq
Positivity of the volume imposes the following constraints on the complex structure and K\"ahler moduli
\beq
\tau_6 > 0 \ , \ t_1,\ t_2,\ t_3  >0 \, .
\eeq

Due to the presence of intersecting sources, the warp factor is set to one and the dilaton to a constant.
By splitting the pure spinor equations \eqref{eq:susyeqIIintro} into forms of fixed degree, it is easy to verify that
supersymmetry implies
\bea
&& \d (\im \Omega) =0  \, , \\
&& \d J=0 \, , \\
&& \d (\re \Omega )= g_s * F_2 \, ,  \\
&& F_6=F_4=F_0=H=0 \, .
\eea
The  only non-zero RR flux reads
\beq
g_s F_2 = \frac{\sqrt{\lambda} \, (q_1 t_1 - q_2 t_2 \tau_3^2)}{\sqrt{t_3}}  (e^3\w e^4-e^1\w e^2) \, ,
\eeq
and it is straightforward to check that its Bianchi identity is satisfied. Let us also recall \cite{GMPT6, KT} the
 transformation the forms should satisfy under the O6-plane involution $\sigma$:
\beq
\sigma(J)=-J \ , \ \sigma(\Omega)=\ov{\Omega} \ , \ \sigma(H)=-H \ , \ \sigma(F)=\lambda(F) \ . \label{Oproj}
\eeq
Given the directions of the sources here, these orientifold projection conditions are clearly verified by the solution.\\

Given the solution above, we want to use the twist action to
produce solutions,
still with O6-planes and D6-branes, on $\Gg_{5.17}^{p,-p,\pm 1} \times S^1$. The
manifolds  $\Gg_{5.17}^{p,-p,\pm 1}  \times S^1$  and
$\Gg_{5.17}^{0,0,\pm 1}  \times S^1$ are related by the twist matrix $A_1$ in \eqref{Amatrix2},
whose adjoint matrix is
\beq
ad_{\partial_5}(\Gn)|_p =  \begin{pmatrix}  a_{12}  &  \\
& a_{34} \\
\end{pmatrix}   \qquad \qquad a_{12} = a_{34}^T =  \begin{pmatrix} 0 & p q_1 \\ p q_2 & 0
\end{pmatrix} \, .
\eeq

The sixth direction being  a trivial circle, we identify $t=x^5$. Then the
twist action on pure spinors,
\beq
\Phi_\pm \mapsto \Phi'_\pm  = O_{\rm tw} \Phi_\pm \, ,
\eeq
can be  rewritten as
\bea
O_{\rm tw} &=& e^{ - p x^5 (q_2 e^1 \w \iota_2 + q_1 e^2 \w \iota_1)} \, e^{- p x^5 (q_1 e^3 \w \iota_4 + q_2 e^4 \w \iota_3)}  \nn \\
&=& O_{12} O_{34} \, ,
\eea
with
\bea
O_{12} &=&  \mathbb{I} + [ \cosh( p \sqrt{q_1 q_2} x^5) -1]
(e^1 \w \iota_1 + e^2 \w \iota_2 + 2 e^1 \w e^2 \w \iota_1 \w \iota_2 ) \nn  \\
&& - \frac{1}{\sqrt{q_1 q_2}}  \sinh( p \sqrt{q_1 q_2} x^5) (q_2 e^1 \w \iota_2
+  q_1 e^2 \w \iota_1)  \, ,\\
O_{34} &=& \mathbb{I} + [ \cosh( p \sqrt{q_1 q_2} x^5) -1]
(e^3 \w \iota_3 + e^4 \w \iota_4 + 2 e^3 \w e^4 \w \iota_3 \w \iota_4 ) \nn \\
&& - \frac{1}{\sqrt{q_1 q_2}}  \sinh( p \sqrt{q_1 q_2} x^5) (q_1 e^3 \w \iota_4
+  q_2 e^4 \w \iota_3) \, .
\eea
Note that unimodularity of the algebra implies
$\det(A)=1$. In comparison to the procedure described in \cite{AMP}, here we do not introduce
a phase in the twist operator, since we do not modify the nature of the fluxes
and sources.

It is straightforward to check that the transformed pure spinors have formally the same expression
as in \eqref{su3ps25} - \eqref{css25} but with the one-forms $e^i$ now given by
\beq
\label{ebasisgen}
e^m = (A_1 A_2)^m_{\, \, \, n} \d x^n \, .
\eeq
Also the metric, which is completely specified by the pure spinors, has the same
form as for the initial solution, but in the new $e^i$ basis
\beq
g =  \textrm{diag} \left( t_1, \lambda t_2 \tau_3^2, t_2 \tau_3^2, \lambda t_1, t_3, t_3 \tau_6^2 \right) \, .
\eeq

In order for the twist transformation to produce new solutions, the transformed pure spinors should again
satisfy the supersymmetry equations
\bea
\label{eq:IInew}
\d_{H^{\prime}} (\Phi_+^{\prime})&=&0 \, , \nn  \\
\d_{H^{\prime}} (\textrm{Re}\Phi_-^{\prime})&=& 0  \, , \nn \\
\d_{H^{\prime}} (\textrm{Im}\Phi_-^{\prime})&=& g_s\ R^{\prime} \, ,
\eea
where $R^{\prime}$ is the new RR field $R=\frac{1}{8} \ast \lambda(F)$ .
The conditions
\beq
\label{newsol}
H'= 0 \qquad \qquad \d J' = 0
\eeq
are automatically satisfied, so that the first two equations in \eqref{eq:IInew} reduce to\footnote{
\label{moregensol}
Note that a slightly more general solution given by $\chi^1= e^1+i \left(\frac{\tau_3}{\tau_4} \lambda \ e^2-\frac{\tau_2}{\tau_4}\ e^3 \right) \ , \ \chi^2= \tau_2\ e^2+\tau_3\ e^3+i \tau_4\ e^4 $ and the same $\chi^3$ leads to the same $\d(\im \Omega^{\prime})$ and to
\beq
\d(J^{\prime}) = -p (\lambda-1)\ \tau_2 \sqrt{\frac{t_1 t_2}{\lambda}} \ (q_2 \ e^1\w e^4 \w e^5+ q_1 \ e^2\w e^3 \w e^5) \ .
\eeq
A supersymmetric solution, requiring $\d(\mathrm{Im}\Omega)=\d J=0$, needs $\lambda=1$. For $\tau_2=0$ we can have non-supersymmetric configurations with a closed $J'$.}
\beq
0= \d(\im \Omega^{\prime})= -p (\lambda-1)\ \tau_3 \tau_6 \sqrt{t_1 t_2 t_3} \ (q_2 \ e^1\w e^4 \w e^5+ q_1 \ e^2\w e^3 \w e^5) \w e^6 \, .
\eeq
From this  we see that, in addition to $p=0$ case, supersymmetric solutions exist
for $p\neq 0$ provided $\lambda=1$.\\

The last equation in (\ref{eq:IInew}) defines the transformed RR field
\beq
\label{RRprime}
g_s R^{\prime}=g_s O_{\rm tw} \cdot  R +  {\rm d}_{H^{\prime}} (O_{\rm tw}) \cdot \im \Phi_- \ .
\eeq
Since the twist operator does not change the degree of forms, it follows from \eqref{RRprime}
that no new RR fluxes have been generated
\beq
F_0 = F_4 = F_6=0 \, ,
\eeq
and (we have already set $\lambda=1$)
\beq
g_s F_2 =  \frac{q_1 t_1 - q_2 t_2 \tau_3^2}{\sqrt{t_3}}  (e^3\w e^4-e^1\w e^2) +
\frac{p (q_1 t_1 + q_2 t_2 \tau_3^2)}{\sqrt{t_3}}  (e^2 \w e^4 + e^1\w e^3) \, .
\eeq

The Bianchi identity for $F_2$ is satisfied
\beq
g_s \d F_2 = c_1 v^1 + c_2 v^2 \ ,
\eeq
with $v^1=t_1\sqrt{t_3}\ e^1 \w e^4 \w e^5$ and
$v^2=t_2 \tau_3^2 \sqrt{t_3}\ e^2 \w e^3 \w e^5$ being the covolumes of the sources in (236) and (146). Let us note that the orientifold projection
conditions (\ref{Oproj}) are again satisfied with such sources. The sign of the charges
\bea
c_1 &=& \frac{2 q_2}{t_3 t_1 } \left[t_1 q_1 (1-p^2)  - (1+p^2)  t_2 q_2 \tau_3^2 \right] \nn\\
c_2 &=& \frac{2 q_1}{t_3 t_2 \tau_3^2} \left[ \tau_3^2 t_2 q_2 (1-p^2)  - (1+p^2) t_1 q_1  \right]
\eea
depends on the parameters, but the sum of the two charges is clearly negative.
This guarantees that the transformed background with $p \neq 0$ and $\lambda =1$
is indeed a solution of the full set of ten-dimensional equations
of motion. In the next section we will use the non-supersymmetric version, with $\lambda \neq 1$,
as starting point for our search for de Sitter solution.

In the literature, de Sitter backgrounds are often given in terms of SU(3) structure torsions,
\bea
\d J&=& \frac{3}{2}\im(\bar{W}_1\Omega)+W_4\w J+W_3\nn\\
\d\Omega&=&W_1 J^2+W_2\w J+\bar{W}_5\w\Omega \, ,
\eea
where $W_1$ is a complex scalar, $W_2$ is a complex primitive $(1,1)$ form, $W_3$ is a real primitive $(2,1)+(1,2)$ form, $W_4$ is a real vector and $W_5$ is a complex $(1,0)$ form. For the more general SU(3) structure solution ($p\neq 0$, $\lambda\neq 1$, $\tau_2 \neq 0$) mentioned in Footnote \ref{moregensol}, we obtain
\bea
W_1&=&\frac{p\,\tau_2\,(A+B)(1-\lambda)}{6(\tau_2^2+\lambda\tau_3^2)\sqrt{t_1 t_2 t_3}}\nn\\
W_2&=&\frac{1}{6(\tau_2^2+\lambda\tau_3^2)\sqrt{t_1 t_2 t_3}}\Bigg[-i t_1\Big(p\tau_2\,(A+B)(\lambda+2)+3\lambda\tau_3(A-B)\Big)\chi^1\w\bar{\chi}^1+\nn\\
&  &+3\sqrt{\lambda t_1 t_2}\Big(\tau_2(B-A)+p\tau_3(\lambda A+B)\Big)\chi^1\w\bar{\chi}^2-3\sqrt{\lambda t_1 t_2}\Big(\tau_2(B-A)+p\tau_3(A+\lambda B)\Big)\chi^2\w\bar{\chi}^1+\nn\\
&  &+i t_2\Big(p\tau_2(A+B)(1+2\lambda)+3\lambda\tau_3(A-B)\Big)\chi^2\w\bar{\chi}^2-ip\tau_2 t_3(A+B)(\lambda-1)\chi^3\w\bar{\chi}^3 \Bigg] \nn\\
W_3&=&\frac{ip\tau_2(\lambda-1)}{8(\tau_2^2+\lambda\tau_3^2)}\Bigg[(A+B)\chi^1\w\chi^2\w\bar{\chi}^3-(A+B)\chi^3\w\bar{\chi}^1\w\bar{\chi}^2+\nn\\
& &-(A-B)(\chi^1\w\chi^3\w\bar{\chi}^2-\chi^1\w\bar{\chi}^2\w\bar{\chi}^3
+\chi^2\w\chi^3\w\bar{\chi}^1-\chi^2\w\bar{\chi}^1\w\bar{\chi}^3)\Bigg]\nn\\
W_4&=&0\nn\\
W_5&=&\frac{ip\sqrt{\lambda}\tau_3(A+B)(\lambda-1)}{4(\tau_2^2+\lambda\tau_3^2)\sqrt{t_1 t_2}}\, \chi^3 \ ,
\eea
with $A=q_1t_1$, $B=q_2 t_2 (\tau_3^2 + \frac{\tau_2^2}{\lambda})$.

\subsubsection{Localizing the sources and warping }
\label{IIA2s25}

The supersymmetric solution discussed in the previous section is global, the warp factor and the dilaton
being constant. It is an interesting question to see whether localised solutions also exist (see e.g. \cite{DK} for a recent discussion about the importance of warping).
The strategy for finding localized solutions used in \cite{GMPT6} was
first to look for a smeared solution at
large volume and then localize it by scaling the  vielbeine, longitudinal and transverse with respect to the source,
with $e^{A}$ and $e^{-A}$, respectively.  This procedure works in a number of cases,
provided only parallel sources are present.  Unfortunately this is not the case for the supersymmetric solution
we took as a departure point for our construction  - the intersecting O6/D6 solution on $s \, 2.5$. \\

It is however possible to find a completely localised solution on $s \, 2.5$ with O6 planes.
The solution has a simpler form in a basis where the algebra is $(25, -15, r 45, -r 35, 0, 0)$, $r^2=1$.
In this basis the O6-plane is along the directions (345).

 The SU(3) structure is constructed as in
\eqref{su3sol25} with \bea
\chi^1&=&e^{-A} e^1 +i e^{A} (\tau_3 e^3+ \tau_4 e^4) \,, \nn\\
\chi^2&=&e^{-A} e^2 +i e^{A} r (-\tau_4 e^3+ \tau_3 e^4) \, , \nn\\
\chi^3&=&e^{A} e^5+i e^{-A} r \tau_6 e^6 \, , \nn\\
\tau_6 &>& 0 \ , \ t_1=t_2,\ t_3  >0 \, ,
\eea
where the non-trivial warp factor, $A$, depends on $x^1,x^2,x^6$. The metric is diagonal
\beq
g =  \textrm{diag} \left( t_1 e^{-2A}, t_1 e^{-2A}, t_1 (\tau_3^2+\tau_4^2) e^{2A}, t_1 (\tau_3^2+\tau_4^2) e^{2A}, t_3 e^{2A}, t_3 \tau_6^2 e^{-2A} \right)  \, ,
\eeq
and the only non-zero flux is the RR two--form
\beq
g_s F_2 =  -r \Big[\tau_6 \sqrt{t_3} \partial_1(e^{-4A})  \ \d x^2\w e^6
- \tau_6  \sqrt{t_3} \partial_2(e^{-4A})  \ \d x^1\w e^6
+   \frac{1}{\tau_6} \sqrt{\frac{t_1^2}{t_3}} \partial_6(e^{-4A})  \ \d x^1\w \d x^2 \Big] \, .
\eeq
Setting the parameters $t_1= t_2$ in the K\"ahler form \eqref{su3sol25}
 allows to have a single source term in the $F_2$ Bianchi identity
\beq
g_s \d F_2  \sim e^{-A} \Delta (e^{-4A}) e^1 \w e^2 \w e^6 \, ,
\eeq
where $\Delta$ is the laplacian with unwarped metric.\\

As $A \rightarrow 0$ this solution becomes fluxless ($s \, 2.5$ can indeed support such solutions), hence it cannot be found following the strategy of localizing the large volume smeared solutions.
Unfortunately this solution does not satisfy the twist to $p\neq 0$, \eqref{g517}, since for $ p \neq 0$
the action of the involution of an O6-plane with a component along direction 5 is not compatible with the algebra.


\subsection{A digression: twist and  non-geometric backgrounds}

We would like to come back to the question of the consistency  of the twist transformation. As explained already, the
transformation is obstructed unless the matrix $A$ is conjugated to an integer-valued matrix. In many cases, the twist
can result in a topology change similar to what is achieved by T--duality. The latter also can be obstructed, and yet these obstructions do not stop us from performing the duality transformation. So what about the obstructed twist?

To keep things simple, let us consider again an almost abelian algebra and the gluing under
$t \rightarrow t+ t_0$. We should have in general
\beq
T_{6}:\left\{
    \begin{array}{l}
        t  \rightarrow t+t_{0}\\
        x^{i}  \rightarrow {\tilde A}_{M}(-t_{0})^{i}_{\phantom{1}j}x^{j}
    \end{array} \right.
\qquad i,j=1,\ldots,5 \, ,
\eeq
where ${\tilde A}_M(-t_0)$ is necessarily an integer-valued matrix for $t_0 \neq 0$. In the case of
compact  solvmanifolds this matrix is given by \eqref{Twistalmab}. For the algebras that do not admit an action of a
lattice,  ${\tilde A}_M(-t_0)$ has nothing to do with the algebra. Then the one forms
$e^{i}=A(t)^{i}_{\phantom{1} j} \d x^{j}$ ($\d x^{6}=\d t$) are defined only locally and have discontinuities
under $t \rightarrow t+ t_0$. These kinds of discontinuity are actually familiar from the situations
when an obstructed T--duality is performed, and are commonly referred to as non-geometric backgrounds.
One way to see this is to work on the generalized tangent bundle and use
local $O(6) \times O(6)$ transformations (for six-dimensional internal manifolds) to
bring the generalized vielbeine to the canonical lower diagonal form (\ref{genviel}). In geometric backgrounds, this is a good
transformation, while in the non-geometric case it involves non-single valued functions \cite{GMPW}. \\

As an example, let us consider the manifold $\Gg^{-p}_{4.2} \times T^2$, where the algebra
$\Gg^{-p}_{4.2}$ is given in Appendix \ref{apcomp}. The corresponding group does not admit a lattice.
For generic $p$ this is very easy to see since the group is not unimodular. For $p=2$, the group is
unimodular but  there still is no lattice. As explained in the Appendix \ref{apcomp}, in this case, the
characteristic polynomial cannot have integer coefficients, and therefore there is an obstruction to the existence of
a lattice.

If we now consider the algebra together with its dual, i.e. examine the existence of a lattice on the generalized
tangent bundle, we should study the $6 \times 6$ matrix
$M(t)=\mathrm{diag}(\mu (t),  \mu (-t)^{T})$ instead of the matrix $\mu(t)$.
One has
\beq\label{M(t)}
M(t)= \left( \begin{array}{cccccc}
e^{pt} & 0&0&0&0&0\\
0&e^{-t}&0&0&0&0\\
0&-te^{-t}&e^{-t}&0&0&0\\
0&0&0&e^{-pt}&0&0\\
0&0&0&0&e^{t}&te^{t}\\
0&0&0&0&0&e^{t}
\end{array} \right) \, .
\eeq
For $t_0=\ln (\frac{3+\sqrt{5}}{2})$ and $p\in\mathbb{N}^*$, $M(t=t_0)$ is conjugated to an integer matrix,
$P^{-1}M(t_0)P=N$, where $N$ is an integer matrix (Theorem $8.3.2$ in \cite{B}):
\beq
P=\left( \begin{array}{cccccc}
1 & 0&0&\frac{18+8\sqrt{5}}{7+3\sqrt{5}}&0&0\\
0&1&0&0&0&\frac{2(2+\sqrt{5})}{3+\sqrt{5}}\\
0&0&\ln (\frac{2}{3+\sqrt{5}})&0&\frac{2(2+\sqrt{5})\ln (\frac{3+\sqrt{5}}{2})}{3+\sqrt{5}}&0\\
1&0&0&\frac{2}{3+\sqrt{5}}&0&0\\
0&0&\ln (\frac{2}{3+\sqrt{5}})&0&-\frac{(1+\sqrt{5})\ln (\frac{3+\sqrt{5}}{2})}{3+\sqrt{5}}&0\\
0&-1&0&0&0&\frac{1+\sqrt{5}}{3+\sqrt{5}}
\end{array} \right) \, ,
\eeq
\beq
N=\left( \begin{array}{cccccc}
a_{11}&0&0&a_{14}&0&0\\
0&2&0&0&0&-1\\
0&2&2&0&1&-1\\
a_{41}&0&0&a_{44}&0&0\\
0&1&1&0&1&-1\\
0&-1&0&0&0&1
\end{array} \right) \, .
\eeq

The piece
\beq
N_4=\left( \begin{array}{cc}
a_{11}&a_{14}\\
a_{41}&a_{44}
\end{array} \right)=\left( \begin{array}{cc}
0&-1\\
1&3
\end{array}\right)^{p}
\eeq
comes from the entries $e^{pt}$ and the result can be obtained\footnote{Another possible
conjugation is given in (\ref{e1adj}). The other part of $N$, the $4 \times 4$ integer matrix, can also be
different, see the change of basis in Proposition $7.2.9$ in \cite{B}.} from (\ref{apBbasis}). We see that on
the generalized tangent bundle
the basic obstruction to the existence of a lattice is easily removed. Moreover it is not hard to
see that, due to putting together the algebra and its dual, even the requirement of unimodularity can be dropped.\\

On the generalized tangent bundle we can therefore obtain a lattice. For non-geometry, one may ask for more: the
integer matrix $N$ being in $O(3,3)$. This question can be decomposed into $N_4 \in O(1,1)$ and the $4 \times 4$
integer matrix in $O(2,2)$. Actually, the latter is true\footnote{Note it is not true for the one given in
Proposition $7.2.9$ of \cite{B}.}. But $N_4 \notin O(1,1)$. Moreover, one can prove  that
$\mathrm{diag}(e^{pt},e^{-pt})$ can only be conjugated to an integer $O(1,1)$ matrix for $t=0$. Indeed,
the eigenvalues of an integer $O(1,1)$ matrices are $\pm1$, and those are not changed by conjugation.

This is reminiscent of the twist construction of the IIB background $n \, 3.14$ discussed in \cite{AMP}.
The internal manifold is a circle fibration over a five manifold $M_5$, which itself is a bundle with a two-torus
fiber, but the only obvious duality seen there is the $O(2,2)$ associated with the two-torus. The solution on
$M_5 \times S^1$ is obtained from IIB solution on $\mathbb{T}^6$ with a self-dual three-form flux, but not
$n \, 3.14$ itself \cite{GMPT6}. \\

By taking $p=0$ in  \eqref{M(t)}, we obtain a different topology. In  $M(t)$  the corresponding direction becomes trivial, and we can forget about it. Up to an $O(1,1)$ action, the non-trivial part of $M(t)$ can still be thought of as corresponding to the algebra on  $T(\varepsilon_{1,1}) \oplus T^*(\varepsilon_{1,1})$. Indeed, $\varepsilon_{1,1}$ has two {\sl local} isometries, and T--duality (the $O(1,1)$ in question)  with respect to any of them will yield a non-geometric background. This can be inferred by simply noticing that the result of the duality in (any direction) is not unimodular;  more detailed discussion of T--duality on $\varepsilon_{1,1}$ can be found in  Appendix \ref{T-dualsolv}.\\

A better understanding of the orientifold planes in generalized complex geometry is needed in order to apply the twist transformation to constructing non-geometric backgrounds. However, the possibility of using  solvable algebras in order to describe (some of) these is interesting.

\section{Supersymmetry breaking and de Sitter vacua}\label{sec:dS}

In the literature on de Sitter backgrounds, O6/D6 models seem to have good chances
at yielding a solution which can be embedded in string theory, at least in the conservative approach of
``geometric'' compactifications. We shall concentrate on the resolution of the ten-dimensional equations of motion
in this conservative set-up, making use of the technology described in the last section and adapting it to the
description of non-supersymmetric configurations.  \\

We consider type IIA supergravity and mostly follow the conventions of \cite{BKORVP,C}; we differ in the
definition of the
Hodge star where we have an extra sign depending on the parity of the forms\footnote{\label{foot:norm} In IIA,
the sign is always positive on RR fields, but not on the odd forms, $H$
and $\d\phi$, hence the sign difference  with respect to \cite{C} for the corresponding terms in the action.
The sign difference is related to the fact we use the Mukai pairing to give the norm (see Footnote \ref{Hodge}):
for a real form $\alpha_i$, we have $\langle * \lambda(\alpha_i), \alpha _i \rangle= |\alpha_i|^2 \times \mathrm{vol}$.
 Note that these conventions are consistent with the SUSY conditions written before. There is a factor of
$2$ difference in the normalization of the RR
 kinetic terms with respect to \cite{Si}, which will result in a difference in the RR quantization conditions.
For a $k$-flux $\alpha$ through a $k$-cycle
 $\Sigma$ (with embedding $i$ into the bulk manifold $M$), we have
\beq
\frac{1}{(2 \pi \sqrt{\alpha^\prime} )^{k-1}}\frac{1}{\mathrm{vol}_M} \int_{\Sigma} i^*\alpha = \frac{1}{(2 \pi \sqrt{\alpha^\prime} )^{k-1}}\frac{1}{\mathrm{vol}_M}\int_M \langle \delta(\Sigma \hookrightarrow M) , \alpha \rangle =n \, ,
\eeq
where $n$ is an integer.}.
In particular,
\beq
F_p \w \hat{\ast} F_p=\d^{10} x\ \sqrt{|g_{10}|} (-1)^{(10-p)p}\ \frac{F_{\mu_1 \dots \mu_p} F^{\mu_1 \dots \mu_p}}{p!} = \d^{10} x\  \sqrt{|g_{10}|} (-1)^{(10-p)p}\ |F_p|^2 \ .
\eeq
We explicitly denote the ten-dimensional Hodge star by $\hat{\ast}$, reserving the symbol $*$  for its
six-dimensional counterpart.\\

In order to derive the ten-dimensional equations of motion, we shall need source terms,
and to this end let us consider the DBI action of only one D$p$-brane in string
frame
\beq
S_s=-T_p \int \d^{p+1}x\ e^{-\phi} \sqrt{|i^{\ast}[g_{10}] + \mathcal{F}|} \ , \qquad
\ T_p^2=\frac{\pi}{\kappa^2} (4\pi^2 \alpha^\prime)^{3-p} \ . \nn
\eeq
Here $T_p$ is the tension of the brane; for an O-plane, one has to replace $T_p$ by $-2^{p-5} T_p$. The open string excitations will not be important for
 our solution, and we shall discard the
$\mathcal{F}$ contribution from now on (note as well that the $B$-field will pull back to zero along the sources).

To derive the equations of motion, a priori, we should take a full variation of the DBI action with respect
 to the bulk metric. For supersymmetry preserving (calibrated) sources, there exists a convenient way of dealing with this. In this case,  one can
think of an expansion of the DBI action around the supersymmetric configuration and, to leading
order, replace the DBI action by a pullback of the calibration form. As discussed around (\ref{calibr}), it is given
 in terms of the non-closed pure spinor discussed in the previous section: $\Phi_-$ in  type IIA.
As shown in \cite{KT}, this allows to prove that, for Minkowski compactifications, the equations of motion follow from the first order pure spinor equations,
 and the flux Bianchi identities. A similar treatment
 of space-time filling sources is also possible for non-supersymmetric Minkowski and $AdS_4$ configurations \cite{LMMT}.
It is worth stressing that, even in these cases, the sources continue being (generalized)
 calibrated and are not responsible for the supersymmetry breaking.
However convenient, as we shall see, these kinds of source are not going to be
helpful in our search for a dS vacuum.

At this point we shall consider an important assumption: inspired by the supersymmetric case just described, we make a proposal for sources
 breaking the bulk supersymmetry. The latter can be applied in the case of an
 internal space with SU(3) structure,
 and the triviality of the canonical bundle is going to be important. We shall assume that,
in analogy with the supersymmetric case, the DBI action can be
replaced to leading order by the pullback of a (poly)form $X$ in the bulk, as discussed around (\ref{calibr-n}).
 The bulk does have invariant forms
and hence pure spinors can be constructed, but $X$ cannot be pure, otherwise the source
would preserve bulk supersymmetry. The form $X$ is expandable in the Hodge diamond defined by the pure spinors.
This amounts to consider forms that are equivalent not to simply the invariant spinor $\eta_+$ (defining the
SU(3) structure) but to a full spinorial basis, $\eta_+$, $\eta_-$,
$\gamma^{\bar{i}} \eta_+$ and $\gamma^i \eta_-$, where $i, \bar{i}=1,...3$  are the internal holomorphic
and antiholomorphic indices\footnote{The covariant derivative on the
invariant spinor contains the same information as the intrinsic
torsions. For the explicit dictionary for SU(3) structure see
\cite{FMT}. In the supersymmetric backgrounds the ($H$-twisted)
derivative on the spinor cancels against the RR contribution
\cite{GMPT4}, and the entire content of that cancellation is
captured by first order equation on the pure spinors
\eqref{eq:susyeqIIintro}. For the non-supersymmetric backgrounds,
the unbalance between the NS and RR contributions results in the
presence of terms that need to be expanded in the full basis (see
e.g. \cite{LMMT}). }. To be concrete we shall consider a generic odd
form \bea \label{X-six}
&&X=\sqrt{|g_4|}\ \d^4x \w X_- = \sqrt{|g_4|}\ \d^4x \w (\re X_- + i \im X_- ) \, ,\nn \\
&&X_-= \re X_- + i \im X_-   =\frac{8}{||\Phi_{-}||} \Big(\alpha_0 \Phi_- +\wt{\alpha}_0 \ov{\Phi}_-
+ \alpha_{m n} \gamma^m \Phi_- \gamma^n + \wt{\alpha}_{m n} \gamma^m \ov{\Phi}_- \gamma^n  \nn \\
&& \qquad\qquad\qquad\qquad\qquad \quad  + \alpha^L_m \gamma^m \Phi_+ + \wt{\alpha}^L_m \gamma^m \ov{\Phi}_+
+\alpha^R_n \Phi_+ \gamma^n + \wt{\alpha}^R_n \ov{\Phi}_+ \gamma^n  \Big) \ ,
\eea
where $\Phi_{\pm}$ are given in (\ref{su3ps25}) and the $\gamma$'s act on even and odd forms via contractions and wedges
\beq
 \gamma^m \Phi_\pm = (g^{m n} \imath_n + \d x^m) \, \Phi_\pm  \, , \qquad \mbox{and}
 \qquad \Phi_\pm \gamma^m = \mp (g^{mn} \imath_n - \d x^m) \, \Phi_\pm \, .
 \eeq

The action for a single source term becomes
\bea\label{defn-X}
S_s&=&-T_p \int_{\Sigma} \d^{p+1}x\ e^{-\phi} \sqrt{|i^{\ast}[g_{10}]|} \nn\\
&=&-T_p \int_{\Sigma} \ e^{-\phi} i^*[\im X] \nn\\
&=&-T_p \int_{M_{10}} \ e^{-\phi} \langle j_p , \im X \rangle \nn\\
&=& T_p \int_{M_{10}} \d^{10}x \sqrt{|g_{10}|} \ e^{-\phi}  \hat{\ast} \langle j_p , \im X \rangle \ ,
\eea
where $i : \Sigma \hookrightarrow M_{10}$ is the embedding of the subspace $\Sigma$ wrapped by the source in the bulk
and $j_p = \delta (\Sigma \hookrightarrow M_{10})$ is
the dimensionless Poincar\'e dual of $\Sigma$.
The change of sign between the last two lines is due to the Lorentzian signature which gives a minus
 when taking the Hodge star.  
For the sum of all sources we then take the action
\beq
S_s =  T_p \int_{M_{10}} \d^{10}x \sqrt{|g_{10}|} \ e^{-\phi}  \hat{\ast} \langle j , \im X \rangle \ , \qquad \qquad
j = \sum_{Dp} j_p -\sum_{Op} 2^{p-5} j_p \, .
\eeq
As discussed after (\ref{calibr-n}), this replacement of the source action is for now only a proposal and we hope to provide a justification for it
in future work. Our interpretation is that sources remain standard D-branes or O-planes, but their embedding into $M$, in particular
 the form which describes the subspace  wrapped by them, is modified from $\im\Phi_-$ to the more general $\im X_-$. As mentioned in the Introduction, a difference
 with the supersymmetric case is that we are not sure anymore that the equations of motion derived from both actions are the same. Our procedure will
 consist in finding solutions to the equations derived from the proposed source action, which are much easier to deal with. We
 will then argue that these solutions are also solutions of the equations derived from the standard source action. Until this is done in Section \ref{sec:check}, we mean
 by solution a solution to the equations of motion derived with our proposed source action.\\

In the following, we will consider solutions where the only non-trivial fluxes are $H$, $F_0$ and $F_2$
on the internal manifold, and the RR magnetic sources are $D6$'s and $O6$'s. The sources will be smeared, so we
take $\delta \rightarrow 1$ and the warp factor $e^{2A}=1$.
The relevant part of the action\footnote{By relevant we mean the parts of the bulk and source actions
that give non-trivial contributions to the Einstein and dilaton equations of motion and to the derivation
of the four-dimensional effective potential of Section \ref{4Dana}. We do not
write down  the Chern-Simons terms of the bulk action  and the Wess-Zumino part of the source action. Indeed
they do not have any metric nor dilaton dependence and,
since we do not allow for non-zero values of RR gauge potentials in the background, they will not
contribute to the vacuum value of the four-dimensional potential either. However,
both terms contribute  the flux e.o.m. and Bianchi identities (in particular, see \cite{KKP, DGKT, C}
for a discussion of the Chern-Simons terms in the presence of non-trivial background fluxes).},
in string frame,  is then
\beq
S=\frac{1}{2\kappa^2} \int \d^{10} x \sqrt{|g_{10}|}\ [ e^{-2\phi} (R_{10} +4|\nabla{\phi}|^2  - \frac{1}{2} |H|^2 )
 - \frac{1}{2}(|F_0|^2 + |F_2|^2 ) +2\kappa^2 T_p \ e^{-\phi}  \hat{\ast} \langle j , \im X \rangle ]\ , \label{action}
\eeq
where $\ 2\kappa^2=(2\pi)^7 (\alpha^\prime)^4$. \\

With the flux ansatz \eqref{RRans}, the flux equations of motion and Bianchi identities
reduce to the six-dimensional equations
\bea
&&\d H=0  \, , \nn\\
&&\d F_0=0\ ,\nn  \\
&&\d F_2-H\w F_0=2\kappa^2 T_p\ j  \ , \nn \\
&&H\w F_2=0 \ , \nn \\
&& \nn\\
&&\d (e^{-2\phi} * H)= - F_0\w *F_2  - e^{- \phi} \ 4\kappa^2 T_p\ j \wedge \im X_1 \, , \nn\\
&&\d( * F_2)=0 \, , \nn
\eea
where $\im X_1$ is the one-form part of $\im X_-$ in \eqref{X-six}\footnote{We refer
to \cite{KT} for a discussion of the last term in the $H$ equation of motion.}. \\

The ten-dimensional Einstein and dilaton equations in string frame now become
\bea
\label{einst}
&& R_{MN}-\frac{g_{MN}}{2} R_{10} = 2g_{MN}(\nabla^2{\phi}-2|\nabla{\phi}|^2)-2\nabla_M\nabla_N \phi + \frac{1}{4} H_{MPQ}H_N^{\ \ PQ}+\frac{e^{2\phi}}{2}F_{2\ MP}F_{2\ N}^{\ \ \ \ P} \nn\\
&& \qquad \qquad \qquad \qquad \qquad \quad  -\frac{g_{MN}}{2} \left(-4|\nabla{\phi}|^2+ \frac{1}{2} |H|^2
+ \frac{e^{2\phi}}{2}(|F_0|^2 + |F_2|^2 )\right)  +e^{\phi}\ \frac{1}{2}T_{MN} \ , \\
&&\nn\\
\label{dil}
&& 8(\nabla^2 \phi - |\nabla \phi|^2 )+ 2 R_{10} -|H|^2 = -e^{\phi} \frac{{T}_0}{p+1} \, .
\eea
Here $T_{MN}$ and $T_0$ are the source energy momentum tensor and its partial trace, respectively\footnote{\label{vielgam} In our conventions
\beq
\frac{1}{\sqrt{|g_{10}|}}\frac{\delta S_s}{\delta \phi}=- \frac{e^{- \phi}}{2 \kappa^2} \frac{T_0}{p+1} \ , \qquad \frac{1}{\sqrt{|g_{10}|}}\frac{\delta S_s}{\delta g^{MN}}=-  \frac{e^{-\phi}}{4 \kappa^2} T_{M N} \, .
\eeq
To derive \eqref{sourceTmn}, we considered the fact that each $\gamma_m$ matrix in the bispinors $\Phi_{\pm}$
carries one vielbein. To derive $C_m^{\, \, n}$ the metric dependence of the full Hodge decomposition
\eqref{X-six} must be taken into account. For supersymmetric cases, the operator $g_{P(M}dx^P \otimes \iota_{N)}$
in $T_{MN}$ is the projector on the cycle wrapped by the source \cite{GKP}.}
\bea
T_{MN}&=& 2\kappa^2 T_p \, \hat{\ast}\langle j , g_{P(M}dx^P \otimes \iota_{N)} \im X \ - \delta^m_{(M}g_{N)n} C_m^{\, \, n} \rangle  \, ,\label{sourceTmn} \\
T_{0} &=& 2\kappa^2 T_p \, \hat{\ast}\langle j , dx^N \otimes \iota_{N} \im X \rangle = (p+1)\ 2\kappa^2 T_p \hat{\ast}\langle j , \im X \rangle \, , \\
T &=& g^{MN}T_{MN} = T_{0}\ - 2\kappa^2 T_p \, \hat{\ast}\langle j , C_m^{\, \, \, m} \rangle \ .
\eea
$m,n$ are real internal indices, $C_m^{\, \, \, n} = \sqrt{|g_4|}\ \d^4x \w c_m^{\, \, \, n}$ and
\bea
c_m^{\, \, n} &=& \frac{8}{||\Phi_{-}||} \im \Big(\alpha^L_m \gamma^n \Phi_+ + \wt{\alpha}^L_m \gamma^n \ov{\Phi}_+
+\alpha^R_m \Phi_+ \gamma^n + \wt{\alpha}^R_m \ov{\Phi}_+ \gamma^n  \nn\\
&& + \alpha_{pm} \gamma^p \Phi_- \gamma^n + \alpha_{mp} \gamma^n \Phi_- \gamma^p
+ \wt{\alpha}_{p m} \gamma^p \ov{\Phi}_- \gamma^n + \wt{\alpha}_{m p} \gamma^n \ov{\Phi}_- \gamma^p \Big) \ .
\eea
For supersymmetric configurations, $\im X_- = 8 \,  {\rm Im}\Phi_-$, $c_m^{\, \, \, n} = 0$,  $T_0$ reduces to the full trace
of the source energy-momentum tensor, $T=T_0$ and  one
recovers the  formulae in \cite{KT}. \\

We can  now split \eqref{einst} into its four and six-dimensional components. Since
for maximally symmetric spaces, $R_{\mu\nu}= \Lambda g_{\mu\nu} =  (R_4/4)  g_{\mu\nu}$, for
constant dilaton, $e^{\phi}=g_s$, the four-dimensional Einstein equation   has only one component and reduces to
\beq
R_4= - 2 R_{6}+  |H|^2 + g_s^2(|F_0|^2 + |F_2|^2 )
- 2 g_s \tilde{T}_{0} = 4\Lambda \ . \label{ls}
\eeq
Not to clutter equations, in the rest of the papers we set $\tilde{T}_0 = T_{0} /(p+1)$.

This equation defines the cosmological constant, $\Lambda$.
Using the dilaton
equation \eqref{dil}, the source contribution can be eliminated and we obtain
\bea
\label{TraceEin4D1}
R_4&=&\frac{2}{3} [-R_6-\frac{g_s^2}{2}|F_2|^2+\frac{1}{2}(|H|^2-g_s^2|F_0|^2) ]  \, , \\
R_{10}&=&\frac{1}{3} [R_6+|H|^2-g_s^2 (|F_0|^2 +|F_2|^2) ] \, .
\eea

We are left with the internal Einstein equation,
\beq
\label{Ein}
R_{mn}- \frac{1}{4} H_{mpq}H_n^{\ \ pq}-\frac{g_s^2}{2}F_{2\ mp}F_{2\ n}^{\ \ \ \ p}
-\frac{g_{mn}}{6} [R_6- \frac{1}{2} |H|^2 - \frac{5}{2} g_s^2 (|F_0|^2 + |F_2|^2 )] =\frac{g_s}{2}T_{mn} \ ,
\eeq
and the dilaton equation
\beq
\label{dilTS0}
g_s \tilde{T}_{0} =\frac{1}{3} [-2R_6+|H|^2+2 g_s^2 (|F_0|^2 +|F_2|^2)] \, .
\eeq

Provided the flux equations of motion and Bianchi identities are satisfied, solving the Einstein
and dilaton equations becomes equivalent to finding the correct energy-momentum tensor for the sources.
We shall now consider an explicit example and see how the non-supersymmetric modifications
to the energy momentum tensor help in looking for de Sitter solutions. In the process we shall establish some properties
of the form $\im X_-$.

\subsection{Solvable de Sitter}
\label{dSsol1}

Our starting point is the solution described in Section \ref{susysol}, based on the algebra
\beq
(q_1(p 25+35), q_2(p 15+45), q_2(p 45-15), q_1(p 35-25), 0, 0) \ . \label{alg}
\eeq

Among the  different O6 projections compatible with the algebra for $p=0$, only those
along $146$ or $236$ are still compatible with the full algebra with $p\neq0$.
In Section \ref{susysol} we showed that, acting with a twist transformation on the supersymmetric solution
with $p=0$ and the right O6 planes, one finds a family of backgrounds characterised by the SU(3) structure
\bea
\label{dSsu3}
&& \Omega = \sqrt{t_1 t_2 t_3} (e^1+i \lambda  \frac{\tau_3}{\tau_4} e^2) \w
(\tau_3\ e^3+i \tau_4\ e^4) \w (e^5-i \tau_6\ e^6) \, , \\
&& J = t_1 \lambda  \frac{\tau_3}{\tau_4} e^1 \w e^2 + t_2  \tau_3 \tau_4 e^3 \w e^4 - t_3 \tau_6
e^5 \w e^6 \, ,
\eea
which satisfy the supersymmetry equations \eqref{eq:susyeqIIintro} only when
the parameter $\lambda = \frac{t_2 \tau_4^2}{t_1}$ is equal to one. One motivation to consider what
happens when supersymmetry is violated comes from the form of the Ricci scalar for
this class of backgrounds\footnote{\label{Ricci} The Ricci tensor of a group manifold is easily computed in frame
indices (where the metric is the unit one) in terms of the group structure constants
\beq
R_{ad}=\frac{1}{2}\left(\frac{1}{2} f_a^{\ \ bc} f_{dbc}-f^c_{\ \ db}f_{ca}^{\ \ b}-f^b_{\ \ ac} f^c_{\ \ db}
\right) \ .
\eeq
In our case, with the appropriate rescaling of the one-forms $e^a$ in \eqref{ebasisgen} and of the structure
constants, we find that the only non-zero components of the Ricci tensor are
\bea
&& R_{11} = - R_{22} = \frac{1}{2 t_1 t_2 t_3 \tau_3^2} \left[ A^2 -B^2
+ \frac{p^2}{\lambda}(A^2 - \lambda^2 B^2) \right] \, , \nn \\
&& R_{33} = - R_{44} =  \frac{1}{2 t_1 t_2 t_3 \tau_3^2} \left[ B^2 -A^2
+ \frac{p^2}{\lambda}(B^2 - \lambda^2 A^2) \right] \, , \nn \\
&& R_{55} = -\frac{1}{t_1 t_2 t_3 \tau_3^2} \left[(A-B)^2 + p^2 \left( \frac{1+\lambda^2}{2 \lambda}
(A^2+ B^2) + 2 A B \right) \right] \, , \\
&& R_{14} = R_{23} = \frac{1}{2 t_1 t_2 t_3 \tau_3^2} \frac{p}{\sqrt{\lambda}} (\lambda - 1) (A^2 -B^2) \, .
\eea
Notice that the curvature only receives contributions from $R_{55}$.}
\beq
\label{R6def}
R_6=-\frac{1}{t_1 t_2 t_3 \tau_3^2} \left[ (A- B)^{2}+p^{2} \left(\frac{(\lambda-1)^{2}}{2\lambda} (A^2+B^2) +
(A+ B)^{2}\right)\right] \ ,
\eeq
where we introduced the following quantities
\beq
A = q_1 t_1 \qquad B = q_2 t_2 \tau_3^2 \ .
\eeq
Indeed, $R_6$ gets more negative when the SUSY breaking parameters $p$ and $|\lambda-1|$ leave their SUSY value $0$.
Therefore, the value $R_4$ as given in (\ref{TraceEin4D1})
is lifted by  SUSY breaking and this is a priori promising for a de Sitter vacuum.\\

The rest of this section is devoted to the search of de Sitter solutions on the class of backgrounds discussed
above. We will take the same SU(3) structure as in \eqref{dSsu3} and metric
\beq
g =  \textrm{diag} \left( t_1, \lambda t_2 \tau_3^2, t_2 \tau_3^2, \lambda t_1, t_3, t_3 \tau_6^2 \right) \label{metric_dS}
\eeq
in the basis of $e^m$ given in \eqref{ebasisgen}. Dilaton and warp factor are still constant: $e^{\phi}=g_s$
and $e^{2 A} =1$.
For the fluxes, beside the RR two-form, we will allow for non-trivial RR zero-form and NS three-form
\bea
H &=& h \, (t_1\sqrt{t_3 \lambda}  \, e^1\w e^4 \w e^5  + t_2 \tau_3^2 \sqrt{t_3 \lambda} \ e^2\w e^3 \w e^5) \, , \label{H-field}\\
g_s F_2 &=& \gamma \sqrt{\frac{\lambda}{t_3}}   \left[ (A-B)  (e^3\w e^4-e^1\w e^2)
+ \, \frac{p}{\lambda} \, (A+B) (\lambda^2\ e^2\w e^4+e^1\w e^3)  \right] \, , \\
g_s F_0 &=& \frac{h}{\gamma} \,  .
\eea
We have introduced here another parameter $\gamma >0$ which is given by the ratio of NS and
RR zero-form fluxes. We consider again D6 or O6 sources along (236) and (146), and one can check that the $SU(3)$ structure forms and the
 fluxes chosen satisfy the orientifold projection conditions (\ref{Oproj}). Note that the NS flux has component along the covolumes\footnote{In 
order not to clutter the notations we did not divide  $v^i$ by $\sqrt{2}$ (and recalibrate the cycles accordingly) with an unfortunate consequence that
 $H$ in the normalization discussed in Footnote \ref{foot:norm} comes out as
even-quantized, and $\gamma$ is rational up to multiplication by $\sqrt{2}$.} of the sources,
 $v^1 =  t_1\sqrt{t_3 \lambda}  \, e^1\w e^4 \w e^5$ and $v^2 = t_2 \tau_3^2 \sqrt{t_3 \lambda} \ e^2\w e^3 \w e^5$.

The SUSY solutions of Section \ref{susysol} are obtained setting
\beq
\lambda=1\ \textrm{or}\ p=0, \ \gamma=1\ , F_0= h=0 \ .
\eeq

\subsubsection{The solution}\label{dSsol2}

We will first consider the four-dimensional Einstein equation \eqref{TraceEin4D1}. Using the  ansatz for the
fluxes we obtain
\bea
g_s^2|F_2|^2&=& \frac{2 \gamma^2}{t_1 t_2 t_3 \tau_3^2} \left[(A-B)^2+ p^{2} (A+B)^{2}
\left( \frac{(\lambda-1)^{2}}{2\lambda}+1 \right) \right] \ , \nn\\
|H|^2&=&2h^2 \ .
\eea
Notice that
\beq
g_s^2|F_2|^2 = 2 \gamma^2 \left[ -R_6 + p^2 \frac{(\lambda-1)^{2}}{\lambda} \, \frac{q_1 q_2}{t_3} \right] \, . \label{relF2R6}
\eeq
This allows to write the four dimensional Ricci scalar as
\beq
\label{R4ans}
R_4=  \frac{2}{3}\left[ (1-2 \gamma^2)(-R_6 -\frac{1}{2}g_s^2 |F_0|^2) +
\gamma^2 \left(-R_6 - \frac{q_1 q_2}{t_3} \ p^2 \frac{(\lambda-1)^{2}}{\lambda} \right) \right] \,  .
\eeq
Since the second bracket is positive (see \eqref{R6def}),
we see  that de Sitter solutions are possible, for instance, for $\gamma^2 \leq\frac{1}{2}$ and small $F_0$. Note also that $R_4$ clearly vanishes in the supersymmetric solution where $\lambda =1$, $\gamma=1$ and $F_0 = 0$.\\

To solve the dilaton and  internal Einstein equations it is more convenient to go to frame indices and take a
unit metric. As already discussed in Footnote \ref{Ricci}, this choice makes the computation of the
Ricci tensor very simple. To simplify notations we introduce the constant
\beq
C=-\frac{1}{6} \left(R_6- \frac{1}{2} |H|^2 - \frac{5}{2} g_s^2 (|F_0|^2 + |F_2|^2 )\right) \ .
\eeq
Then the dilaton equation becomes
\beq
\label{dileqfin}
g_s \tilde{T}_{0} =4C-\frac{h^2}{\gamma^2}- \frac{2 \gamma^2}{t_1 t_2 t_3 \tau_3^2}
\left[(A-B)^2+ p^{2} (A+B)^{2}\left( \frac{(\lambda-1)^{2}}{2\lambda}+1 \right) \right] \, .
\eeq
For the internal Einstein equations, only some components are non-trivial
\bea
g_s  T_{14} &=& \frac{1}{t_1 t_2 t_3 \tau_3^2} \frac{p}{\sqrt{\lambda}}
(A^2-B^2)(\lambda-1) (1-\gamma^{2}) \, , \nn\\
g_s  T_{23} &=& \frac{1}{t_1 t_2 t_3 \tau_3^2} \frac{p}{\sqrt{\lambda}} (A^2-B^2)(\lambda-1) (1-\gamma^{2}) \, , \nn \\
g_s  T_{11} &=& \frac{1}{t_1 t_2 t_3 \tau_3^2} \left[ A^2- B^2 + \frac{p^2}{\lambda} (A^2-B^2\lambda^2)- \gamma^2
 ((A- B)^2 + \frac{p^2}{\lambda} (A+B)^2) \right]- h^2 + 2C \, , \nn \\
g_s  T_{22} &=& \frac{1}{t_1 t_2 t_3 \tau_3^2} \left[B^2-A^2 + \frac{p^2}{\lambda} (B^2\lambda^2-A^2)- \gamma^2
( (A- B)^2 + p^2 \lambda (A+B)^2) \right] -h^2 + 2C \, , \nn \\
g_s  T_{33} &=& \frac{1}{t_1 t_2 t_3 \tau_3^2} \left[B^2-A^2 + \frac{p^2}{\lambda} (B^2-A^2 \lambda^2)- \gamma^2
( (A- B)^2 + \frac{p^2}{\lambda} (A+B)^2 ) \right] - h^2 + 2C \, , \nn \\
g_s  T_{44} &=& \frac{1}{t_1 t_2 t_3 \tau_3^2} \left[A^2-B^2 + \frac{p^2}{\lambda} (A^2 \lambda^2- B^2)- \gamma^2 \left( (A- B)^2 + p^2 \lambda (A+B)^2 \right) \right] - h^2 + 2C \, , \nn \\
g_s  T_{55} &=& - \frac{2}{t_1 t_2 t_3 \tau_3^2} \left[ (A-B)^2 + p^2 \left( \frac{(\lambda^2+1)}{2\lambda} (A^2+B^2) +2AB \right) \right]- 2h^2 + 2C  \, , \nn \\
g_s  T_{66} &=& 2 C \, . \label{TabEinst}
\eea
The remaining components set to zero the corresponding source term $T_{ab} = 0$. \\

To solve these equations we need the explicit expressions for the source energy momentum tensor,
\eqref{sourceTmn}. In six-dimensional frame indices we have
\bea
T_{ab}&=& 2\kappa^2 T_p \hat{\ast} \langle j , \delta_{c(a} e^c \otimes \iota_{b)} \im X \
- \delta^c_{(a}\delta_{b)d} C_c^d \rangle \nn\\
&=& 2\kappa^2 T_p \hat{\ast}  \left(\sqrt{|g_4|}\ \d^4x\w  \langle j , \delta_{c(a} e^c \otimes \iota_{b)} \im X_-
\ - \delta^c_{(a} \delta_{b)d} c_c^d \rangle \right) \nn\\
&=& 2\kappa^2 T_p \frac{1}{\sqrt{|g_6|}}\ \left[j\w \left( \delta_{c(a} e^c \otimes \iota_{b)} \im X_3  \
- \delta^c_{(a} \delta_{b)d} c_c^d|_3 \ \right) \right]_{1\dots6} \nn\\
&=& \frac{1}{\sqrt{|g_6|}}\ \left[\left(\d F_2- HF_0\right)\w \left( \delta_{c(a} e^c \otimes \iota_{b)} \im X_3
 \ - \delta^c_{(a} \delta_{b)d} c_c^d|_3\ \right) \right]_{1\dots6} \ .
\eea
Since, in our case,  the source $j$ is a three-form,
\beq
2 \kappa^2 T_p\ j = \d F_2 - H F_0 \, ,
\eeq
only the three-form parts $\im X_3$ and $c_c^d|_3$ of $\im X_-$  and $c_c^d$ contribute to the equations.

In the same way, we obtain
\beq
\label{simpTS0}
g_s \tilde{T}_{0} = g_s\ 2\kappa^2 T_p \, \hat{\ast} \langle j , \im X \rangle
=\frac{1}{\sqrt{|g_6|}}\ \left[g_s\left(\d F_2- HF_0\right)\w \im X_3  \right]_{1\dots6} \ .
\eeq

Combining \eqref{X-six} and the explicit expression for SU(3) pure spinors, it is easy to see that
$\im X_-$ decomposes into a one-form, a three-form and a five-form piece\beq
\im X_- = \im X_1 + \im X_3 + \im X_5 \, ,
\eeq
where\footnote{We have not imposed \eqref{defn-X} yet, and shall return to it later.}
\bea
\im X_1 &=&  (a_k^{i \, L} + a_k^{i\, R})  \d x^k - (a_k^{r \, L} - a_k^{r \, R}) g^{kj} \iota_j J
+  (g^{km}g^{jl} \iota_m \iota_l ) [ - a^r_{kj} \re \Omega + a^i_{kj} \im \Omega ] \, , \nn \\
\im X_3 &=&  -(a_k^{r \, L} +  a_k^{r \, R} ) \d x^k\w J
- (a_k^{i \, L} - a_k^{i\, R}) \,  g^{kj} \iota_j J \w J\nn\\
&&- [a_0^r  - a^r_{kj} \, (g^{kj}-(g^{kl} \d x^j+g^{jl} \d x^k)\iota_l) ]
\re \Omega  \nn \\
&&+ [a_0^i  - a^i_{kj} (g^{kj}-(g^{kl} \d x^j+g^{jl} \d x^k)\iota_l) ] \im \Omega   \, , \nn \\
\im X_5&=& - \frac{1}{2} [(a_k^{i \, L} + a_k^{i\, R}) ) \d x^k
- (a_k^{r \, L} - a_k^{r \, R}) ) g^{kj} \iota_j J ] \w J^2 \nn \\
&&- \d x^k\w \d x^j\w [ - a^r_{kj}  \re \Omega + a^i_{kj}  \im \Omega ] \, . \label{polyX}
\eea
The superscripts $r$ and $i$ indicate real and imaginary parts:
\bea
a^r_{0} = \re (\alpha_0 - \wt{\alpha}_0) \, ,  &\qquad &  a^r_{jk} = \re (\alpha_{jk} - \wt{\alpha}_{jk}) \, ,\nn \\
a^i_{0} = \im (\alpha_0 + \wt{\alpha}_0) \, ,  &\qquad &  a^i_{jk} = \im (\alpha_{jk} + \wt{\alpha}_{jk}) \, .
\eea
and
\bea
a^{r \, L}_k = \re (\alpha_k^L - \wt{\alpha}_k^L) \, ,& \qquad \qquad &
a^{r \, R}_k = \re (\alpha_k^R -\wt{\alpha}_k^R ) \, ,\nn  \\
a^{i \, L}_k =  \im (\alpha_k^L + \wt{\alpha}_k^L) \, , & \qquad \qquad &  a^{i \, R}_k  = \im (\alpha_k^R + \wt{\alpha}_k^R) \, .
\eea

As already discussed, only the three-form parts of $\im X_-$ and $c_c^d$
contribute to the equations. Then, for simplicity, we choose to set to zero $\im X_1$ and $\im X_5$.
This amounts to setting
\beq
a^{r \, L}_k = a^{i \, L}_k = a^{r \, R}_k =  a^{i \, R}_k = 0 \, ,
\eeq
and choosing $a^r_{jk}$ and $a^i_{jk}$ symmetric. Then, in frame indices, $\im X_3$
becomes
\bea
\label{Xisimp}
\im X_3  &=&  [ a^i_0 - Tr(a^i_{b c}) +  a^i_{b c}(\delta^{bd} e^c  +\delta^{cd} e^b )\iota_d] \im  \Omega \nn \\
&&- [a^r_0 - Tr(a^{r}_{b c}) + a^r_{b c}(\delta^{bd} e^c  +\delta^{cd} e^b )\iota_d ] \re \Omega   \, .
\eea

Similarly, we find that the three-form part of $c_a^b$ is given by
\bea
c_a^b|_3 &=& 2 a^i_{ac}  [-\delta^{bc} + (\delta^{cd} e^b  +\delta^{bd} e^c )\iota_d ] \im \Omega \nn \\
&&- 2 a^r_{ac} [-\delta^{bc} + (\delta^{cd} e^b  +\delta^{bd} e^c )\iota_d ] \re \Omega \, .
\eea

The coefficients in $\im X_3$ are free parameters which should be fixed by solving
 the dilaton and internal Einstein equations. \\

The equations  $T_{mn}=0$ are satisfied by choosing\footnote{The parameters  $a^i_{12}$,
$a^i_{13}$, $a^i_{24}$, $a^i_{34}$, $a^i_{56}$ are not fixed by any equation. For simplicity,
we decide to put them to zero.}
\bea
&& a^i_{0} =  0 \qquad \qquad  \, \,  a=1,\ldots, 6 \, ,\nn \\
&& a_{bc}^i=0  \qquad  \qquad  b, c = 1,\ldots, 6  \,  , \nn \\
&& a_{bc}^r=0 \qquad   \qquad  (bc) \notin \{(bb),\ (14),\ (23) \} \, . \label{0para}
\eea

The Einstein and dilaton equations, \eqref{TabEinst} and \eqref{dileqfin}  fix the other parameters
\bea
\label{para}
a^r_{0} &=&- g_s\ \frac{\tilde{T}_{0}+ T_{55}+ T_{66}-x_0}{2(c_1+c_2)} \, ,  \nn \\
a^r_{14} &=& g_s\ \frac{T_{14}}{2(c_2-c_1)} \, ,  \nn\\
a^r_{23} &=& g_s\ \frac{T_{23}}{2(c_1-c_2)} \, ,\nn\\
a^r_{11} &=& g_s\ \frac{1}{2 (c_2-c_1)} \, \left[T_{11}-\frac{c_2 \tilde{T}_0}{c_1+c_2} +
\frac{ x_0 c_1c_2}{(c_1^2-c_2^2)} \right] \, , \nn \\
a^r_{22} &=& g_s\ \frac{1}{2 (c_1-c_2)} \, \left[T_{22}-\frac{c_1 \tilde{T}_0}{c_1+c_2} +
\frac{ x_0 c_1c_2}{(c_2^2-c_1^2)} \right] \, ,\nn \\
a^r_{33} &=& g_s\ \frac{1}{2 (c_1-c_2)} \, \left[T_{33}-\frac{c_1 \tilde{T}_0}{c_1+c_2} +
\frac{ x_0 c_1c_2}{(c_2^2-c_1^2)} \right] \, , \nn \\
a^r_{44} &=& g_s\ \frac{1}{2 (c_2-c_1)} \, \left[T_{44}-\frac{c_2 \tilde{T}_0}{c_1+c_2} +
\frac{ x_0 c_1c_2}{(c_1^2-c_2^2)} \right]  \, , \nn \\
a^r_{55} &=& -g_s\ \frac{T_{55}}{2(c_1+c_2)} \, , \nn \\
a^r_{66} &=& g_s\ \frac{T_{66}- \tilde{T}_0}{2(c_1+c_2)} \, ,
\eea
where $x_0=2 \tilde{T}_0-(T_{11}+T_{22}+T_{33}+T_{44})$ and
$T_{ab}$ are given by \eqref{TabEinst}. The coefficients $c_1$ and $c_2$ appear in the source
term of the Bianchi identity for $F_2$
\beq
g_s (\d F_2 - H F_0) = c_1\ v^1 +c_2\ v^2 \, ,
\eeq
where $v^1$ and $v^2$ are covolumes of sources in the directions $(146)$ and $(236)$ and
\bea
\label{ccoeff}
c_1 &=& -\frac{h^2}{\gamma} + \frac{q_1 q_2}{ A t_3} \, \gamma \left[2 (A-B) - p^2 \frac{\lambda^2+1}{\lambda} \, (A+ B) \right] \, ,  \nn \\
c_2 &=& -\frac{h^2}{\gamma} + \frac{q_1 q_2}{ B t_3} \, \gamma \left[ 2 (B-A) - p^2 \frac{\lambda^2+1}{\lambda} \, (A+ B) \right]
\, .
\eea
In agreement with our quantization conventions (see Footnote
\ref{foot:norm}), we impose that $(c_1 + c_2)$ is an integer. We emphasize once more, that the overall tension of
 the intersecting sources is always negative (and so is $c_1 + c_2$), but depending on the parameters of the solution the
 individual sources may be either O6 planes or D6 branes.\\

So far, we have solved the external and internal Einstein equations, the dilaton equation of motion, and checked that the Bianchi identity for $F_2$ is
 satisfied. As far as the bulk fields are concerned, we should also solve the equations of motion and the remaining Bianchi identities for the
 fluxes. These are actually automatically satisfied by our ansatz for the fluxes, provided $j\w \im X_1=0$. As a matter of fact, our choice of
 the parameters $a$ in \eqref {0para} already sets $\im X_1$ to zero, so we are done with the bulk fields. 

As a last step in the construction of a de Sitter solution
 (we recall we mean here a solution to the equations derived from our proposed action for the sources), we need to check the source fields equations of
 motion. One should vary our source action with respect to the world-volume coordinates and the gauge fields. The latter is trivially satisfied, since
we do not consider any gauge field here, and the pullback of the $B$-field giving \eqref{H-field} vanishes. For the
world-volume coordinates, from our action $-T_p \int_{\Sigma} \ e^{-\phi} i^*[\im X]$ and WZ, one can derive, as discussed in the Introduction,
an equation of motion of the form
\beq
\del_{[i_1} (e^{-\phi} \im X_3)_{i_2 i_3] {\alpha}} \sim (*F_2)_{[i_1 i_2 i_3] {\alpha}} \ ,
\eeq
where $i_k$ label world-volume directions, and $\alpha$ is orthogonal. One can check that pulling back any three indices of the four-form $*F_2$ to the
 world-volume gives zero, as discussed after (\ref{condstab}). The left-hand side also vanishes (see \eqref{dImX}), and so we conclude that the world-volume equations of motion
are satisfied.\\

This concludes our resolution of all equations of motion derived from the action (\ref{action}) which contains our proposal for sources breaking 
bulk supersymmetry. Provided one chooses the free parameters as discussed below (\ref{R4ans}), one can obtain a de Sitter solution. In the next
section, we come back to the question of generalizing first order differential equations to the non-supersymmetric case. This will fix for us the
free parameters to values which indeed give a de Sitter solution. In Section \ref{sec:check} we will argue that the solution we found
here is also a solution to the equations derived with the standard source action.

\subsection{More on the polyform $X$}\label{MoreX}

In this section, we will try to provide further justification for our choice of polyform $X_-$.

In supersymmetric compactifications,
the imaginary part of the non-closed pure spinor, $\Phi_-$ in type IIA, on one side, defines the
calibration for the sources and, on the other, gives the bulk RR fields in the supersymmetry equations \eqref{eq:susyeqIIintro}.  We will
show that, for our de Sitter solution, the polyform $X_-$  satisfies the same equations $\Phi_-$ satisfies in the
 supersymmetric case
\bea
(\d-H) \re X_- &=& 0 \, ,  \nn\\
(\d-H) \im X_- &=& c_0 \ g_s * \lambda(F) \, , \label{collect}
\eea
where the constant $c_0$ can a priori be different from $1$.\\

Keeping only the parameters $a$ that are non-zero in the de Sitter solution \eqref{para}, it is easy to compute
\bea
\d (\im X_-) & =&  [ (a_{0}^r+a_{66}^r-a_{55}^r ) [ p (q_1 + q_2)  (e^1 \w e^3 + e^2 \w e^4 )  \nn \\
&&-  (q_1 - q_2)  (e^1 \w e^2 - e^3 \w e^4) ] \w e^5 \w e^6 \nn \\
&&- (a_{11}^r+a_{44}^r-a_{22}^r-a_{33}^r) [ p (q_1 - q_2) (e^1 \w e^3 + e^2 \w e^4 ) \nn \\
&&- (q_1 + q_2)  (e^1 \w e^2 - e^3 \w e^4 ) ]  \w e^5 \w e^6  \, , \label{dImX}
\eea
and
\beq
H\w \im X_- =  -2 h \ (a_{0}^r+a_{66}^r-a_{55}^r) \ e^1 \w e^2 \w
e^3 \w e^4 \w e^5 \w e^6 \, .
 \eeq

In order to have $\d(\im X_-)$ proportional to $g_s *F_2$, one must impose the relation
\beq a_{11}^r+a_{44}^r-a_{22}^r-a_{33}^r=0 \ .
\label{Constraint}
\eeq
Then, one has
\bea
 \d(\im X_-) &=& -c_0\ g_s *F_2 \ , \nn \\
 H\w \im X_- &=& -2\gamma^2\ c_0\ g_s *F_0 \ ,
\eea
with
\beq
\label{ceq}
c_0=\frac{a_{0}^r+a_{66}^r-a_{55}^r}{\gamma} = - g_s \frac{\tilde{T}_0}{\gamma (c_1 + c_2)} \ .
\eeq
To obtain the second equality, we used the explicit expression \eqref{para}, \eqref{TabEinst} for the parameters $a$, while $c_1$ and
$c_2$ are defined in \eqref{ccoeff}.
Also, using \eqref{para}, it is easy to show that the constraint (\ref{Constraint}) reduces to
\beq
x_0=2 \tilde{T}_0-(T_{11}+T_{22}+T_{33}+T_{44})=0 \qquad \Leftrightarrow \qquad (2 \gamma^2-1)\ h^2 = 0 \ .
\eeq
Therefore, for\footnote{Clearly also $h=0$ (no NS flux) is a solution to this constraint.
It would be interesting to explore the possibility of having de Sitter or non-supersymmetric Minkowski solution
with $h=0$. Notice that, in this case, the condition of having $F_0 \neq 0$ \cite{HSUV}, necessary to avoid de Sitter
no-go theorems \cite{HKTT}, is not required.}
\beq
\gamma^2=\frac{1}{2}
\eeq
we can write a differential equation for $\im X_-$
\beq
(\d-H) \im X_- =c_0\ g_s * \lambda(F) \, ,
\eeq
which is the analogue of the supersymmetry equations\footnote{Notice that from the equation for $\im X_-$  we recover the condition $T_0 >0$ \eqref{dilTS0}.
Indeed, as in \cite{GMPT6}, starting from (\ref{simpTS0}) we have
\beq
\frac{T_0}{p+1} \int_M {\rm vol}_{(6)} =
-\int_M \langle \d_H F , \im X_- \rangle \nn\\
= -\int_M \langle F , \d_H \im X_- \rangle \nn\\
= c_0\ g_s\ \int_M   \langle * \lambda(F) , F \rangle > 0 \ .
\eeq} for $\im \Phi_-$. In addition, fixing the value $\gamma^2 = 1/2$ gives a de Sitter solution, according to the condition (\ref{R4ans}).

The value of the constant $c_0$ is also fixed by the solution. Indeed, in order for $X_-$ to reproduce the correct Born-Infeld
action \eqref{defn-X} on-shell, we get from our solution that a combination of coefficients of $X_-$ has to be one: $a_{0}^r+a_{66}^r-a_{55}^r=1$. Out of
\eqref{ceq}, we deduce that we have to impose $c_0 \, \gamma = 1$. This relation is automatically satisfied for supersymmetric backgrounds, where $c_0 = \gamma = 1$ and the pullback of
 $\re \, \Omega$ agrees with the DBI action on the solution. In our non-supersymmetric solution, the condition $c_0 \, \gamma = 1$ fixes
 the value of the constant, $c_0 = \sqrt{2}$.

More generally, requiring the two actions being equal on-shell can be formulated as $-g_s \tilde{T}_0 =c_1 + c_2$, where the right-hand side
 is given by the sum of the source charges. Indeed, as we can see in \eqref{simpTS0}, if $\im X$ gives the sum of the source volume forms on-shell,
 and $j$ or $\d_H F$ gives the sum of the charges times the covolumes (Bianchi identity), then $\tilde{T}_0$ should be given by the sum of the charges;
this sum is negative, hence the minus sign. We can verify that this condition is equivalent for our solution to the condition $c_0 \, \gamma = 1$, given
the second equality in \eqref{ceq}. Finally, let us note that such a relation would fix one of the three parameters $h, \gamma, \lambda$ in terms of
 the others and the moduli. In particular, for $\lambda=1$, one gets
\beq
h^2=\frac{(A-B)^2 + p^2 (A+B)^2}{t_1 t_2 t_3 \tau_3^2}\ \frac{(\gamma-1)(1-2\gamma) \gamma^2}{\gamma^2 - 3 \gamma + 1} \ .\label{hl1}
\eeq
Note one clearly recovers the supersymmetric case with $\gamma=1$. For our de Sitter solution, one should impose instead
 $\gamma=\frac{1}{\sqrt{2}}$, and then $h \neq 0$.\\


We can now show that $\d_H$ - closure can be imposed on  $\re X_-$. Indeed, the three-form part of $\re X_-$ can be written as
\bea
\re X_3&=&- [b_{0}^r - Tr(b_{kj}^r) + b_{kj}^r (g^{kl} dx^j+g^{jl} dx^k)\iota_l ] \re \Omega \nn\\
&& +  [b_{0}^i - Tr(b_{kj}^i) + b_{kj}^i (g^{kl} dx^j+g^{jl} dx^k)\iota_l ] \im \Omega \nn\\
&& +  [(b_{k}^{i \, R} -b_{k}^{i \, L}) \ dx^k +g^{kl} (b_{k}^{r R} - b_{k}^{r L}) \ \iota_l J ] \w J \ ,
\eea
where, as for $\im X_3$,  we have defined
\bea
b^r_{0} = \im (\wt{\alpha}_0 - \alpha_0) \, \,   &\qquad &  b^r_{kj} = \im (\wt{\alpha}_{kj} - \alpha_{kj}) \ ,\nn \\
b^i_{0} = \re (\wt{\alpha}_0 + \alpha_0)  \, \, &\qquad &  b^i_{kj} = \re (\wt{\alpha}_{kj} + \alpha_{kj}) \ ,\nn \\
b^{r \, L}_{k} =\re(\wt{\alpha}_{k}^{L}+\alpha_{k}^{L}) & \qquad & b^{r \, R}_{k} = \re (\alpha_{k}^{R}+ \wt{\alpha}_{k}^{R}) \ ,\nn\\
b^{i \, L}_{k} = \im(\wt{\alpha}_{k}^{L}-\alpha_{k}^{L}) & \qquad & b^{i \, R}_{k} = \im (\alpha_{k}^{R}- \wt{\alpha}_{k}^{R}) \ .
\eea
Consistently with (\ref{0para}), we can choose
\bea
&& b^r_{0} =0 \, ,  \nn \\
&&  b^{r \, L}_{k} = b^{r \, L}_{k} = b^{i \, L}_{k}= b^{i \, R}_{k} = 0 \qquad \forall \ k=1,\ldots 6 \, \nn \\
&& b_{j k}^r=0  \, \qquad \qquad \qquad \qquad \qquad  \forall \ j,k =  1,\ldots 6  \, \nn \\
&& b_{jk}^i=0 \qquad {\rm for}\ (kj) \notin \{(kk),\ (14),\ (23),\ (41),\ (32)\} \, .
\eea
Furthermore, choosing
\beq
\frac{b_{14}^i}{t_1}=-\frac{b_{23}^i}{t_2 \tau_3^2} \quad , \quad \frac{b_{11}^i}{t_1} +\frac{b_{33}^i}{t_2 \tau_3^2} -\frac{b_{22}^i}{t_2 \tau_3^2 \lambda} -\frac{b_{44}^i}{t_1 \lambda} =0 \ ,
\eeq
we obtain
\beq
\d_H (\re X_3)=\sqrt{t_1 t_2 t_3}\ \tau_3 \tau_6 \ p (1-\lambda) \ \left(b_0^i+\frac{b^i_{66}}{t_3 \tau_6^2}-\frac{b^i_{55}}{t_3}\right) (q_2\ e^1 \w e^4 +q_1\ e^2\w e^3)\w e^5 \w e^6 \ ,
\eeq
which is zero either in the SUSY solution, or by further setting
\beq
b_0^i = - \frac{b^i_{66}}{t_3 \tau_6^2}+\frac{b^i_{55}}{t_3} \, .
\eeq
\\

While these equations are derived in the vanishing warp factor and constant dilaton limit, their extension to the general case
is natural\footnote{Just like $\Phi_-$, $X_-$ is globally defined, and both $B$-field and the dilaton are needed in order to
define an isomorphism between such forms and the positive and negative helicity spin bundles $S^{\pm}(E)$ \cite{GMPW}. The
dilaton assures the correct transformation under $GL(6)$, making
the (non-pure) spinor $e^{-\phi}e^{-B} \, X_-$  the natural variable for the first order equations \eqref{dirty-one}. }
\bea \label{dirty-one}
&& {\rm d}_H(e^{2A -\phi}  \re X_- ) =  0 \, ,\nn\\
&& {\rm d}_H(e^{4A -\phi} \im X_-  ) =  c_0 e^{4A} \ast \lambda(F)  \, .
\eea
In general the odd form $X_-$ should receive contribution from both pure spinors, but in our solution we have chosen to ``decouple'' the even pure spinor completely. Note that any two objects in the trio of the even and odd compatible pure spinors and the metric determine the third. Here we have worked with the almost complex structure and the metric.
In the supersymmetric backgrounds it is clearly more convenient to solve the first order equations for the pure spinors rather than the Einstein equation for the metric. Hence it is natural to ask if and when it might be possible to find an even-form counterpart to \eqref{dirty-one}, $X_+$ , so that $X_-$ and $X_+$ (together with flux Bianchi identities) imply the solution to the Einstein equations. However it is not yet clear to us what the correct generalization of the notion of compatibility is, and what algebraic properties $X_+$ should satisfy.  Hoping for a symmetry with the supersymmetric solutions (and the possibility of having a solution to some variational problem) one may construct $X_+$ satisfying
\beq \label{dirty-two}
{\rm d}_H(e^{3A -\phi} X_+  ) =  0  \, .
\eeq
Assuming $X_+$ has an expansion similar to that of $X_-$,
which does not receive contributions from $\Omega$,  this amounts to finding a closed two-form on
$\Gg_{5.17}^{p,-p,\pm 1} \times S^1$. It is indeed not hard to construct such a form for our solution, since
the symplectic form itself is closed, provided $\tau_2=0$ (even if $\lambda\neq1$, see \eqref{newsol}). Even if we do not take $\tau_2=0$, finding a conformally closed $X_+$ of this form is always possible, since the manifold is symplectic.
A better understanding of such first order equations applicable to non-supersymmetric backgrounds
is a work in progress and we hope to return to it in a future publication.

\subsection{A solution for the standard source action?}\label{sec:check}

In this paper we made a proposal of an action for sources breaking bulk supersymmetry. As discussed in the
 Introduction and at  the beginning of Section \ref{sec:dS}, we cannot conclude (as one would do in the supersymmetric case) whether the equations of 
 motion derived from the action (\ref{action}) are the same
as those derived from the standard source action DBI + WZ. Our proposal is to be considered as an 
assumption with interesting consequences, we are not able to prove such an equivalence, but hope to provide a better justification of it in future work. What can be done
is to verify that the solution found in our example is indeed a solution to the equations of motion derived from the standard source action. Let us discuss now
in practice what should be checked, starting with the world-volume equations of motion.\\

There are two equations to consider, coming from the variation of DBI + WZ action
 with respect to the world-volume coordinates and the gauge fields  (for a general form of these equations see \cite{ST}).
 The latter is easier, and we shall consider it first. In our solution the dilaton is constant and the world-volume gauge fields
 vanish. Moreover we recall that the pullback of the $B$-field computed from \eqref{H-field} also vanishes. Then the
 equation reads
\beq
\partial_i \left(e^{-\phi} \sqrt{|i^*[g]|}\ (i^*[g])^{[ij]} \right) \sim \epsilon^{jkl} \left(i^* [*F_4]\right)_{kl} \ ,
\eeq
where $i,j,k,l$ are indices along the brane world-volume. Since our solution has no RR four-form flux, both sides vanish trivially.
 The variation of the world-volume action with respect to the world-volume coordinates
 (again, in presence of constant dilaton and vanishing pullback of $B$) connects the trace of the second fundamental form
 $\mathcal{S}^{\alpha}_{ij}$ to the RR fluxes (${\alpha}$ spans normal directions). It reads
\beq
e^{-\phi}(i^*[g])^{ij} \mathcal{S}^{\alpha}_{ij}  \sim \epsilon^{jkl} (*F_2)^{\alpha}_{jkl} \ .
\eeq
One can check\footnote{This check is analoguous to that of the corresponding equation of motion derived from our proposed source action, as discussed
 at the end of Section \ref{dSsol2}.} that pulling back any three indices of the four-form $*F_2$ to the world-volume gives zero.
 For our intersecting configuration, we need to worry only about ${\alpha}=5$, and may use the relation of the second fundamental form with the
 (components of) the spin connection $\omega^{\alpha}_i = \mathcal{S}^{\alpha}_{ij} e^j$. We can check that while
 the second fundamental form does not vanish (the embedding is not geodesic), it has no diagonal element. However the metric \eqref{metric_dS} in the
 basis \eqref{ebasisgen} is diagonal, and $(i^*[g])^{ij} \mathcal{S}^{\alpha}_{ij} $ vanishes. Thus the world-volume equations of motion are satisfied.\\
 
Let us now consider the bulk field equations of motion. As mentioned at the end of Section \ref{dSsol2}, the ansatz chosen for the fluxes guarantees that their equations of motion and Bianchi
 identities are satisfied. Let us also emphasize the following details: first we do not have any $B$-field along the sources and therefore a correction
term due to the source in its equation of motion could be discarded; second the proposed generalization of the first order equations \eqref{dirty-one}, satisfied
by our solution, guarantees that the RR equations of motion are satisfied. Therefore, for the bulk fields, only the internal Einstein equation and the dilaton
equation of motion remain to be checked. 

The dependence of the dilaton equation on the source action is simply through $\tilde{T}_0$ (see for instance \eqref{dilTS0}), which is proportional to the
 source action on-shell. Therefore, as long as the standard source action and our proposed action match on-shell, the dilaton equations of motion are the
 same. As discussed in the previous section, this equality amounts in general to the condition $-g_s \tilde{T}_0 =c_1 + c_2$, which for our solution
is equivalent to $c_0 \gamma=1$. This fixes one of the three parameters $h, \gamma, \lambda$ in terms of the others and the moduli (see for instance
 \eqref{hl1}). Provided this condition is enforced, the dilaton equation of motion derived from DBI is therefore satisfied by our solution.

We are now left with the internal Einstein equation. An explicit check can be done for the family given by:
\beq
  \lambda=1 \qquad F_0\neq0\ , \ h\neq0 \ , \ \mathrm{given} \ {\rm by}\ \eqref{hl1} \ ,
\eeq
with particular interest in the non-supersymmetric value $\gamma=\frac{1}{\sqrt{2}}$ giving our de Sitter solution. Solving the Einstein equation amounts
to match the values of the energy-momentum tensor $T_{ab}$ given by \eqref{TabEinst}. In the supersymmetric case, one can derive
from the standard source action that the non-zero components of $T_{ab}$ of one source are the diagonal ones along the source directions, and are all equal.
We recover this situation in the family we consider by simply taking $\gamma=1$. For our non-supersymmetric solution,
the supersymmetry breaking will manifest itself as $T_{55}\neq0$ and $T_{66}\neq T_{11}+T_{22}$. Then, in order to match the results, one needs to consider
 a non-trivial dependence of the embedding functions on the metric moduli. The computation is rather involved and not particularly enlightening, thus
 we will not present it here. However, let us note that this non standard embedding corresponds to our interpretation of the proposed action, as discussed
in the Introduction. We can also obtain a perturbative solution (the perturbation parameter is $\epsilon=\lambda-1$) where the deviation from the
 SUSY solution is more severe due to $T_{14}$ and $T_{23}$ not being zero as opposed to their supersymmetric value.\\

Let us end this section by adding few words about the stability of our solution. Solving all the equations of motion of course means extremizing the
 energy density of the bulk plus brane system, but we cannot be sure that the solution is a minimum for
arbitrary values of the parameters. The problem is currently under study. For the time being we can try to give some heuristic justification of the
 fact we believe our non-supersymmetric solution is stable. For $\lambda = 1$ and $\gamma =1$ the manifold admits the
 supersymmetric solution described in Section \ref{susysol}. By keeping $\lambda=1$ and setting $\gamma = 1/\sqrt{2}$
 we obtain a non-supersymmetric solution with the same internal geometry as in the
SUSY case, meaning the metric is not changed and the directions wrapped are the same. The pullback of $\im X_-$ does coincide with the pullback
  of the (generalized) calibrating form $\re \Omega$. In a sense the brane is still wrapping a minimal volume cycle (even if this is done with a different
embedding), and we can imagine
the parameters, other than $\gamma$, can be chosen in such a way to have small contributions to the potential from
 the supersymmetry breaking term, and the energy density of combined bulk and brane system at the minimum.

\subsection{Four-dimensional analysis}\label{4Dana}

In this section we do a partial study of the stability of our solution  by analyzing the four dimensional effective
 potential with respect to two moduli.

 The search for de Sitter vacua, or for no-go theorems against their existence, has generally
 been performed from a four-dimensional point of view \cite{IRW, HKTT, Si, HSUV, DHSV, CKKLWZ, FPRW, CGM},
analysing the behaviour of the four dimensional effective potential with respect to its moduli dependence.
In this section, we want to make contact with this approach and show that our solution has
the good behaviour one expects to find for de Sitter vacua, as far as the volume and the dilaton are concerned.
 We use in this section the ten-dimensional action (\ref{action}) which contains our proposal for sources breaking bulk supersymmetry.
We will show that this proposal gives rise to interesting new terms in the potential.

\subsubsection{Moduli and $4d$ Einstein frame}

Let us consider the ten-dimensional action (\ref{action}). By Kaluza-Klein reduction on
the internal manifold, we obtain a four-dimensional effective action for the moduli. In particular, in addition to the kinetic terms, the
four-dimensional action will contain a potential for the moduli fields. Their number and the way they enter the potential
will depend on the peculiar features of the single model.\\

A de Sitter solution of the four-dimensional effective action will correspond to a positive valued minimum of the potential. Determining the
minima of the potential is in general rather difficult, since, a priori one should extremize along all the directions in the moduli space.
This complicated problem is generally solved only by numerical analysis, because of the large number of variables.
However, some information can be extracted by restricting the analysis to  a subset of the moduli fields.

For whatever choice of the manifold on which
the compactification is performed, we are always able to isolate two universal moduli: the internal volume and
the four-dimensional dilaton. Their appearance in the effective potential at tree-level is also universal. We will then only focus on these two moduli.  We define the internal volume as
\beq
\int_M \d^6 x \sqrt{|g_6|}=\frac{L^6}{2}=\frac{L_0^6}{2}\rho^3 \ ,
\eeq
where the factor of $\frac{1}{2}$ is due to the orientifold and the vacuum value is  $\rho=1$.
Defining the ten-dimensional dilaton fluctuation as
$e^{- \tilde{\phi}}  = g_s e^{-\phi} $,  the four-dimensional dilaton is given by
\beq
\sigma = \rho^{\frac{3}{2}} \, e^{- \tilde{\phi}}  \, .
\eeq

Then reducing the action (\ref{action}), we obtain the four-dimensional effective action for gravity,  4$d$ dilaton
and volume modulus in the string frame
\beq
S = \frac{1}{2\kappa^2} \int \d^4 x \sqrt{|g_{4}|} \left[ \frac{L^6}{2} e^{-2\phi} (R_{4}+4|\nabla \phi|^2) - 2 \kappa^2 U \right]
\label{actionS} \, ,
\eeq
with $U(\rho, \sigma)$ the four-dimensional potential. To derive the explicit form of the potential, we need to determine how the internal
Ricci scalar,  fluxes and source terms scale with the volume. For $R_6$ and the fluxes this is easily computed
\beq
R_6 \rightarrow   \rho^{-1} \, R_6 \, , \quad  |H|^2 \rightarrow   \rho^{-3} \, |H|^2\, , \quad    |F_k|^2 \rightarrow \rho^{-k}  \, |F_k|^2 \, .
\eeq
The source term requires some more attention. As shown in (\ref{simpTS0}),
\beq
2\kappa^2 T_p \, \hat{\ast} \langle j , \im X \rangle
=\frac{\left[\left(\d F_2- HF_0\right)\w \im X_3  \right]_{1\dots6}}{\sqrt{|g_6|}} \ .
\eeq
The terms in  $\im X_3$ in (\ref{polyX}) appearing with $a_0$, $a_{jk}$ and $a_{k}^{(L,R)}$  scale differently with the volume. Let us
denote them  by  $X_0$, $X_\Omega$ and $X_{J}$, respectively
\beq
\im X_3 =X_0+X_{\Omega}+X_J \ .
\eeq
Their $\rho$ dependence is determined by the scaling of the forms $J$ and $\Omega$
\beq
J \rightarrow \rho J \ , \ \Omega \rightarrow \rho^{\frac{3}{2}}\Omega \ ,
\eeq
and by the metric factors in the gamma matrices  of  (\ref{polyX})
\beq
X_0 \rightarrow \rho^{\frac{3}{2}} X_0 \ ,\ X_{\Omega} \rightarrow \rho^{\frac{1}{2}} X_{\Omega} \ , \ X_J \rightarrow \rho X_J \ .
\eeq
Then, the source term scales as
\beq
\label{relbi}
\frac{\left[(\d F_2-HF_0) \w \im X_3 \right]_{1\dots6}}{\sqrt{|g_{6}|}} \rightarrow \rho^{-\frac{3}{2}}\ \left(b_0 + b_1\ \rho^{-1} + b_2\ \rho^{-\frac{1}{2}} \right)\ ,
\eeq
where
\bea
 \label{bidef}
&& b_0= \frac{\left[(\d F_2-HF_0) \w X_0 \right]_{1\dots6}}{\sqrt{|g_{6}|}} \, ,\nn \\
&& b_1= \frac{\left[(\d F_2-HF_0) \w X_{\Omega} \right]_{1\dots6}}{\sqrt{|g_{6}|}} \, , \nn \\
&&  b_2= \frac{\left[(\d F_2-HF_0) \w X_{J} \right]_{1\dots6}}{\sqrt{|g_{6}|}} \, ,
\eea
are vacuum values.
Then the four-dimensional potential for $\rho$ and $\sigma$ becomes
\bea
U &=&  \frac{1}{2\kappa^2} \int_M \d^6 x \sqrt{|g_6|} [ e^{-2\phi} (-R_{6} + \frac{1}{2} |H|^2 )
+ \frac{1}{2}(|F_0|^2 + |F_2|^2 ) -2\kappa^2 T_p \ e^{-\phi}  \,\hat{\ast}\langle j , \im X \rangle ]  \nn  \\
&=&  \frac{L_0^6}{4 g_s^2 \kappa^2}\ \sigma^{2} \,
[ ( - \frac{R_{6}}{ \rho}  + \frac{|H|^2}{2 \rho^3})
- \frac{g_s}{\sigma} \,   (b_0 +  \frac{b_1}{\rho} + \frac{b_2}{\sqrt{\rho}} )   +  \frac{g_s^2 \, \rho^3}{2 \sigma^2} (|F_0|^2 +  \frac{|F_2|^2}{ \rho^2} ) ] \,  .
\eea
Note that the terms in $b_1$ and $b_2$ are purely non-supersymmetric contributions of the source. They are due to the new metric dependence of the source action with respect to the supersymmetric case.\\

In order to correctly identify the cosmological constant, but also to perform the study of the moduli dependence, we need to go to the four-dimensional Einstein frame
\beq
g_{\mu\nu\ E} = \sigma^2\ g_{\mu\nu} \, .
\eeq
The four-dimensional Einstein-Hilbert term transforms as\footnote{Under a conformal rescaling of the four dimensional metric we have
\beq
g_{\mu\nu}\rightarrow e^{2\lambda} g_{\mu\nu} \qquad\Rightarrow \qquad
\sqrt{|g_4|} \rightarrow e^{4\lambda} \sqrt{|g_4|} \ , \ R_4 \rightarrow e^{-2\lambda} R_4 \ .
\eeq}
\bea
\frac{1}{2\kappa^2} \int \d^4 x \sqrt{|g_{4}|} \frac{L^6}{2} e^{-2\phi} R_{4}
&=& \frac{L_0^6}{2g_s^2\ 2\kappa^2} \int \d^4 x \sqrt{|g_{4}|}\ \sigma^2 R_{4} \nn \\
&=& M^2_4 \int \d^4 x \sqrt{|g_{4E}|} R_{4E} \, , \nn
\eea
where we denote Einstein frame quantities by $E$, and we introduced $M^2_4 = \frac{L_0^6}{2g_s^2\ 2\kappa^2}$, the squared four-dimensional Planck mass.
Similarly, the four-dimensional potential in the Einstein frame becomes
\beq
 \label{4Dpotential}
U_E = \sigma^{-4}\ U  =4\kappa^4 M_4^4 \frac{e^{4\phi}}{(\frac{L^6}{2})^2} \ U \ ,
\eeq
and we can write the Einstein frame action as
\beq
S = M_4^2 \int \d^4 x \sqrt{|g_{4E}|} \left( R_{4E}+{\rm kin} - \frac{1}{M_4^2} U_E \right) \, .
\eeq

The cosmological constant, \eqref{ls}, is then related to the vacuum value of the potential
\beq
\Lambda =\frac{1}{2 M_4^2} U_E|_0 \ .
\eeq

\subsubsection{Extremization and stability}

In order to find a solution, one should determine the minima of the potential. For our choice of moduli, $\rho$ and $\sigma$, one  has
\bea
&& \frac{\partial U_E}{\partial \sigma} = - \frac{M_4^2}{\sigma^{5}}  [ 2 g_s^2 \,  (|F_0|^2 \rho^3 + |F_2|^2 \rho) + 2  \sigma^2\ ( - \frac{R_6}{ \rho}  +
\frac{  |H|^2 }{2 \rho^3})  -3\sigma\ g_s (b_0+ \frac{b_1}{\rho} +\frac{ b_2}{ \sqrt{\rho}} ) ] \ , \label{ders}\\
&& \frac{\partial U_E}{\partial \rho} =  \frac{M_4^2}{ \sigma^2 }\ [ (\frac{R_6 }{\rho^2}  - \frac{3 |H|^2}{2 \rho^4})
+ \frac{g_s}{\sigma} \ ( \frac{b_1}{\rho^2}  + \frac{b_2}{2 \, \sqrt{\rho^3}} ) + \frac{g_s^2}{2 \sigma^2}
( 3 |F_0|^2 \rho^2 + |F_2|^2)]  \, . \label{derr}
\eea
In our conventions, the extremization conditions are
\beq
\frac{\partial U_E}{\partial \sigma}|_{\sigma=\rho=1} =0 \quad , \quad \frac{\partial U_E}{\partial \rho}|_{\sigma=\rho=1} =0 \ , \label{extre}
\eeq
where  $\sigma=\rho=1$ are the values of the moduli on the vacuum. Actually, the  conditions (\ref{extre}) are equivalent to the ten-dimensional dilaton e.o.m. and the trace of internal Einstein equation.  Combining the dilaton equation \eqref{dilTS0} and the trace of the internal Einstein equation, (\ref{Ein}), we can write the six-dimensional Ricci scalar as
\beq
 \label{TraceEin6D}
R_6= \frac{3}{2} |H|^2 - \frac{g_s^2}{2} (3 |F_0|^2 +|F_2|^2) - \frac{g_s}{2} (T_0-T) \, ,
\eeq
where
\bea
T_0-T &=& 2\kappa^2 T_p \hat{\ast}\langle j , C_m^m \rangle = \frac{\left[(\d F_2-HF_0) \w (X_J + 2 X_{\Omega})  \right]_{1\dots6}}{\sqrt{|g_{6}|}} \nn \\
&=&  2 b_1 +  b_2 \, .
\eea
In the last line we used (\ref{bidef}).  With this expression for $T_0 -T$, it is immediate to verify that
\eqref{TraceEin6D} is indeed equal to
the $\partial_\rho U_E$ in (\ref{extre}). Similarly, one can see that using  (\ref{derr}), (\ref{relbi}), (\ref{simpTS0}) and  (\ref{extre}), the dilaton equation (\ref{dilTS0})  reduces to $\partial_\sigma  U_E$ in (\ref{extre}). \\

From the equivalence of the ten-dimensional equations and \eqref{extre} we see that the ten-dimensional solution discussed in the previous sections does indeed  satisfy the extremization conditions (\ref{extre}). The next step is to see whether such extremum correspond to
a minimum of the potential and  whether, furthermore, it is stable.

Let us consider (\ref{derr}) and discuss the $\rho$ dependence of the potential. It is convenient to define the function
\beq
P(\rho^2)= \frac{\partial U_E}{\partial \rho}\ \frac{\sigma^{2} \rho^{4}}{M_4^2} \ .
\eeq
It is easy to check that $P(\rho^2)$ is negative for $\rho=0$ and positive for $\rho \rightarrow \infty$.
Hence there must be a real positive root and this is a minimum of $U_E$.
A priori, $P(\rho^2)$ could have other zeros. Let us focus only on the situation in which $b_2=0$, which, in particular, is the case for our ten-dimensional solution. In that case,  $P(\rho^2)$ has two other roots which are either complex conjugate\footnote{Since the polynomial is real, they come in conjugate pairs.}, or real and negative, according to the value of the parameters. Indeed, studying $\partial_{\rho^2} P$, one can show that $P(\rho^2)$ can be $0$ only once. Therefore, at least for $b_2=0$, there is only one extremum of $U_E$ in $\rho$ and it is a minimum. So satisfying the extremization in $\rho$ is enough for the stability. \\

Let us now analyze the $\sigma$ dependence of (\ref{4Dpotential}). It is easy to see that the potential admits an extremum  for
\beq
\sigma_\pm=\frac{1}{4a}\left(3b\pm \sqrt{8b^2\left(\frac{9}{8}-\frac{4ac}{b^2} \right)}\right)  \qquad \qquad
\frac{4ac}{b^2}<\frac{9}{8} \ ,
\eeq
where for simplicity we introduced
\bea
&& a = - R_6 \rho^{-1} + \frac{1}{2} |H|^2 \rho^{-3} \, , \nn\\
&& b = g_s (b_0 + b_1 \rho^{-1} + b_2 \rho^{- \frac{1}{2}} ) \, , \nn \\
&& c =  \frac{g_s^2}{2} \rho^3 (|F_0|^2 + |F_2|^2 \rho^{-2})   \, .
\eea
In our case, asking for $\sigma=1$ and using the extremization in $\sigma$ in $(\ref{extre})$, which can be written as $2a-3b+4c=0$, we find that the minimum in $\sigma_-$ corresponds to
\beq
a-2c <0 \ .
\eeq
This condition is satisfied by our solution choosing $\gamma^2=\frac{1}{2}$, as we can see from (\ref{relF2R6}). Therefore, our solution is at the minimum in $\sigma$, and it is then stable both in the volume and the dilaton moduli.\\

It is easy to see that the four-dimensional potential takes a positive value
at the minimum, and, hence,
the minimum corresponds to a de Sitter vacuum. In \cite{Si}, it has been shown that the potential has a strictly positive minimum in $\sigma$ for
\beq
1<\frac{4ac}{b^2}<\frac{9}{8} \ , \label{bound}
\eeq
where the lower bound comes from asking the potential to be never vanishing (strictly positive). This condition is satisfied by our solution.\\

In addition, we can actually compute the value of the potential at $\sigma=\rho=1$. Starting from (\ref{4Dpotential}) and using the two equations of (\ref{extre}), we obtain
\beq
\frac{U_E}{M_4^2} = \frac{1}{3} \left(\frac{g_s}{2} (T_0-T) +g_s^2 |F_{0}|^2 - |H|^2 \right) \ . \label{UEmin}
\eeq
Using (\ref{TraceEin4D1}) and (\ref{TraceEin6D}), one can show that the four-dimensional Ricci scalar
is proportional to \eqref{UEmin}, $R_4 = 2 U_E/ M_4^2$. For $\gamma^2=1/2$, $R_4$ is positive (see
the discussion below \eqref{R4ans}), and hence so is the value of the potential at the minimum.

Note also that, for $\gamma^2 = 1/2$,  the last
two terms in \eqref{UEmin} cancel each other and the entire contribution to the cosmological constant comes
from sources, $(T_0 - T)$. For supersymmetry breaking branes, this contribution is never vanishing but, for
generic situations, we do not know what its sign is.
It would be nice to have a model independent argument to determine whether, for this
mechanism of supersymmetry breaking, the resulting four-dimensional space is always de Sitter. \\

As a further check of the existence of a de Sitter minimum for our solution,
we can plot the four-dimensional potential $U_E$ as a function of
$\sigma$ and $\rho$ for  some values of the parameters
\bea
&& t_{1}=t_{2}=t_{3}=\tau_{3}=\tau_{6}=1 \, , \nn \\
&& q_{1}=1\ ,\ q_{2}=3\ , \ p=\frac{\cosh^{-1}(2)}{\pi} \, , \nn \\
&& \lambda=5\ ,\ \gamma=\frac{1}{\sqrt{2}}\ ,\ h=4 \ .
\eea

\begin{figure}[H]
\begin{center}
\begin{tabular}{cc}

\includegraphics[height=4cm,width=6cm]{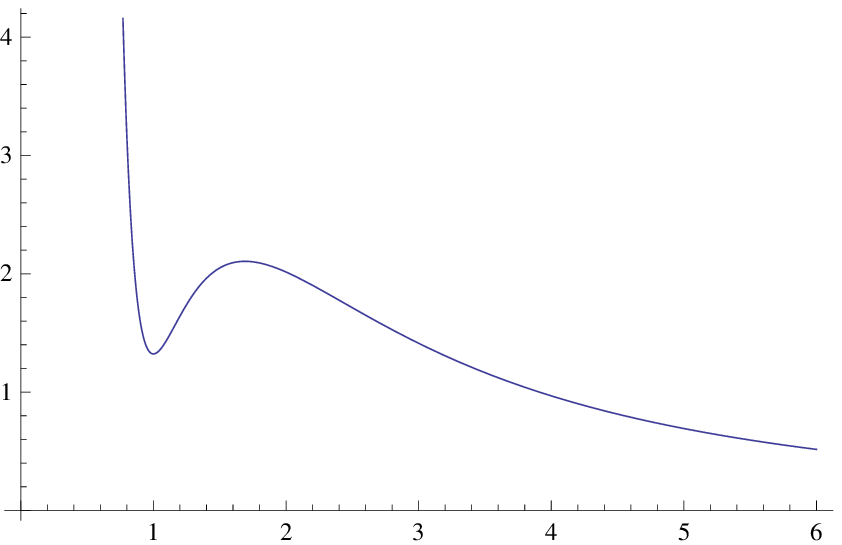}
&
\includegraphics[height=4cm,width=6cm]{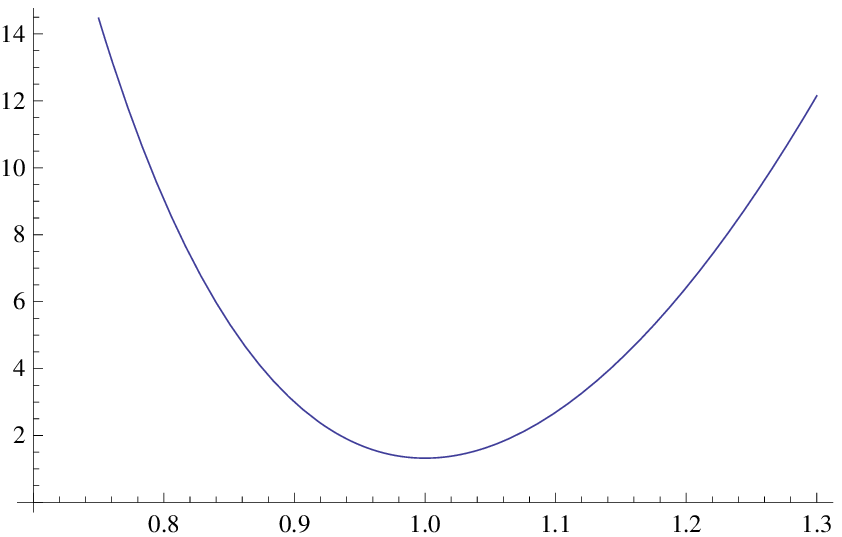}\hspace{1cm}
\\
& \\
$\frac{1}{M_4^2} U_E(\sigma,\rho=1)$
& $\frac{1}{M_4^2} U_E(\sigma=1,\rho)$

\end{tabular}\caption{Dependence of the potential on dilaton and volume modulus}\label{Plot}
\end{center}
\end{figure}


\section* {Acknowledgements}

We would like to thank Ch. Bock, A. Tomasiello and D. Tsimpis for numerous useful discussions; helpful discussions
with D. Cassani, A. Dabholkar, U. Danielsson, M. Goodsell, A. Kashani-Poor, P. Koerber, S. Rollenske, M. Rubin,  W. Schulgin, S. Theisen and T. Van Riet are also gratefully acknowledged.   R.M. would like to thank Max Planck Institute for gravitational physics at Potsdam  for hospitality and A. von Humboldt  foundation for support. This work is supported in part by  ANR grants BLAN05-0079-01 (DA and MP) and BLAN06-3-137168 (EG and RM).

\newpage

\appendix

\section{Solvable algebras and the geometry of solvmanifolds}\label{apmath}

\subsection{Algebraic aspects}\label{apalg}


We consider a connected and simply-connected real Lie group $G$ of identity element $e$. $H$, $N$ and $\Gamma$ will
be subgroups of $G$. We denote the associated Lie (sub)algebras of $G$, $H$, $N$ by $\Gg$, $\Gh$, $\Gn$.
Connected and simply-connected (sub)groups are in one-to-one correspondence with the corresponding (sub)algebras.
Many properties of the (sub)algebras will have their counterpart in the (sub)groups and vice versa.\\

The ascending series $(G_k)_{k\in \mathbb{N}}$, the descending series $(G^k)_{k\in \mathbb{N}}$ and the derived
series $(D^k G)_{k\in \mathbb{N}}$ of subgroups of $G$ are defined as
\bea
&& G_0=\{e\} \ , \ G^0=D^0 G=G \ ,\nn\\
&& G_k=\{g\in G | [g,G]\subset G_{k-1} \} \ , \ G^k=[G,G^{k-1}] \ , \ D^k G=[D^{k-1} G, D^{k-1} G] \ ,\nn
\eea
where the commutator of two group elements $g$ and $h$ is $[g,h]=ghg^{-1}h^{-1}$. We define in the same way the
ascending, descending and derived series of $\Gg$ or its subalgebras, by using the Lie bracket instead of the
commutator, and $0$ instead of $e$.

$G$ is {\it nilpotent} respectively {\it solvable} if there exist $k$ such that $G^k=\{e\}$ respectively
$D^k G =\{e\}$. We define the same notions for the algebra $\Gg$ replacing $0$ with $e$. Lie (sub)algebras
corresponding to nilpotent/solvable groups are nilpotent and solvable, respectively. The converse is also true.
All nilpotent Lie algebras/groups are solvable (the converse is not true). \\

An ideal $\mathfrak{i}$ of $\Gg$ is a subspace of $\Gg$ stable under the Lie bracket:
$[\Gg, \mathfrak{i}] \subset \mathfrak{i}$. Obviously $\mathfrak{i}$ is also a subalgebra.
The subalgebras given in the previously defined series are all ideals.

The nilradical $\Gn$ of the algebra $\Gg$ is the biggest nilpotent ideal of $\Gg$. The nilradical is unique
\cite{Au, CF} as will be the corresponding subgroup $N$ of $G$, also named nilradical.

To ideals of $\Gg$ will correspond normal subgroups of $G$. We recall that a subgroup $N$ is said normal if
$\forall g \in G$, $gNg^{-1} \subset N$, i.e. it is invariant under conjugation (inner automorphisms).
This property is necessary in order to be able to define a group structure on the quotient $G/N$.
Note that the nilradical $N$ of a solvable Lie group $G$ as well as the subgroups $D^{k}G$ of the derived
serie are normal subgroups.

\subsubsection{The adjoint action}\label{apad}

Let $V$ be a vector space over a field $\mathbb{K}$ and let $\Gg$ be a Lie algebra over the same field.
A representation of $\Gg$ is a map $\pi:\Gg\rightarrow\mathrm{End}(V)$ such that:
\begin{enumerate}
\item $\pi$ is linear\ ;
\item $\pi\left([X,Y]\right)=\pi(X)\pi(Y)-\pi(Y)\pi(X)$ \ .
\end{enumerate}
There is a natural representation of a Lie algebra over itself called the adjoint representation:
\begin{eqnarray}
ad&:&\Gg\rightarrow\mathrm{End}(|\Gg|)\nn\\
&& X\mapsto ad(X)=ad_{X}\nn  \ ,
\end{eqnarray}
where $|\Gg|$ means the underlying vector space of the Lie algebra $\Gg$, $\mathrm{End}(|\Gg|)$ the space of all linear maps on it\footnote{These maps do not necessarily respect the Lie bracket, or in other words, are not necessarily algebra morphisms. In particular, for $X \in \Gg$, $ad_{X}$ is not an algebra morphism.}, and
\bea
{\rm for}\ X\in \Gg\ , \ ad_{X}&:&\Gg\rightarrow\Gg \nn\\
&& Y\mapsto ad_{X}(Y)=[X,Y] \ . \nn
\eea

We can obtain a matrix form of the adjoint representation from the structure constants in a certain basis of the Lie algebra. Let $\{E_{a}\}_{a=1,\dots,d}$ be a basis of a Lie algebra $\Gg$, and the structure constants in that basis given by
\beq
[E_{b},E_{c}]=f^{a}_{\ \ bc}E_{a} \ .
\eeq
Then the matrices ($a$ is the row index and $c$ is the column index)
\beq
(M_{b})^{a}_{\ c}=f^{a}_{\ \ bc}
\eeq
provide a representation of the Lie algebra $\Gg$.\\

A unimodular algebra $\Gg$ is such that $\forall X\in \Gg$, $tr(ad_X)=0$. In view of what has been discussed, this is equivalent to $ \sum_a f^a_{\ \ ba} =0$, $\forall b$ .\\

Let $G$ be a Lie group and let $V$ be a (real) vector space. A representation of $G$ in $V$ is a map
$\pi:G\rightarrow\mathrm{Aut}(V)$ such that:
\begin{enumerate}
\item $\pi(e)=Id$ \ ;
\item $\pi(g_{1}g_{2})=\pi(g_{1})\pi(g_{2}) \ , \ \forall\,g_{1},g_{2}\in G$ \ .
\end{enumerate}
There is a natural representation of the group over its algebra called the adjoint representation:
\begin{eqnarray}
Ad&:&G\rightarrow\mathrm{Aut}(\Gg)\nn\\
&& g\mapsto Ad(g)=Ad_g \ ,\nn
\end{eqnarray}
where $Ad_g=exp^{Aut(|\Gg|)}(ad_{X_{g}})$ for $X_{g}\in\Gg\ , \ exp^{G}(X_{g})=g$. Actually one can show the following relations between the representations:

\medskip\centerline {\xymatrix{
  G  \ar[r]^-{Ad} & \mathrm{Aut}(\Gg) \\
  \Gg \ar[u]^{exp^{G}} \ar[r]^-{ad} & \mathrm{End}(|\Gg|) \ar[u]_{exp^{Aut(|\Gg|)}}}}\medskip

The map $ad$ then turns out to be the derivation\footnote{It is the derivative with respect to the parameters of the group element $g$, taken at the identity.} of $Ad$. At the level of the single elements, they act according to the following diagram:

\medskip\centerline{\xymatrix{ g \ar[r]^-{Ad} & Ad(g)=Ad_{g} \\
                               X_{g}\ar[u] \ar[r]^-{ad} & ad(X_{g})=ad_{X_{g}}\ar[u]}}\medskip

One can show as well that the derivation of the inner automorphism $I_g$ for $g\in G$ (the conjugation) is actually the adjoint action $Ad_g$:
\beq
d(I_g)=Ad_g \ .
\eeq
Furthermore, for $\varphi:G\rightarrow G$ an automorphism, the following diagram is commutative:

\medskip\centerline{\xymatrix{ G\ar[r]^{\varphi} & G \\
                               \Gg\ar[u]^{exp^{G}}\ar[r]^-{d\varphi }& \Gg\ar[u]_{exp^{G}} }}\medskip

A Lie group is said to be exponential (the case for us) if the exponential map is a diffeomorphism. Denoting its inverse as $log^{G}$, then we deduce
\beq
I_{g}=exp^{G}\circ Ad_{g}\circ log^{G} \ .\label{map}
\eeq

\subsubsection{Semidirect products}\label{apsemi}

Most of the solvable groups we are interested in are semidirect products, we recall here some definitions.\\

Let us consider two groups $H$ and $N$ and a (smooth) action $\mu:H\times N\rightarrow N$ by (Lie) automorphisms.
The semidirect product of $H$ and $N$ is the group noted $H \ltimes_{\mu} N$, whose underlying set is $H \times N$
and the product is defined as
\beq
(h_{i=1,2},n_{i=1,2})\in H\times N \ , (h_1,n_1) \cdot (h_2,n_2) = (h_1 \cdot h_2, n_1\cdot \mu_{h_1}(n_2)) \ .
\eeq
\\
The semidirect product of Lie algebras can be defined in a similar way. Let $\mathfrak{d}(\Gh)$ be the derivation
algebra of an algebra $\Gh$ (for instance $ad \in \mathfrak{d}(\Gg)$). Let $\sigma:\Gg\rightarrow\mathfrak{d}(\Gh)\ ,\ X \mapsto \sigma_{X}$ be a representation of the Lie algebra $\Gg$ in $|\Gh|$. Then we can define the semidirect
product $\Gg\ltimes_{\sigma}\Gh$ of the two Lie algebras with respect to $\sigma$ in the following way:
\begin{itemize}
\item the vector space is $|\Gg| \times| \Gh|$
\item the Lie bracket is $\left[(X_{1},Y_{1}),(X_{2},Y_{2})\right]=\left([X_{1},X_{2}]_{\Gg},[Y_{1},Y_{2}]_{\Gh}+\sigma_{X_{1}}(Y_{2})-\sigma_{X_{2}}(Y_{1})\right)$.
\end{itemize}
This provides a Lie algebra structure to the vector space $|\Gg| \times| \Gh|$. Note that the fact $\sigma$ is
a derivation is important to verify the Jacobi identity for the new bracket.

If we denote $\Gg'=\Gg\times\{0\}$ and $\Gh'=\{0\}\times\Gh$ then $\Gh'$ is an ideal of the new algebra and $\Gg'$
is a subalgebra of it. Furthermore
\beq
\Gg'+\Gh'=\Gg\ltimes_{\sigma}\Gh \ ,\ \Gg'\cap\Gh'=0 \ .
\eeq
There is a unique decomposition of an element of $|\Gg| \times| \Gh|$ as a sum of an element of $|\Gg|$ and one of
$|\Gh|$, thus we can think of it as the couple in $|\Gg| \times| \Gh|$ or as an element of a direct sum of
vector spaces.\\

Let us consider a Lie group $G$ and two subgroups $H$ and $N$ with $N$ normal. If every element of $G$ can be uniquely
written as a product of an element in $H$ and one in $N$, then one can show that $G \approx H \ltimes_{\mu} N$ with
$\mu$ being the conjugation\footnote{In particular it is the case for a group $G = H \ltimes_{\nu} N$ with $\nu$
being not the conjugation.}. This point of view will be important for us. As discussed previously, the conjugation can
be given in terms of the restriction of the adjoint action of $H$ over $\Gn$ as in (\ref{map}), so we are able to
determine $\mu$ in terms of $Ad_H(N)$. For exponential groups, as we consider here, the corresponding Lie algebra of
$G = H \ltimes_{\mu} N$ is then clearly $\Gg = \Gh \ltimes_{ad_{\Gh}(\Gn)} \Gn$ (we just write $ad$ in the following
for simplicity).

Let us now consider a group $G$ with a normal subgroup $N$ of codimension $1$. The Lie algebra $\Gg$ has two components,
 $\mathbb{R}$ and $\Gn$. We want to show that $\Gg$ is isomorphic to $\mathbb{R} \ltimes_{ad} \Gn$, and then, as
discussed, we get that $G \approx \mathbb{R} \ltimes_{\mu} N$ with $\mu$ the conjugation. At level of the algebra,
in terms of vector spaces, the isomorphism is obviously true. What needs to be verified is that the Lie brackets
coincide. The Lie bracket of two elements of $\mathbb{R}$ or of $\Gn$ clearly coincide with those of the corresponding
two elements of $\mathbb{R} \ltimes_{ad} \Gn$. Let us now take $X \in \mathbb{R}, \ Y \in \Gn$. We have for
$\mathbb{R} \ltimes_{ad} \Gn$:
\beq
[(X,0), (0,Y)]=(0, 0+ ad_X (Y) - ad_0 (0))=(0, [X,Y]) \, ,
\eeq
 which clearly coincides with the bracket $[X,Y]$ for $\Gg$. We can conclude that $\Gg$ is isomorphic to
$\mathbb{R} \ltimes_{ad} \Gn$ and thus the group is isomorphic to $\mathbb{R} \ltimes_{\mu} N$.

\subsubsection{Solvable groups}\label{apsolv}

According to Levi's decomposition, any real finite dimensional
Lie algebra is the semidirect sum of its largest solvable ideal called the radical, and a semi-simple
subalgebra. So solvable and nilpotent algebras do not enter the usual Cartan classification.
Solvable algebras $\Gg$ are classified with respect to the dimension of their nilradical $\Gn$. One can show
\cite{M1,B} that $\textrm{dim}\ \Gn \geq \frac{1}{2} \textrm{dim}\ \Gg$. Since we are interested in six dimensional
 manifolds we will consider $\textrm{dim}\ \Gn= 3, \dots, 6$. If $\textrm{dim}\ \Gn= 6$, $\Gn=\Gg$ and the algebra is
nilpotent (they clearly are a subset of the solvable ones).  There are $34$ (isomorphism) classes of
six-dimensional nilpotent algebras
(see for instance \cite{GMPT6, Sal} for a list), among which $24$ are indecomposable. Among the $10$ decomposable
algebras, there is of course the abelian one, $\mathbb{R}^6$. There are $100$ indecomposable solvable algebras with
$\textrm{dim}\ \Gn= 5$ ($99$ were found in \cite{M3}, and \cite{T} added $1$, see \cite{CS} for a complete and
corrected list), and $40$ indecomposable solvable algebras with $\textrm{dim}\ \Gn= 4$ \cite{T}. Finally, those with
$\textrm{dim}\ \Gn= 3$ are decomposable into sums of two solvable algebras. There are only $2$ of them, see Corollary
1 of \cite{M4}. In total, there are $164$ indecomposable six-dimensional solvable algebras.
For a list of six-dimensional
indecomposable unimodular\footnote{See Appendix \ref{apad} for a definition.} solvable algebras, see \cite{B}.\\

Most of the solvable groups are semidirect products. For $G$ a solvable group and $N$ its nilradical, we consider
the following definitions:
\begin{itemize}
\item If $G=\mathbb{R} \ltimes_{\mu} N$, $G$ is called almost nilpotent. All three and four-dimensional solvable
groups are of that kind \cite{B}.
\item If furthermore, the nilradical is abelian (i.e. $N=\mathbb{R}^k$), $G$ is called almost abelian.
\end{itemize}
The result at the end of the previous section applies here: any solvable group for which
${\rm dim}\ N={\rm dim}\ G-1$ is almost nilpotent. In fact $N$ is a normal subgroup of $G$.
Let us label the $\mathbb{R}$ direction with a parameter $t$, which  we can take as a coordinate,
with the corresponding algebra element being $\partial_t$. According to (\ref{map}), we then have
\beq
\mu(t)=exp^{N}\circ Ad_{e^{t\partial_t}}(\Gn)\circ log^{N} \ , \
Ad_{e^{t\partial_t}}(\Gn)=e^{ad_{t\partial_t}(\Gn)}=e^{t\ ad_{\partial_t}(\Gn)} \ .
\eeq
Furthermore, for the almost abelian case, we can identify $N$ and $\Gn$, so the $exp$ and $log$ correspond to the
identity. Then, we obtain the simpler formula
\beq
\label{apmu}
\mu(t)= Ad_{e^{t\partial_t}}(\Gn)=e^{t\ ad_{\partial_t}(\Gn)} \ .
\eeq

We will mainly focus on solvable algebras with $\textrm{dim}\ \Gn= 5$ (to which correspond almost nilpotent solvable groups) because, as we will discuss further, the compactness question is simpler to deal with.

\subsection{Compactness}
\label{apcomp}

We recall here that according to the definition\footnote{Let us emphasize the non-trivial result that, according to
our (restrictive) definition solvmanifolds, these are always parallelizable (see \cite{OT} for a proof).} we adopt
in this paper (Section \ref{susyback}) a solvmanifold is a compact homogeneous space $G/\Gamma$ obtained by
the quotient of a connected, simply-connected solvable group and a discrete cocompact subgroup $\Gamma$, the
lattice \cite{B, OT}. The main result concerning the geometry of these manifolds is the Mostow bundle, and we
refer to Section \ref{susyback} for its discussion (see in particular diagram (\ref{Mostowbundle}) and
\cite{Mostow} for the original reference). In this appendix, we come back to the problem of the existence of a lattice.

Whether a lattice exists or not, and so whether the manifold can be made compact is not always an easy question
for non-nilpotent solvable groups. There is a simple necessary condition for a  manifold to be compact,
namely that the algebra has to be unimodular. Sufficient conditions
are on the contrary more difficult to establish.

A theorem by Malcev \cite{M} states that a connected and
simply-connected nilpotent Lie group $G$ admits a lattice if and only if there exists a basis for the Lie algebra $\Gg$
such that the structure constants are rational numbers. This condition is always satisfied for
all the $34$ classes of nilpotent six dimensional algebras. For the non-nilpotent cases, several criteria have been
proposed. The first is due to  Auslander \cite{Au}. Despite its generality the criterion is difficult to use in concrete
situations and we will not refer to it in our search for lattices. Details about it can be found in
the original paper \cite{Au} and in \cite{B}. Another criterion, which is closer to the one we use in this paper,
is due to Sait\^o \cite{S}. It is less general than Auslander's because it applies to solvable groups that
 are algebraic subgroups of $GL(n,\mathbb{R})$ for some $n$. The criterion deals with the adjoint action of
the group $G$ over the nilradical $\Gn$ of its algebra $\Gg$. For an illustration, see \cite{GMPT6}.

The criterion we adopt in this paper follows \cite{B} and it applies to almost abelian solvable groups. As discussed
above almost abelian solvable groups are characterized  by the map $\mu(t)$ \eqref{apmu}. Then the criterion
states the group $G$ admits a lattice if and only if it exists a $t_0 \neq 0$ for which $\mu(t_0)$ can be conjugated
to an integer matrix. This criterion is very useful in practice since we have a simple formula (\ref{apmu})
for $\mu(t)$.

In \cite{B}, some almost nilpotent (not almost abelian) cases were also proved to admit a lattice,
thanks to some further technique that we will not consider here. \\

In Section \ref{susyback} we applied the compactness criterion mentioned above to the two algebras $\varepsilon_2$ and
$\varepsilon_{1,1}$ (corresponding to $\Gg^0_{3.5}$ and $\Gg^{-1}_{3.4}$ given in the Table 1, respectively).
Here we will review the argument for $\varepsilon_{1,1}$, using a change of basis closer to \cite{B}.
The algebra $\varepsilon_{1,1}$ is defined by
\beq
\label{apE1basis1}
[E_1,E_3]=E_1\ ,\ [E_2,E_3]=-E_2 \ .
\eeq
We have $\Gn=\{E_1,E_2\}$ and $\partial_t=E_3$. Then, in the $(E_1,E_2)$ basis,
\beq
ad_{\partial_t}(\Gn)=\left( \begin{array}{cc} -1 & 0 \\ 0 & 1 \end{array} \right) \ , \ \mu(t)=e^{t \ ad_{\partial_t}(\Gn)}=\left( \begin{array}{cc} e^{-t} & 0 \\ 0 & e^t \end{array} \right) \ .
\eeq
It is not possible to have $\mu(t_0)$ being an integer matrix for $t_0 \neq 0$. To check if
the group admits a lattice, we have to find another basis where the matrix $\mu(t_0)$ can be integer.
Let us consider the particular change of basis given by
\beq
P=\left( \begin{array}{cc} 1 & c \\ 1 & \frac{1}{c} \end{array} \right) \ , \ P^{-1}=\frac{1}{c-\frac{1}{c}}\left( \begin{array}{cc} -\frac{1}{c} & c \\ 1 & -1 \end{array} \right) \ ,
\eeq
where $c=e^{-t_1}$ and $t_1\neq 0$. Then:
\beq
\hat{\mu}(t)=P^{-1} \left( \begin{array}{cc} e^{-t} & 0 \\ 0 & e^{t} \end{array} \right) P= \left( \begin{array}{cc} \frac{\sinh(t_1-t)}{s_1} & -\frac{\sinh(t)}{s_1} \\ \frac{\sinh(t)}{s_1} & \cosh(t)+c_1 \frac{\sinh(t)}{s_1} \end{array} \right) \ ,
\eeq
with $s_1=\sinh(t_1)$ and $c_1=\cosh(t_1)$. For $t=t_1$, we get
\beq
\hat{\mu}(t=t_1)=\left( \begin{array}{cc} 0 & -1 \\ 1 & 2c_1 \end{array} \right) \ . \label{apBbasis}
\eeq
The conjugated matrix $\hat{\mu}(t)$ can have integers entries for some non-zero $t=t_1$ when
$2\cosh(t_1)$ is integer. In \cite{B}, $2\cosh(t_1)=3$.\\

Let us now describe an example for which there is no lattice. We consider the algebra $\Gg^{-p}_{4.2}$
\beq
[E_1,E_4]=-p E_1\ ,\ [E_2,E_4]=E_2\ ,\ [E_3,E_4]=E_2 +E_3 \ , \ p\neq 0 \ .
\eeq
It is easy to check that the algebra is unimodular only for $p=2$. This is a necessary condition for
compactness, we can exclude all other values of $p$.

We have $\Gn=\{E_1,E_2,E_3\}$ and $\partial_t=E_4$ (the algebra is almost abelian). Then, in the $(E_1,E_2,E_3)$ basis,
\beq
ad_{\partial_t}(\Gn)=\left( \begin{array}{ccc} p & 0 & 0 \\ 0 & -1 & 0 \\ 0 & -1 & -1 \end{array} \right) \ , \ \mu(t)=e^{t \ ad_{\partial_t}(\Gn)}=\left( \begin{array}{ccc} e^{pt} & 0 & 0 \\ 0 & e^{-t} & 0 \\ 0 & -t e^{-t} & e^{-t} \end{array} \right) \ .
\eeq
Following \cite{B}, we are going to prove that this matrix cannot be conjugated to an integer matrix\footnote{A na\"ive
reason one could think of would be that it is due to the off-diagonal piece, but as we are going to show, this
piece actually does not contribute.} except for $t=0$. A way to verify if the matrix $\mu(t)$
can be conjugated to an integer one is to look at the coefficients of its characteristic polynomial $P(\lambda)$. This is
independent of the basis in which it is computed, and hence, for the criterion to be satisfied it
should have integer coefficients. Here we have:
\beq
P(\lambda)=(\lambda-e^{2t})(\lambda-e^{-t})^2=\lambda^3-\lambda^2 (2e^{-t}+e^{2t})+\lambda (e^{-2t}+2e^{t})-1 \ .
\eeq
The coefficients are given by sums and products of roots. We can use Lemma $(2.2)$ in \cite{H}.
Let
\beq
P(\lambda)=\lambda^{3}-k\lambda^2+l\lambda-1\in\mathbb{Z}[\lambda] \, .
\eeq
Then $P(\lambda)$ has a double root $\lambda_{0}\in\mathbb{R}$ if and only if $\lambda_{0}=+1$ or $\lambda_{0}=-1$
for which $P(\lambda)=\lambda^{3}-3\lambda^{2}+3\lambda-1$ or $P(\lambda)=\lambda^{3}+\lambda^{2}-\lambda-1$ respectively.

In our case, we find the double root $e^{-t}$. This means the only way to have this polynomial with integer
coefficients is to set $t=0$. Then we can conclude there is no lattice.\\


\subsubsection{Algebras admitting a lattice}
\label{aptables}

We present here a list of indecomposable solvable, non-nilpotent unimodular algebras that admit a lattice (at least
for certain values of the parameters $p,q,r$, for instance those chosen in table \ref{apglobdef}.).
For dimension up to four the algebras are almost nilpotent or almost abelian. For dimension 5 and 6,
only almost abelian algebras have been considered.
For the other six-dimensional indecomposable algebras, we do not know if a lattice exists.

\begin{center}
\begin{table}[H]
\begin{center}
\begin{tabular}{|c|c|c|}
\hline
Name & Algebra & \\
\hline
\hline
 $\Gg_{3.4}^{-1}$ & $[X_1,X_3]=X_1, [X_2,X_3]=-X_2$ & alm. ab.  \\
\hline
 $\Gg_{3.5}^{0}$ & $[X_1,X_3]=-X_2, [X_2,X_3]=X_1$ & alm. ab.   \\
\hline
\hline
$\Gg_{4.5}^{p,-p-1}$ & $[X_1,X_4]=X_1, [X_2,X_4]=p X_2, [X_3,X_4]=-(p+1) X_3$, $-\frac{1}{2} \leq p < 0$ & alm. ab.  \\
\hline
 $\Gg_{4.6}^{-2p,p}$ & $[X_1,X_4]=-2p X_1, [X_2,X_4]=p X_2-X_3, [X_3,X_4]=X_2+pX_3$, $p > 0$ & alm. ab.  \\
\hline
 $\Gg_{4.8}^{-1}$ & $[X_2,X_3]=X_1, [X_2,X_4]=X_2, [X_3,X_4]=-X_3$ & alm. nil.\\
\hline
 $\Gg_{4.9}^{0}$ & $[X_2,X_3]=X_1, [X_2,X_4]=-X_3, [X_3,X_4]=X_2$ & alm. nil. \\
\hline
\end{tabular}
\end{center}
\caption{Indecomposable non-nilpotent solvable unimodular algebras up to dimension $4$, that admit a lattice}
\end{table}
\end{center}

\begin{center}
\begin{table}[H]
\begin{center}
\begin{tabular}{|c|c|}
\hline
Name & Algebra \\
\hline
\hline
 $\Gg_{5.7}^{p,q,r}$ & $[X_1,X_5]=X_1, [X_2,X_5]=p X_2, [X_3,X_5]=q X_3, [X_4,X_5]=r X_4$, \\
 & $-1 \leq r\leq q \leq p \leq 1 \ , \ pqr \neq 0 \ , \ p+q+r+1=0$ \\
\hline
 $\Gg_{5.8}^{-1}$ & $[X_2,X_5]=X_1, [X_3,X_5]=X_3, [X_4,X_5]=-X_4$ \\
\hline
 $\Gg_{5.13}^{-1-2q,q,r}$ & $[X_1,X_5]=X_1, [X_2,X_5]=-(1+2q) X_2, [X_3,X_5]=q X_3 -r X_4, [X_4,X_5]=r X_3 +q X_4$, \\
 & $-1 \leq q \leq 0 \ , \ q \neq -\frac{1}{2} \ , \ r \neq 0$ \\
\hline
 $\Gg_{5.14}^{0}$ & $[X_2,X_5]=X_1, [X_3,X_5]=-X_4, [X_4,X_5]=X_3$ \\
\hline
 $\Gg_{5.15}^{-1}$ & $[X_1,X_5]=X_1, [X_2,X_5]=X_1 +X_2, [X_3,X_5]=-X_3, [X_4,X_5]=X_3 -X_4$ \\
\hline
 $\Gg_{5.17}^{p,-p,r}$ & $[X_1,X_5]=pX_1 -X_2, [X_2,X_5]=X_1 +pX_2, [X_3,X_5]=-pX_3 -rX_4, [X_4,X_5]=rX_3 - pX_4$, \\
 & $r \neq 0$ \\
\hline
 $\Gg_{5.18}^{0}$ & $[X_1,X_5]=-X_2, [X_2,X_5]=X_1, [X_3,X_5]=X_1-X_4, [X_4,X_5]=X_2+X_3$ \\
\hline
\end{tabular}
\end{center}
 \caption{Indecomposable solvable unimodular almost abelian algebras of dimension $5$, that admit a lattice}
\end{table}
\end{center}

\begin{center}
\begin{table}[H]
\begin{center}
\begin{tabular}{|c|c|}
\hline
Name & Algebra \\
\hline
\hline
 $\Gg_{6.3}^{0,-1}$ & $[X_2,X_6]=X_1, [X_3,X_6]=X_2, [X_4,X_6]=X_4, [X_5,X_6]=-X_5$ \\
\hline
 $\Gg_{6.10}^{0,0}$ & $[X_2,X_6]=X_1, [X_3,X_6]=X_2, [X_4,X_6]=-X_5, [X_5,X_6]=X_4$ \\
\hline
\end{tabular}
\end{center}
\caption{Indecomposable solvable unimodular almost abelian algebras of dimension $6$, for which we know a lattice exists}
\end{table}
\end{center}

\subsection{Algorithmic construction of the one-forms of a solvable group}\label{apA}

Let us consider a connected and simply-connected six-dimensional solvable group $G$. As a manifold, its tangent
bundle at the identity is given by $T_e G\approx\Gg$, and has a basis of vectors $E_{a}$ ($a=1 \dots 6)$ satisfying
\beq
[E_b, E_c]= f^a_{\ \ bc} E_a  \ .
\eeq
We will focus on the dual basis of one-forms $e^{a}$ on the cotangent bundle $g^*\approx T_e G^*$, which
verify the Maurer-Cartan equation
\beq
{\rm d}e^a =-\frac{1}{2} f^a_{\ \ bc} e^b\w e^c = - \sum_{b<c} f^a_{\ \ bc} \ e^b \wedge e^c \ .
\eeq
We want to consider a transformation  $A$ relating the one-forms of $\mathbb{R}^6$ to those of $G$:
\beq
A \ \left( \begin{array}{c} {\rm d}x^1  \\ \vdots  \\ {\rm d}x^6  \end{array} \right) = \left( \begin{array}{c} e^1  \\ \vdots  \\ e^6  \end{array} \right) \ . \label{const}
\eeq
Clearly the one-forms in \eqref{const} must satisfy the
corresponding\footnote{Whether the exterior derivative is defined on these new forms will not be treated (see Footnote \ref{footchev}): we will just define it as the exterior derivative of $\mathbb{R}^6$ acting on the left-hand side of (\ref{const}).} Maurer-Cartan equation.

The matrix $A$ should reproduce the different fibrations of the solvable group (the bundle structure is manifest in
the Maurer-Cartan equations).
Given the general form of solvable groups (a nilradical subgroup $N$ and an abelian left over subgroup $G/N=\mathbb{R}^{{\rm dim}\ G-{\rm dim}\ N}$), we will consider $A$ to be a product of two pieces:
\beq
A= \left( \begin{array}{c|c} A_N & 0 \\ \hline 0 & \mathbb{I}_{6-\textrm{dim}\ N}  \end{array} \right) \left( \begin{array}{c|c} A_M & 0 \\ \hline 0 & \mathbb{I}_{6-\textrm{dim}\ N}  \end{array} \right) \ ,
\eeq
where we take $A_M$ and $A_N$ to be $\textrm{dim}N \times \textrm{dim}N$ matrices, and we put the abelian directions
of $\mathbb{R}^{{\rm dim}\ G-{\rm dim}\ N}$ in the last entries. $A_M$ will provide the non-trivial fibration of
$N$ over $\mathbb{R}^{{\rm dim}\ G-{\rm dim}\ N}$, the Mostow bundle fibration of the solvmanifold for the compact
case, see Section \ref{susyback}. In turn,  $A_N$ will provide fibrations inside $N$, the fibrations within the
nilmanifold piece for the compact case. If the solvable group is nilpotent, then we take $A_M$ to be the identity.

To explicitly construct the matrices $A_M$ and $A_N$ we will now restrict ourselves to
$G=N$ (nilpotent) or $G=\mathbb{R} \ltimes_{\mu} N$ (almost nilpotent).

\subsubsection{Mostow bundle structure: $A_M$}\label{A-mostow}

We focus on the case of an almost nilpotent group. We identify the  $\mathbb{R}$ subalgebra with the direction
$x^6$. Then we take $\partial_t= \partial_6$ the basis for the $\mathbb{R}$ subalgebra,
and the corresponding one-form ${\rm d}x^6= {\rm d}t$. Then we define
\beq
\label{twistaa}
A_M=Ad_{e^{-t \partial_t}}(\Gn)=e^{-t\ ad_{\partial_t} (\Gn)} \ ,
\eeq
and
\beq
e^i = (A_M)^i_{\ k} \, \d x^k \, .
\eeq
Let us prove that this action will give forms which do verify the Maurer-Cartan equation.
Consider first the simpler case of an almost abelian group, i.e. with $N=\mathbb{R}^5$, which has $A_N=\mathbb{I}_N$. Then
\bea
 \label{apMC}
\d e^i&=& \d (e^{-t\ ad_{\partial_t}})^i_{\ \ k} \w \d x^k \nn\\
&=& -\d t \w (ad_{\partial_t} e^{-t\ ad_{\partial_t}})^i_{\ \ k} \d x^k \nn\\
&=& -\d t \w (ad_{\partial_t})^i_{\ \ j} (e^{-t\ ad_{\partial_t}})^j_{\ \ k} \d x^k \nn\\
&=& -\d t \w (ad_{\partial_t})^i_{\ \ j} e^j \nn\\
\d e^i &=& -f^i_{\ \ tj}\ \d t \w e^j \ .
\eea
The fact that we used the adjoint action allows to easily verify the Maurer-Cartan equations. \\

Expression \eqref{twistaa} for the matrix $A_M$ holds also for the more general case of almost nilpotent algebras.
In this case the Maurer-Cartan equations have component in  direction $\d t$ and also in the directions of the
nilradical. The $t$ dependence is always determined by $A_M$ and hence it is not modified by the presence of
a non-trivial nilradical. The form of the nilradical matrix, $A_N$, is given in the subsection below.

\subsubsection{Nilmanifold fibration structure: $A_N$}\label{A-nil}

The matrix $A_N$ should reproduce the iterated fibration structure of $N$. The iterated structure is related to the descending serie of $\Gn$ noted:
\beq
\Gn^{k=0 \dots p} \textrm{ with } \Gn^0=\Gn \ , \ \Gn^p=\{0\} \ .\nn
\eeq
Every $\Gn^k$ is an ideal of $\Gg$, so $\forall k\geq 1 \ ,\ \Gn^k=[\Gn,\Gn^{k-1}]\subset [\Gg,\Gn^{k-1}] \subset \Gn^{k-1}$. Let us now define another serie:
\beq
\textrm{For } 1 \leq k \leq p,\ s^k=\{E \in \Gn^{k-1} \ \textrm{with} \ E\notin \Gn^k\} \ .
\eeq
Let us prove some property of this serie. Assume that $\exists X \in s^p \bigcap s^q\ , \ p>q$ with $X\neq 0$. Then $X \in \Gn^{p-1} \subset \Gn^{p-2} \subset \dots \subset \Gn^q \subset \Gn^{q-1}$. So $X\in \Gn^{q-1}$ and $X\in \Gn^q$, so $X\notin s^q$, which is a contradiction. So $s^p \bigcap s^q=\{0\}$ for $p\neq q$. Furthermore, we always have $s^p=\Gn^{p-1}$. So $s^{p-1} \bigcup s^p= s^{p-1} \bigcup \Gn^{p-1}=\Gn^{p-2} \bigcup \Gn^{p-1} = \Gn^{p-2}$. Assume that $s^k \bigcup s^{k+1} \bigcup \dots \bigcup s^{p-1} \bigcup s^p=\Gn^{k-1}$. Then $s^{k-1} \bigcup s^k \bigcup \dots \bigcup s^{p-1} \bigcup s^p=s^{k-1} \bigcup \Gn^{k-1}=\Gn^{k-2} \bigcup \Gn^{k-1}=\Gn^{k-2}$. So by recurrence, we get that $\bigcup_{k=1 \dots p} s^k = \Gn$. In other words, each element of $\Gn$ appears in one and only one element of the serie $s^{\{k\}}$.

Let us give an example: consider the five-dimensional solvable algebra $(0,31,-21,23,24)$ (notations of Section \ref{susyback}). We have
\bea
\Gg=\{1,2,3,4,5 \} &,& \Gn=\{2,3,4,5 \} \ , \ \Gn^1=\{4,5 \} \ , \ \Gn^2=\{5 \}\ , \ \Gn^3=\{0 \} \nn\\
&& s^1=\{2,3 \} \ , \ s^2=\{4 \} \ , \ s^3=\{5 \} \ .\nn
\eea
The descending serie of $\Gn$ is known to be related to the fibration structure of the nilpotent group: each element gives a further fibration. Now we understand that the serie $s^{\{k\}}$ gives us what directions are fibered at each step. The correspondence between basis, fibers and series for a general iteration is given in the following diagram (of course it should be understood in terms of group elements instead of algebra elements as given here, see \cite{R}):
\beq
\begin{array}{ccc}
\fff^{p-1}=s^p & \hookrightarrow & \mmm^{p-1}=\Gn \nn\\
 & & \downarrow \nn\\
\fff^{p-2}=s^{p-1} & \hookrightarrow & \mmm^{p-2}=\bbb^{p-1} \nn\\
 & & \downarrow \nn\\
 & & \vdots \nn\\
 & & \downarrow \nn\\
\fff^2=s^3 & \hookrightarrow & \mmm^2=\bbb^3 \nn\\
 & & \downarrow \nn\\
\fff^1=s^2 & \hookrightarrow & \mmm^1=\bbb^2 \nn\\
 & & \downarrow \nn\\
 & & \bbb^1=s^1 \nn\\
\end{array}
\eeq
We see the unique decomposition of $\Gn$ into the serie $s^{\{k\}}$. We have $\bbb^i=\bigcup_{k=1 \dots i} s^k$ and $\fff^i=s^{i+1}$.\\

The matrix giving a single fibration was worked out in \cite{AMP}, we recover this result here. In the general case of an iteration, we consider a product of several operators, each of them giving one fibration of the iteration:
\bea
A_N=A_{p-1} \dots A_1 \ , \ A_i=e^{-\frac{1}{2} f_i}\qquad \textrm{(for $p=1$, $\Gn=\mathbb{R}^5$ and $A_N=1$)} \ , \nn
\eea
with $f_i\in \textrm{End}(\Gn)$:
\bea
\textrm{For}\ i=1\dots p-1\ , \ f_i : \Gn &\rightarrow& \Gn \nn\\
X &\mapsto& Y = ad_{\bbb^i} (X)\ \textrm{if}\ X\in \bbb^i \ \textrm{and}\ ad_{\bbb^i} (X) \in \fff^i \ , \nn\\
&& Y =0 \ \textrm{otherwise}\ . \label{apfi}
\eea
We choose to give a basis of $\Gn$ in the order given by $s^1, s^2, \dots, s^p$, and in each $s^k$ we can choose some order for the elements. Then in that basis, $f_i$, as a matrix, is an off-diagonal block with lines corresponding to $\fff^i=s^{i+1}$ and columns to $\bbb^i=\bigcup_{k=1 \dots i} s^k$. Then $A_i$ is the same plus the identity. Furthermore, the block depends on parameters $a^j$ of a generic element $a^j E_j$ of $\bbb^i$, and we have $ad_{a^j E_j \in \bbb^i}=a^j ad_{E_j \in \bbb^i}$. So for instance for the previous algebra, we get:
\beq
A_1= \left(\begin{array}{cccc} 1&0&0&0 \\ 0&1&0&0 \\ \frac{1}{2} a^3&-\frac{1}{2} a^2&1&0 \\ 0&0&0&1 \end{array}\right) \quad A_2= \left(\begin{array}{cccc} 1&0&0&0 \\ 0&1&0&0 \\ 0&0&1&0 \\ \frac{1}{2} a^4&0&-\frac{1}{2} a^2&1 \end{array}\right)\ .
\eeq
The parameters $a^j$ can be understood as a coordinate along $E_j$, so they are such that $\d a^j=e^j$, dual of $E_j$.\\

Let us prove that the operator $A_r$ gives the fibration of directions of $\fff^r$ over a base $\bbb^r$, and the correct corresponding Maurer-Cartan equation. As explained, an element of $A_r$ is given by:
\beq
(A_r)^i_{\ \ k}=\delta^i_k -\frac{1}{2} \sum_{j \in \bbb^r} a^j (ad_{E_j})^i_{\ \ k} \ \Theta(i\in \fff^r) \Theta(k\in \bbb^r)=\delta^i_k -\frac{1}{2} \sum_{j,k \in \bbb^r} a^j f^i_{\ \ jk} \ \Theta(i\in \fff^r) \ . \label{Ar}
\eeq
The forms on which we act with $A_r$ at the step $r$ of the iteration are labelled $e^k$, and they become after the operation $\tilde{e}^i$:
\beq
\tilde{e}^i=(A_r)^i_{\ \ k}\ e^k \ .
\eeq
The directions we fiber with $A_r$ are initially not fibered, so $e^{k\in \fff^r}=\d x^k$. All the other directions are not modified by $A_r$, so in particular $\tilde{e}^{i \in \bbb^r}=e^{i \in \bbb^r}$. So the Maurer-Cartan equations of the forms not in $\fff^r$ are not modified at this step. Their equation is then only modified at the step when they are fibered, so we don't have to consider it here. For the directions $\fff^r$, we get:
\bea
\tilde{e}^{i \in \fff^r}&=&e^{i \in \fff^r} -\frac{1}{2} \sum_{j,k \in \bbb^r} a^j f^i_{\ \ jk}\ e^{k}\nn\\
&=& \d x^{i} -\frac{1}{2} \sum_{j,k} a^j f^i_{\ \ jk}\ e^{k} \ ,\nn
\eea
where we dropped the restriction $j,k \in \bbb^r$ because due to the iterated structure, for $i\in \fff^r$, $f^i_{\ \ jk}=0$ if $k$ or $j \notin \bbb^r$. This operation then gives the fibration structure, since we can read the
connection. We can verify that we have the correct Maurer-Cartan equation:
\bea
{\rm d}\tilde{e}^{i \in \fff^r}&=& -\frac{1}{2} f^i_{\ \ jk}\ \d a^j \w e^{k} \nn\\
&=&-\frac{1}{2} f^i_{\ \ jk}\ e^j \w e^{k}\nn\\
{\rm d}\tilde{e}^i&=& -\frac{1}{2} f^i_{\ \ jk}\ \tilde{e}^j \w \tilde{e}^{k} \ .\nn
\eea

\subsection{Six-dimensional solvmanifolds in terms of globally defined one-forms}
\label{apglobdef}

In the following table we present the solvmanifolds that we considered in this paper.  They have the form
$G/\Gamma = H_1/\Gamma_1 \times H_2/\Gamma_2$, i.e. they are products of (at most) two solvmanifolds. Each of these
two solvmanifolds are constructed from the algebras given in the previous Tables (see Appendix \ref{aptables})
and the three-dimensional nilpotent algebra $\Gg_{3.1} : (-23, 0,0)$.
In particular, these are indecomposable solvable algebras for which the group admits a lattice.
The difference with respect
to Section \ref{aptables} is that the algebras are given here in terms of a basis of  globally defined forms
(see discussion in Section \ref{twistc}).
They are related by isomorphisms to the algebras given in the Tables of \ref{aptables}.
The fact the forms are globally defined is important for studying the compatibility of orientifold planes
with the manifold and for finding solutions. For $\Gg_{4.5}^{p,-p-1} \oplus \mathbb{R}^2$ and $\Gg_{4.6}^{-2p,p} \oplus \mathbb{R}^2$, we were not able to find such a basis, even if a priori we expect it to exist. \\

The column Name indicates the label of the  algebra and the corresponding solvmanifold. The column Algebra gives the
corresponding six-dimensional algebra in terms of exterior derivative acting on the dual basis of globally defined
one-forms (see Section \ref{susyback}). The next two columns give the O5 and O6 planes that
are compatible with the manifold. The column $Sp$ indicates by a $\checkmark$ when the manifold is symplectic,
according to \cite{B,CS}. Notice that the same results can be obtained as conditions for
the twisted pure spinors to solve the supersymmetry equations. In particular, for the even SU(3) pure spinor
$\Phi_+ = \frac{1}{8} e^{- i J}$
the condition
\beq
\d (O_{tw}) \Phi_+ = 0
\eeq
is equivalent to the requirement that the manifold is symplectic. \\

There is an additional subtlety for not completely solvable manifolds, when one looks for solutions on them. This is due to the
lack of isomorphism between the cohomology groups $H^{*}(\Gg)$ and $H_{dR}^{*}(G/\Gamma)$ for
not completely solvable manifolds (see Footnote \ref{footchev}).
In other words, the Betti numbers for the  Lie algebra cohomology give only
the lower bound for the corresponding numbers for de Rham cohomology. When looking for e.g.
symplectic manifolds,  we have considered only the forms in $H^{2}(\Gg)$, and hence might have missed some
candidate two-forms in $H_{dR}^{2}(G/\Gamma)$.

\newpage

\begin{landscape}

\begin{center}
\begin{table}[H]
\begin{center}
{\small
\begin{tabular}{|c|c|c|c|c|}
\hline
Name & Algebra & O5 & O6 & Sp \\
\hline
\hline
 $\Gg_{3.4}^{-1} \oplus \mathbb{R}^3$ & $(q_1 23 , q_2 13 , 0 , 0 , 0 , 0)$ $\quad q_1,q_2>0$ & 14, 15, 16, 24, 25,  & 123, 145, 146, 156, 245,  & $\checkmark$ \\
 & & 26, 34, 35, 36 & 246, 256, 345, 346, 356 & \\
\hline
 $\Gg_{3.5}^{0} \oplus \mathbb{R}^3$ & $(-23 , 13 , 0 , 0 , 0 , 0)$ & 14, 15, 16, 24, 25,  & 123, 145, 146, 156, 245,  & $\checkmark$ \\
 & & 26, 34, 35, 36 & 246, 256, 345, 346, 356 & \\
\hline
 $\Gg_{3.1} \oplus \Gg_{3.4}^{-1}$ & $(-23 , 0 , 0 , q_1 56 , q_2 46 , 0)$ $\quad q_1,q_2>0$ & 14, 15, 16, 24, 25,  & - & $\checkmark$ \\
 & & 26, 34, 35, 36 & & \\
\hline
 $\Gg_{3.1} \oplus \Gg_{3.5}^{0}$ & $(-23 , 0 , 0 , -56 , 46 , 0)$ & 14, 15, 16, 24, 25,  & - & $\checkmark$ \\
 & & 26, 34, 35, 36 & & \\
\hline
 $\Gg_{3.4}^{-1} \oplus \Gg_{3.5}^{0}$ & $(q_1 23 , q_2 13 , 0 , -56 , 46 , 0)$ $\quad q_1,q_2>0$ & 14, 15, 16, 24, 25,  & - & $\checkmark$ \\
 & & 26, 34, 35, 36 & & \\
\hline
 $\Gg_{3.4}^{-1} \oplus \Gg_{3.4}^{-1}$ & $(q_1 23 , q_2 13 , 0 , q_3 56 , q_4 46 , 0)$ $\quad q_1,q_2,q_3,q_4>0$ & 14, 15, 16, 24, 25,  & - & $\checkmark$ \\
 & & 26, 34, 35, 36 & & \\
\hline
 $\Gg_{3.5}^{0} \oplus \Gg_{3.5}^{0}$ & $(-23 , 13 , 0 , -56 , 46 , 0)$ & 14, 15, 16, 24, 25,  & - & $\checkmark$ \\
 & & 26, 34, 35, 36 & & \\
\hline
\hline
 $\Gg_{4.5}^{p,-p-1} \oplus \mathbb{R}^2$ & ? &  &  & - \\
\hline
 $\Gg_{4.6}^{-2p,p} \oplus \mathbb{R}^2$ & ? &  &  & - \\
\hline
 $\Gg_{4.8}^{-1} \oplus \mathbb{R}^2$ & $(-23 , q_1 34 , q_2 24 , 0 , 0 , 0)$ $\quad q_1,q_2>0$ & 14, 25, 26, 35, 36 & 145, 146, 256, 356 & - \\
\hline
 $\Gg_{4.9}^{0} \oplus \mathbb{R}^2$ & $(-23 , -34 , 24 , 0 , 0 , 0)$ & 14, 25, 26, 35, 36 & 145, 146, 256, 356 & - \\
\hline
\hline
 $\Gg_{5.7}^{1,-1,-1} \oplus \mathbb{R}$ & $(q_1 25 , q_2 15 , q_2 45 , q_1 35 , 0 , 0)$ $\quad q_1,q_2>0$ & 13, 14, 23, 24, 56 & 125, 136, 146, 236, 246, 345 & $\checkmark$ \\
\hline
 $\Gg_{5.8}^{-1} \oplus \mathbb{R}$ & $(25 , 0 , q_1 45 , q_2 35 , 0 , 0)$ $\quad q_1,q_2>0$ & 13, 14, 23, 24, 56 & 125, 136, 146, 236, 246, 345 & $\checkmark$ \\
\hline
 $\Gg_{5.13}^{-1,0,r} \oplus \mathbb{R}$ & $(q_1 25 , q_2 15 , -q_2 r 45 , q_1 r 35 , 0 , 0)$ $r \neq 0,\ q_1,q_2>0$ & 13, 14, 23, 24, 56 & 125, 136, 146, 236, 246, 345 & $\checkmark$ \\
\hline
 $\Gg_{5.14}^{0} \oplus \mathbb{R}$ & $(-25 , 0 , -45 , 35 , 0 , 0)$ & 13, 14, 23, 24, 56 & 125, 136, 146, 236, 246, 345 & $\checkmark$ \\
\hline
 $\Gg_{5.15}^{-1} \oplus \mathbb{R}$ & $(q_1 (25 -35) , q_2 (15-45) , q_2 45 , q_1 35 , 0 , 0)$ $\quad q_1,q_2>0$ & 14, 23, 56 & 146, 236 & $\checkmark$ \\
\hline
 $\Gg_{5.17}^{p,-p,r} \oplus \mathbb{R}$ & $(q_1(p 25+35) ,q_2(p 15+45) , q_2(p 45-15) , q_1(p 35-25) , 0 , 0)$ & 14, 23, 56 & 146, 236 & $\checkmark$ \\
 & $r^2=1,\ q_1,q_2>0$  & $p=0$: 12, 34 & $p=0$: 126, 135, 245, 346 &  \\
\hline
 $\Gg_{5.18}^{0} \oplus \mathbb{R}$ & $(-25 -35 , 15 -45 , -45 , 35 , 0 , 0)$ & 14, 23, 56 & 146, 236 & $\checkmark$ \\
\hline
\hline
 $\Gg_{6.3}^{0,-1}$ & $(-26 , -36 , 0 , q_1 56 , q_2 46 , 0)$ $\quad q_1,q_2>0$ & 24, 25 & 134, 135, 456 & $\checkmark$ \\
\hline
 $\Gg_{6.10}^{0,0}$ & $(-26 , -36 , 0 , -56 , 46 , 0)$ & 24, 25 & 134, 135, 456 & $\checkmark$ \\
\hline
\end{tabular}
}
\end{center}
\caption{Six-dimensional solvmanifolds considered in this paper, in terms of globally defined one-forms}
\end{table}
\end{center}

\end{landscape}

\newpage

\section{T--dualising solvmanifolds}
\label{T-dualsolv}

T--duality has been extensively used in flux compactifications in order to obtain solutions on nilmanifolds. Being iterations of torus bundles, these
are obtainable from torus solutions with an appropriate  $B$-field  (the contraction of $H$ with the isometry vectors  should be a closed
horizontal two-form that can be thought as a curvature of the dual torus bundle.). Correspondingly, the structure constants
$f^a_{\ \ bc}$ have also a T--duality friendly form. For any upper index there is a well-defined isometry vector
$\partial_a$ with respect to which one can perform an (un-obstructed) T--duality.

In this section we would like to study some aspects of T-duality for solvmanifolds.
In this case, the situation is more complicated. For instance,
it can happen that  the structure constants have the same index in the upper and lower position $f^a_{\ \ ac}$  and are
not fully antisymmetric. Put differently, most of our knowledge about the global aspects of T--duality comes form the study of its action on (iterations of) principal $U(1)$ bundles.  Since the Mostow bundles are not in general principal, the topology of the T--dual backgrounds is largely unexplored. We shall not attempt to do this here, but rather illustrate some of novel features by considering T--duality on the simplest cases of almost abelian manifolds.

Requiring that T--duality
preserves supersymmetry imposes that the Lie derivatives with respect to any isometry vector $v$ vanish,
$\mathcal{L}_{v} \Psi_{\pm}=0$ \cite{GMPW}.
For the simple case of almost abelian solvmanifolds,
it is not hard to check that all vectors $v_i = \partial_i$, where, in the basis chosen in this paper,  $i=1,...,4,6$, satisfy this condition.
However, these vectors are defined only locally\footnote{As discussed, on the compact solvmanifolds there exists a set of globally defined one forms $\{e\} = \{A_M \d x\}$ and the dual basis $\{E\} = \{(A_M^{-1})^T \partial \}$ is made of globally defined vectors. However, the Lie derivative of the pure spinors with respect to these does not vanish.}, since they transform non-trivially under $t \sim t+t_0$. Hence, in general, the result of
 T--duality will be non-geometric. We shall see that there are subtleties even for the case when the supersymmetry-preserving isometries $ \partial_i$ are well defined. \\

We shall consider the action of  T--duality on  two solvmanifolds,
$\Gg_{5.17}^{0,0,\pm 1} \times S^1 $ ($s\ 2.5$) and  $\Gg_{5.7}^{1,-1,-1} \times S^1$.
For $s \, 2.5$, following  \cite{GMPT6}, we write the algebra as $(25, -15, r 45, -r 35, 0, 0)$, $r^2=1$.
The twist matrix $A(t)$ is made of periodic functions of $t = x^5$,
\beq
\label{A-ncs}
A = \left( \begin{array}{ccc}
R_{r=1} & & \\
 & R_r & \\
 & & \mathbb{I}_2
\end{array} \right) \, ,  \quad  \quad R_r=\left( \begin{array}{cc}
\cos x^5 & -r \sin x^5 \\
r \sin x^5 & \cos x^5
\end{array} \right) \, ,
\eeq
and T--duality is un-obstructed. The various supersymmetric solutions found in \cite{GMPT6,A}
are all related by two  T--dualities

\bigskip\hspace{0.7cm}\centerline{\xymatrix@R=12pt@C=37pt{ & & \!\!\!\!\!\!\!\!\!\!\!\!\!\!\!\!\!\!\!\!\!\!\!\!\!\!\!\!\!\!\!\!\!\!\!\!\!\!\!\!\!\!\!\!\!\!\!\!\!\!\!\!\!\!\!\!\!\!\!\!\mathrm{IIB}& & & \!\!\!\!\!\!\!\!\!\!\!\!\!\!\!\!\!\!\!\!\!\!\!\!\!\!\!\!\!\!\!\!\!\!\!\!\!\!\!\!\!\!\!\!\!\!\!\!\!\!\!\!\!\!\!\!\!\!\!\!\!\!\!\!\!\!\!\!\!\!\!\!\!\!\!\!\!\!\!\!\!\!\!\!\!\!\!\!\!\!\!\!\!\!\!\!\!\!\!\!\!\!\!\!\!\!\!\!\!\!\!\!\!\!\!\!\!\!\!\!\!\!\!\!\!\!\!\!\!\!\mathrm{IIA} & \\
                            & \,\,\mathrm{t\!:\!30} & \!\!\!\!\!\!\!\!\!\!\!\mathrm{t\!:\!12} & \qquad\,\,\mathrm{t\!:\!30} & \,\,\mathrm{t\!:\!12} & \\
                           & \ar@{}[dr]|*+{\!\!\!\!\!\mathrm{T}_{12}} (13+24) \ar@{<->}[r]   & (14+23) \ar@{<->}[r] \ar@{}[dr]|*+{\!\!\mathrm{T}_{6}}\qquad  &\qquad \ar@{}[dr]|*+{\,\,\,\,\,\,\mathrm{T}_{12}} (136+246) \ar@{<->}[r]   & (146+236)\\           & (14+23) \ar@{<->}[r]   & (13+24)\ar@{<->}[r]\qquad  &\qquad  (146+236) \ar@{<->}[r]   &  (136+246)  }}\bigskip

In the table we labelled each solution by the dominant O-plane charge. The sources are labelled by their longitudinal directions,
e.g. $(13+24)$ stands for a solution with two sources (one O5 and one D5) along directions $e^1\w e^3$ and $e^2\w e^4$.
T--dualities (the subscripts indicate the directions in which they are performed) exchange the columns in the table;
lines are exchanged by relabellings (symmetries of the algebra).

The T--dualities are type\footnote{A pure spinor $\Psi$ can always be written as
\beq
\Psi = e^{B+ i j} \Omega_k \, ,
\eeq
where $\Omega_k$ is a holomorphic $k$-form, $B$ and $j$ are real two-forms. The degree of  $\Omega_k$ is the type of the pure spinor.}
changing, meaning  a pair of type $0$ and $3$ (t:30) pure spinors is exchanged with a pair of type $1$ and $2$ (t:12) and vice versa. \\

It is natural to see what will it be the effect of a single T-duality. To be precise we take as starting point Model 3 of
\cite{GMPT6}. We shall concentrate on the NS sector and discuss the topology changes under T--duality. The NS flux is zero
and the metric, in the $\d x^i$ basis is
\bea \label{g-t-g}
\d s^2 &=& \frac{t_1^2}{t_2} (\tau_2^1)^2 G (\d x^1+ \aaa \d x^2)^2 +
\frac{t_1}{G} (\d x^2)^2 + t_1 (\tau_2^1)^2  G (\d x^3+ r \aaa \d x^4)^2 \nn \\
&& + \frac{t_2 }{G} (\d x^4)^2 + t_3 (\d x^5)^2 + t_3 (\d x^6)^2
\eea
with
\beq
G=\cos^2 (x^5) + \frac{t_2}{t_1 (\tau_2^1)^2}  \sin^2 (x^5) \qquad \qquad  \ \aaa=
 \frac{t_2 - t_1 (\tau_2^1)^2}{2 G \, t_1 (\tau_2^1)^2 } \ \sin(2 x^5)   \, .
\eeq
A single T-duality along $x^1$ yields the manifold $T^3 \times \varepsilon_2$ ($\varepsilon_2 : (-23,13,0)$)
with O7-D5 (or D7-O5) and an $H$-flux given by
\beq\label{h-t-h}
H=- \d \aaa \w \d x^1 \w \d x^2 \ .
\eeq
Note that the $H$-flux \eqref{h-t-h} allows for topologically different choices of $B$-field. Being not
completely solvable (see Footnote \ref{footchev}), $s\ 2.5$ can yield manifolds of different topology (different
Betti numbers). Correspondingly, the results of T--duality should vary as well, and the application of the local Buscher rules
might be ambiguous. The choice of $B$-field in \eqref{h-t-h}, $ B=  - \aaa \d x^1 \w \d x^2$,
corresponding to the application of the local rules to
\eqref{g-t-g}, is globally defined due to $\aaa(x^5 + l) = \aaa(x^5)$. There is a less trivial choice
with $ B= -  x^1 \, \partial_5 \aaa \, \d x^2 \w \d x^5$ which however does not arise from the application of local T--duality rules to \eqref{g-t-g} since the metric does not have off-diagonal elements between $x^2$ and $x^5$.

A further T-duality along $x^2$ gives back $s\ 2.5$ with O6-D6 sources, but the supersymmetry now is captured by a different pair of pure spinors.  \\

For the manifold $\Gg_{5.7}^{1,-1,-1} \oplus \mathbb{R}$, the twist matrix  is
\beq
A(x^5) = \begin{pmatrix}
 R (x^5) & & \\
 & R (-x^5) & \\
 & & \mathbb{I}_2
\end{pmatrix}  \qquad \ R (x^5)=  \begin{pmatrix} {\rm ch} & -\eta {\rm sh} \\ -\frac{1}{\eta} {\rm sh} & {\rm ch} \end{pmatrix} \ ,
\eeq
where we set
\beq
{\rm ch}=\cosh(\sqrt{q_1 q_2} x^5) \ , \ {\rm sh}=\sinh(\sqrt{q_1 q_2} x^5) \ , \ \eta=\sqrt{\frac{q_1}{q_2}} \ .
\eeq
Then it is straightforward to check that the isometry vectors $v_i = \partial_i$ are local. Any  T--duality along these
is thus obstructed, and hence the O6-D6 solution
of \cite{CFI, GMPT6} does not have geometric T--duals. For this case we shall adopt the method applied
to nilmanifolds  in \cite{GMPW}, and  work out the action of T--duality  on the generalized vielbeine.

The generalized vielbeine on $\Gg_{5.7}^{1,-1,-1} \oplus \mathbb{R}$ can be obtained using twist transformation (see (\ref{twistactv})) from the generalized vielbeine of the torus (on which we take for simplicity the identity metric)
\beq
\mathcal{E}=\left( \begin{array}{c|c}
\mathbb{I}_6 & 0_6\\
\hline
0_6 & \mathbb{I}_6
\end{array} \right) \left( \begin{array}{c|c}
A & 0_6\\
\hline
0_6 & A^{-T}
\end{array} \right) \ .
\eeq
To work out their T--duals, we act by
\beq
\mathcal{E}_T= O_T \times \mathcal{E} \times O_T \ ,
\eeq
where $O_T$ is the O($d,d$) matrix for T--duality. The $O_T$ on the right is the regular action of T--duality, while the $O_T$
on the left assures that the map has no kernel (see \cite{GMPW}). The T--duality is done in the $x^1$ direction,
so the $O_T$ is
\beq
O_T=\left( \begin{array}{ccc|ccc}
T_1 & & & T_2 & & \\
 & \mathbb{I}_2 & & & 0_2 & \\
 & & \mathbb{I}_2 & & & 0_2 \\
\hline
T_2 & & & T_1 & & \\
 & 0_2 & & & \mathbb{I}_2 & \\
 & & 0_2 & & & \mathbb{I}_2
\end{array} \right) \ , \  T_1=\left( \begin{array}{cc}
0 & \\
 & 1
\end{array} \right) \ , \ T_2=\left( \begin{array}{cc}
1 & \\
 & 0
\end{array} \right) \ ,
\eeq
and then
\beq
\mathcal{E}_T=  \left( \begin{array}{ccc|ccc}
C_1  & & & B_1 & & \\
 & R(-x^5) & & & 0_2 & \\
 & & \mathbb{I}_2 & & & 0_2 \\
\hline
B_2  & & & C_2 & & \\
 & 0_2 & & & R(x^5)^T & \\
 & & 0_2 & & & \mathbb{I}_2
\end{array} \right) \ , \label{Td}
\eeq
with
\beq
C_1=C_2={\rm ch} \ \mathbb{I}_2 \ , \ B_1= -\frac{1}{\eta} {\rm sh} \ \epsilon \ , \ B_2= \eta {\rm sh} \ \epsilon \ , \ \epsilon=\left( \begin{array}{cc} 0 & -1 \\ 1 & 0 \end{array} \right) \ .
\eeq

The generalized vielbeine $\mathcal{E}_T$ can be brought to the canonical lower diagonal form (\ref{genviel}) by a
local $O(d)\times O(d)$ transformation. When such a transformation cannot be made single-valued, we talk about non-geometric
backgrounds (where the action of a non-trivial $\beta$ cannot be gauged away). The result of the $O(d)\times O(d)$ transformation
is
\beq
\mathcal{E}^\prime=\left( \begin{array}{ccc|ccc}
O_1 & & & O_2 & & \\
 & \mathbb{I}_2 & & & 0_2 & \\
 & & \mathbb{I}_2 & & & 0_2 \\
\hline
O_2 & & & O_1 & & \\
 & 0_2 & & & \mathbb{I}_2 & \\
 & & 0_2 & & & \mathbb{I}_2
\end{array} \right) \times \mathcal{E}_T = \left( \begin{array}{ccc|ccc}
O_1C_1+O_2B_2 & & & O_1B_1+O_2C_2 & & \\
 & R_2 & & & 0_2 & \\
 & & \mathbb{I}_2 & & & 0_2 \\
\hline
O_2C_1+O_1B_2 & & & O_2B_1+O_1C_2 & & \\
 & 0_2 & & & R_2^{-T} & \\
 & & 0_2 & & & \mathbb{I}_2
\end{array} \right) \ ,
\eeq
where the non-trivial $O(d)\times O(d)$ components are
\beq
O_{1/2}=\frac{1}{2} (O_+ \pm O_-) \qquad  \ O_{\pm} \in O(2) \ .
\eeq
By solving $O_1B_1+O_2C_2=0$,  we can obtain $O_2$ and express $O_{\pm}$  in terms of $O_1$:
\bea
&& O_{\pm}=O_1 (\mathbb{I}_2 \pm u \ \epsilon) \ , \,\, \ u=\frac{{\rm sh}}{\eta {\rm ch}} \ , \nn\\
&& O_{\pm}^T O_{\pm}= \mathbb{I}_2 \quad \Leftrightarrow \quad O_1^T O_1= \frac{1}{1+u^2} \ \mathbb{I}_2 \ .
\eea
A simple solution is given by
\beq
O_1=\frac{1}{\sqrt{1+u^2}} \ \mathbb{I}_2 \quad \Rightarrow\ \quad O_2=\frac{u}{\sqrt{1+u^2}} \ \epsilon \ .
\eeq
Thus we can indeed bring $\mathcal{E}_T$ to a lower-diagonal form, but with an $O(d)\times O(d)$ transformation that is not globally
defined. It is not hard to see that replacing the $x^1$ direction by others does not change much. Hence any
T--dual to $\Gg_{5.7}^{1,-1,-1} \times S^1$ is non-geometric. \\

A similar analysis for $s\ 2.5$ shows that one can easily solve the constraint  $O_1B_1+O_2C_2=0$
with $O_1$ and $O_2$ being globally defined (this is easy since the functions entering are all periodic).

\end{document}